\documentclass{aa}

\usepackage{color,graphicx}
\usepackage{txfonts}
\usepackage{natbib}
\usepackage{aalongtable}
\usepackage{lscape}
\usepackage{morefloats}
\newcommand\Lsun{{L$_{\sun}$}}
\newcommand\Msun{{M$_{\sun}$}}

\newcommand\micron{{$\mu$m}}

\newcommand\kms{{km~s$^{-1}$}}

\newcommand\cfluxcapfit[1]{Measured correlated flux spectra (solid
  black line and error bars) for #1, in order of increasing projected
  baseline.  The caption shows the projected baseline and position
  angle of each observation. The dashed blue line shows a continuum fit
  to the red points, and the solid blue line shows the best fit to the
  absorption spectrum (see Sec.~\ref{sec_absfit}).}

\newcommand\cfluxcapcont[1]{Measured correlated flux spectra for #1,
  in order of increasing projected baseline.  The caption shows the
  projected baseline and position angle of each observation.  The
  dashed blue line shows a continuum fit to the red points.}

\newcommand\cfluxcap[1]{Measured correlated flux spectra for #1, in
  order of increasing projected baseline.  The caption shows the
  projected baseline and position angle of each observation.}

\begin{document}

\title{The VLTI/MIDI survey of massive young stellar
  objects\thanks{Based on observations with the Very Large Telescope
    Interferometer of the European Southern Observatory.  The reduced
    interferometric data presented here are available in electronic
    form at the CDS via anonymous ftp to cdsarc.u-strasbg.fr
    (130.79.128.5) or via
    http://cdsweb.u-strasbg.fr/cgi-bin/qcat?J/A+A/.}}

\subtitle{Sounding the inner regions around intermediate- and
  high-mass young stars using mid-infrared interferometry}

\author{Paul A. Boley\inst{1,2}
  \and Hendrik Linz\inst{1}
  \and Roy van Boekel\inst{1}
  \and Thomas Henning\inst{1}
  \and Markus Feldt\inst{1}
  \and Lex Kaper\inst{3}
  \and Christoph Leinert\inst{1}
  \and Andr\'e M\"uller\inst{4}
  \and Ilaria Pascucci\inst{5}
  \and Massimo Robberto\inst{6}
  \and Bringfried Stecklum\inst{7}
  \and L.~B.~F.~M. Waters\inst{8,3}
  \and Hans Zinnecker\inst{9}
}

\institute{Max Planck Institute for Astronomy, K\"onigstuhl 17,
  Heidelberg 69117, Germany
  \and
  Max Planck Institute for Radio Astronomy, Auf dem H\"ugel 69,
  Bonn 53121, Germany
  \and
  Astronomical Institute Anton Pannekoek, Science Park 904, Amsterdam
  1098 XE, The Netherlands
  \and
  European Southern Observatory, Alonso de Cordova 3107, Vitacura,
  Santiago, Chile
  \and
  Lunar and Planetary Laboratory, University of Arizona, Tucson AZ 85721,
  United States
  \and
  Space Telescope Science Institute, 3700 San Martin Dr., Baltimore MD
  21212, United States
  \and
  Th\"uringer Landessternwarte, Sternwarte 5, Tautenburg 07778,
  Germany
  \and
  SRON Netherlands Institute for Space Research, Sorbonnelaan 2,
  Utrecht 3584 CA, The Netherlands
  \and
  SOFIA Science Center, NASA Ames Research Center, Mail Stop N232-12,
  Moffet Field CA 94035, United States
}

\date{\today}

\abstract{Because of inherent difficulties involved in observations
  and numerical simulations of the formation of massive stars, an
  understanding of the early evolutionary phases of these objects
  remains elusive.  In particular, observationally probing
  circumstellar material at distances $\la 100$~AU from the central
  star is exceedingly difficult, as such objects are rare (and thus,
  on average, far away) and typically deeply embedded.  Long-baseline
  mid-infrared interferometry provides one way of obtaining the
  necessary spatial resolution at appropriate wavelengths for studying
  this class of objects; however, interpreting such observations is
  often difficult due to sparse spatial-frequency coverage.} {We aim
  to characterize the distribution and composition of circumstellar
  material around young massive stars, and to investigate exactly
  which physical structures in these objects are probed by
  long-baseline mid-infrared interferometric observations.}{We used
  the two-telescope interferometric instrument MIDI of the Very Large
  Telescope Interferometer of the European Southern Observatory to
  observe a sample of 24 intermediate- and high-mass young stellar
  objects in the $N$ band (8-13~\micron).  We had successful fringe
  detections for 20 objects, and present spectrally-resolved
  correlated fluxes and visibility levels for projected baselines of
  up to 128~m.  We fit the visibilities with geometric models to
  derive the sizes of the emitting regions, as well as the orientation
  and elongation of the circumstellar material.  Fourteen objects in
  the sample show the 10~\micron{} silicate feature in absorption in
  the total and correlated flux spectra.  For 13 of these objects, we
  were able to fit the correlated flux spectra with a simple
  absorption model, allowing us to constrain the composition and
  absorptive properties of the circumstellar material.}{Nearly all of
  the massive young stellar objects observed show significant
  deviations from spherical symmetry at mid-infrared wavelengths.  In
  general, the mid-infrared emission can trace \emph{both} disks and
  outflows, and in many cases it may be difficult to disentangle these
  components on the basis of interferometric data alone, because of
  the sparse spatial frequency coverage normally provided by current
  long-baseline interferometers.  For the majority of the objects in
  this sample, the absorption occurs on spatial scales larger than
  those probed by MIDI.  Finally, the physical extent of the
  mid-infrared emission around these sources is correlated with the
  total luminosity, albeit with significant scatter.}{Circumstellar
  material is ubiquitous at distances $\la 100$~AU around young
  massive stars.  Long-baseline mid-infrared interferometry provides
  the resolving power necessary for observing this material directly.
  However, in particular for deeply-embedded sources, caution must be
  used when attempting to attribute mid-infrared emission to specific
  physical structures, such as a circumstellar disk or an outflow.}


\maketitle

\section{Introduction}

Circumstellar disks and outflows are essential components in, and
natural byproducts of, the process of star formation.  They are
commonly detected around low-mass T~Tauri stars and intermediate-mass
Herbig Ae/Be stars.  For ``massive'' ($\ga10$~\Msun) stars, extended
outflows have been detected around a number of massive young stellar
objects, or MYSOs \citep[e.g.][]{Mitchell91,deBuizer09}.  For a few
MYSO candidates, a $K$-band spectrum has been obtained
\citep{Hanson97,Hanson02,Blum04,Bik06}; these objects show a red
continuum, likely due to hot dust, and an emission-line spectrum
that includes Br$\gamma$ and CO 2.3~\micron{} bandhead emission.  The
latter can be modeled as being produced by a Keplerian rotating disk
\citep[e.g.][]{Bik04,Wheelwright10,Ilee13}; however, direct
observations of material in circumstellar disks around MYSOs are
complicated by the typically large distances ($\ga1$~kpc) to these
objects, as the required spatial resolution ($\la$~0\farcs1) is
difficult to achieve at infrared wavelengths, where thermal disk
emission is expected to dominate.

Mid-infrared interferometry provides an important tool for achieving
the high spatial resolution required to study the circumstellar
material around young stars.  The spatial scales probed by
interferometric measurements are given roughly by $\lambda / 2B$,
where $\lambda$ is the wavelength ($8-13$~\micron{} for the
observations presented here), and $B$ is the baseline between the
telescopes forming the interferometer projected on the sky (up to
128~m for the observations presented here).  For a typical projected
baseline of 50~m, this corresponds to angular scales of tens of
milliarcseconds at a wavelength of 10~\micron{}, or to linear
distances of tens of AU at typical distances to MYSOs ($\sim1$ to
several kpc).  Furthermore, due to the sharp cut-off of the Wien tail,
material emitting thermally in the $N$ band must be warmer than
roughly 200~K.

Even with state-of-the-art facilities like the Very Large Telescope
Interferometer (VLTI) on Cerro Paranal, performing interferometric
observations is a time-consuming and challenging process.
Consequently, obtaining sufficient $uv$ coverage to perform true
model-independent image reconstruction remains difficult at
near-infrared wavelengths, and effectively impossible at mid-infrared
wavelengths.  Practically speaking, $N$-band observations today are
limited to a handful of visibility amplitude measurements, with either
no or very little information available about the phase of the complex
visibility.  Consequently, interpreting such measurements is
challenging.

In the present work, we report the results of a campaign to observe
intermediate- and high-mass YSO candidates with the two-telescope
mid-infrared interferometric instrument MIDI \citep{Leinert03} on the
VLTI.  This survey was begun in 2004, and uses time within the
guaranteed time for observations (GTO) programs of the MIDI consortium
and the Max Planck Society.  We present long-baseline ($\sim5-130$~m),
spatially- and spectrally-resolved ($\lambda/\Delta \lambda \approx
35$) observations for these sources.  Where available, we combine
these long-baseline observations with the aperture-masking
measurements made with the Keck telescope and presented earlier by
\citet{Monnier09}.

We analyze these observations using two approaches.  First, we examine
the entire sample in terms of geometric models.  This approach makes
it possible to assess interferometric observations with sparse $uv$
coverage, and extract fundamental parameters (size, orientation,
elongation, brightness distribution) about the sources.  Second, for
selected sources, we also fit the silicate absorption feature seen in
the correlated flux spectra, which provides information about the
composition and distribution of absorbing material.

\section{Sample}

\begin{table*}
\begin{center}
\caption{Source properties}
\label{tab_sourceprops}
\begin{tabular}{l c c c c c c}
\hline \hline
Source & RA (J2000) & Dec (J2000) & Distance & $\log_{10} \frac{L}{L_\odot}$ & Outflow & Disk \\
       & (h:m:s)    & (d:m:s)     & (kpc)    &   & PA (deg) & PA (deg)       \\
\hline
\object{AFGL 2136} & 18:22:26.38 & -13:30:12.0 & $\sim2^{1}$ & $5.0^{2}$ & $\sim135^{3}$ & $ 46\pm   3^{4}$\\
\object{AFGL 4176} & 13:43:01.70 & -62:08:51.2 & $5.34^{5}$ & $5.5^{6}$ & &\\
\object{G305.20+0.21} & 13:11:10.45 & -62:34:38.6 & $4.84\pm1.70^{7}$ & & &\\
\object{IRAS 13481-6124} & 13:51:37.86 & -61:39:07.5 & $3.62^{5}$ & $4.5^{8}$ & $26\pm9^{8}$ & $120^{8}$\\
\object{IRAS 17216-3801} & 17:25:06.51 & -38:04:00.4 & $3.08\pm0.60^{7}$ & & &\\
\object{M8E-IR} & 18:04:53.18 & -24:26:41.4 & $\sim1.25^{9}$ & $3.8^{10}$ & $\sim140^{11}$ &\\
\object{M17 SW IRS1} & 18:20:19.48 & -16:13:29.8 & $1.98\pm0.13^{12}$ & $4.3^{13}$ & &\\
\object{M17 UC1} & 18:20:24.82 & -16:11:34.9 & $1.98\pm0.13^{12}$ & & &\\
\object{Mon R2 IRS2} & 06:07:45.80 & -06:22:53.5 & $0.83\pm0.05^{14}$ & $3.8^{15}$ & & $162^{16}$\\
\object{Mon R2 IRS3 A} & 06:07:47.84 & -06:22:56.2 & $0.83\pm0.05^{14}$ & $4.1^{15}$ & & $147^{16}$\\
\object{Mon R2 IRS3 B} & 06:07:47.86 & -06:22:55.4 & $0.83\pm0.05^{14}$ & & &\\
\object{MWC 300} & 18:29:25.69 & -06:04:37.2 & $1.8\pm0.2^{17}$ & $5.1\pm0.1^{17}$ & & $  4\pm   3^{18}$\\
\object{NGC 2264 IRS1} & 06:41:10.16 & +09:29:33.7 & $0.913\pm0.110^{19}$ & $3.60^{20}$ & $\sim353^{21}$ &\\
\object{Orion BN} & 05:35:14.10 & -05:22:22.9 & $0.414\pm0.007^{22}$ & $3.9-4.3^{23}$ & & $ 36^{24}$\\
\object{R CrA} & 19:01:53.65 & -36:57:07.6 & $0.138\pm0.016^{25}$ & & & $185\pm   5^{26}$\\
\object{R Mon} & 06:39:09.95 & +08:44:10.7 & $\sim0.8^{27}$ & $3.13^{27}$ & $\sim170^{28}$ &\\
\object{S255 IRS3} & 06:12:54.02 & +17:59:23.6 & $1.59\pm0.07^{29}$ & $4.7^{30}$ & $ 63^{31}$ &\\
\object{V921 Sco} & 16:59:06.78 & -42:42:08.3 & $1.15\pm0.15^{32}$ & $4.0\pm0.3^{32}$ & & $145.0\pm1.3^{33}$\\
\object{V1028 Cen} & 13:01:17.80 & -48:53:18.8 & $1.66\pm0.23^{34}$ & $3.22\pm0.08^{34}$ & &\\
\object{VY Mon} & 06:31:06.92 & +10:26:04.9 & $0.9\pm0.2^{35}$ & $3.07^{36}$ & &\\
\multicolumn{7}{l}{\it Sources without fringe detections:} \\
\object{GGD 27} & 18:19:12.08 & -20:47:30.9 & $1.9\pm0.8^{7}$ & $ 4.4^{37}$ & & \\
\object{IRAS 16164-5046} & 16:20:11.31 & -50:53:25.3 & $3.6^{38}$ & $ 5.6^{38}$ & & \\
\object{IRAS 17136-3617} & 17:17:02.29 & -36:21:08.2 & $2.0^{39}$ & $ 5.2^{39}$ & & \\
\object{Orion SC3} & 05:35:16.34 & -05:23:22.6 & $0.41\pm0.02^{40}$ & $ $ & & \\

\hline
\end{tabular}
\end{center}
\tablebib{(1)~\citet{Kastner92}; (2)~\citet{deWit11}; (3)~\citet{Kastner94}; (4)~\citet{Minchin91}; (5)~\citet{Fontani05}; (6)~\citet{Boley12}; (7)~This work; (8)~\citet{Kraus10}; (9)~\citet{Prisinzano05}; (10)~\citet{Linz09}; (11)~\citet{Mitchell91}; (12)~\citet{Xu11}; (13)~\citet{Follert10}; (14)~\citet{Herbst76}; (15)~\citet{Henning92}; (16)~\citet{Yao97}; (17)~\citet{Miroshnichenko04}; (18)~\citet{Wang12}; (19)~\citet{Baxter09}; (20)~\citet{Grellmann11}; (21)~\citet{Schreyer03}; (22)~\citet{Menten07}; (23)~\citet{deBuizer12}; (24)~\citet{Jiang05}; (25)~\citet{Ortiz10}; (26)~\citet{Kraus09}; (27)~\citet{Cohen84}; (28)~\citet{Brugel84}; (29)~\citet{Rygl10}; (30)~\citet{Longmore06}; (31)~\citet{Tamura91}; (32)~\citet{Borges07}; (33)~\citet{Kraus12}; (34)~\citet{Verhoeff12}; (35)~\citet{Casey90}; (36)~\citet{Henning98}; (37)~\citet{Aspin92}; (38)~\citet{BikPhD}; (39)~\citet{Tapia09}; (40)~\citet{Kraus09b}.}\end{table*}

The initial aim of our survey was to resolve MYSOs which are both
bright at mid-infrared wavelengths, and were unresolved in previous
imaging with 4--8~m-class telescopes.  In particular, the first target
selection focused largely on Becklin-Neugebauer (BN)-type objects
\citep[e.g.][]{Henning90}, including several massive disk candidates.
Later, we extended the sample to include lower-luminosity sources,
namely several early B-type stars with evidence for circumstellar
disks (e.g. Herbig Be stars).  Thus, the luminosities of the sample of
24 intermediate- to high-mass\footnote{Note, the mass of the central
  object in these systems is generally estimated from the luminosity,
  as direct measurement of the photospheric spectrum
  \citep[e.g.][]{Testi10} is very difficult.} YSOs presented here
range from $\sim10^3$ to $\sim10^5$~\Lsun{}.

In Table~\ref{tab_sourceprops}, we show the sources which make up the
target sample.  The coordinates shown are our best estimate of the
infrared position, and were generally taken from the 2MASS point
source catalog, when possible.  The last four sources shown in
Table~\ref{tab_sourceprops} are objects for which fringes were not
found, despite successful target acquisition by both telescopes; these
objects are discussed in more detail in Sec.~\ref{sec_nondetect}.

Where available, we have compiled values for distance and luminosity,
as well as disk and/or outflow orientation, collected from the
references listed.  We emphasize that these properties range from
being very well determined, to virtually unknown\footnote{For example,
  the distance to the Orion star-forming region, which hosts Orion~BN,
  has been measured very precisely by \citet{Menten07} using parallax
  measurements.  At the same time, the distance to R~Mon has never
  been measured, yet is generally quoted as being 800~pc, based on
  distance measurements to a star cluster which is \emph{one full
    degree} away from the source on the sky.  We regard such
  determinations as highly uncertain, but defer comments on specific
  sources to Sec.~\ref{sec_individual}.}.  The distances selected from
the literature and listed in Table~\ref{tab_sourceprops} represent, in
our judgment, the best estimates available for the objects in our
sample.  Furthermore, the luminosities listed here were taken from
works which adopt distances similar to those provided in the table.
For disk orientation, we list the position angle of the major axis,
and limit ourselves to near-infrared (generally interferometric and/or
polarimetric) detections.  For the outflow orientation, we use a
variety of signatures and wavelengths, ranging from optical detections
of Herbig-Haro objects to radio maps of dense gas.

For three sources where a distance estimate was not available, we used
the Milky Way rotation curve of \citet{Reid09} together with
radial velocity measurements from molecular line data.  Specifically,
for G305.20+0.21, \citet{Hindson10} found
$V_\mathrm{LSR}=-42.0$~\kms{} from observations of the NH$_3$(3,3)
line at 24~GHz, which corresponds to a kinematic distance of $4.84 \pm
1.70$~kpc, with no near/far distance ambiguity.  For IRAS~17216-3801,
however, no thermal molecular line data have been published to date.
Several studies list OH maser detections
\citep[e.g.][]{Cohen95,Argon00}, but the measured $V_\mathrm{LSR}$
velocities cover a range from $-18$ to $-24$~\kms{}.  We therefore
adopt a velocity of $-22.0$~\kms{}, based on recently-conducted APEX
measurements in the H$^{13}$CO$^+$(4--3) line (Linz et al., in
preparation).  This corresponds to a near distance of $3.1 \pm
0.6$~kpc, which we adopt in this work, and a far distance of $13 \pm
0.6$~kpc.  Finally, GGD~27 has a radial velocity of
$V_\mathrm{LSR}\approx12.0$~\kms{} \citep{Gomez03,Fernandez11}, which
corresponds to a near distance of $1.9 \pm 0.8$~kpc and a far distance
of $14.6 \pm 0.8$~kpc.

\section{Observations and data reduction}

\begin{figure*}
  \begin{center}
    \includegraphics[width=180mm]{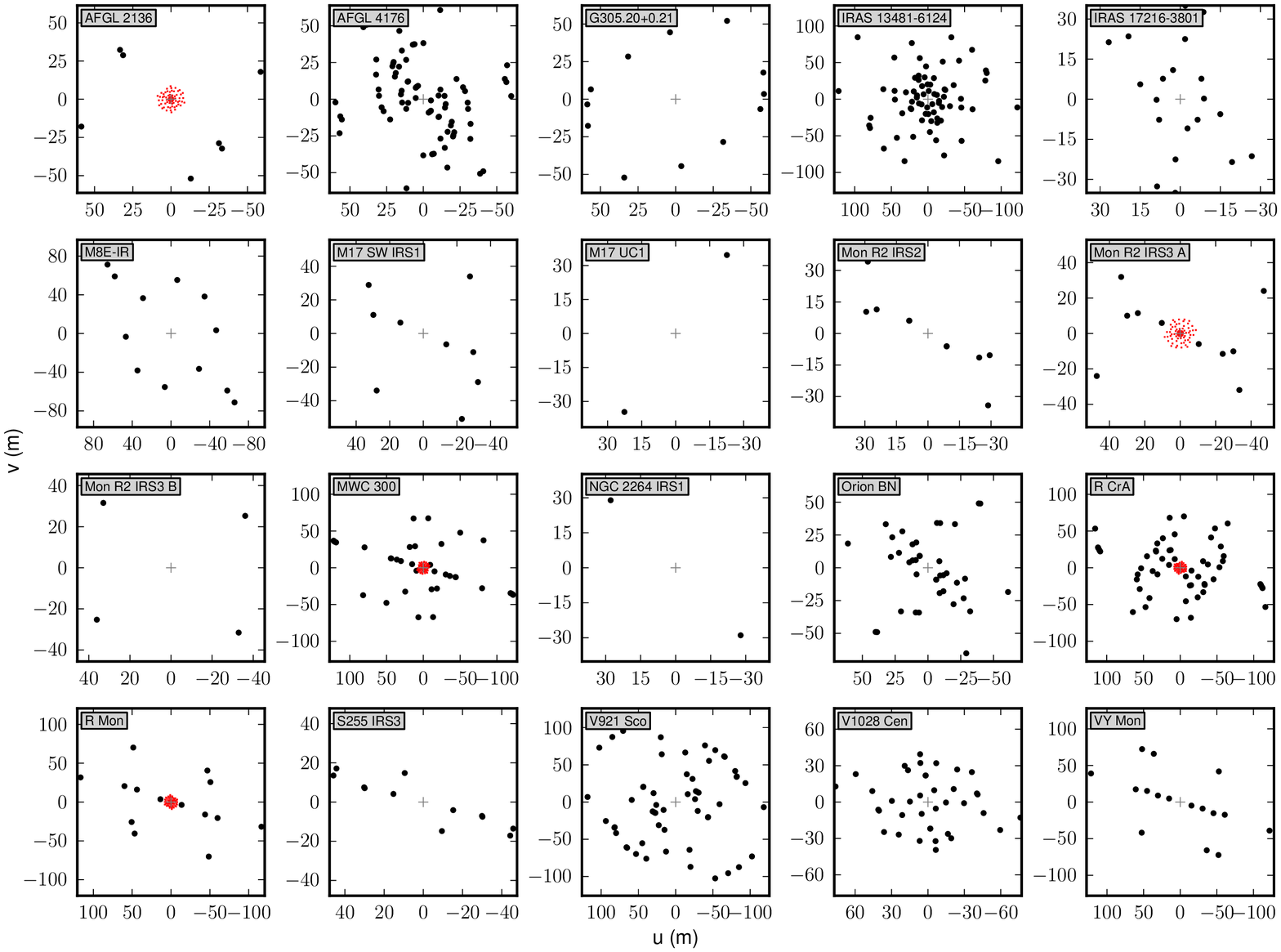}
    \caption{The $uv$ coverage for each source.  The observations from
      MIDI are shown in black, while the Keck data from
      \citet{Monnier09} are shown in red (only available for select
      sources).}
    \label{fig_uvplot}
  \end{center}
\end{figure*}

\begin{figure*}
  \begin{center}
    \includegraphics[width=180mm]{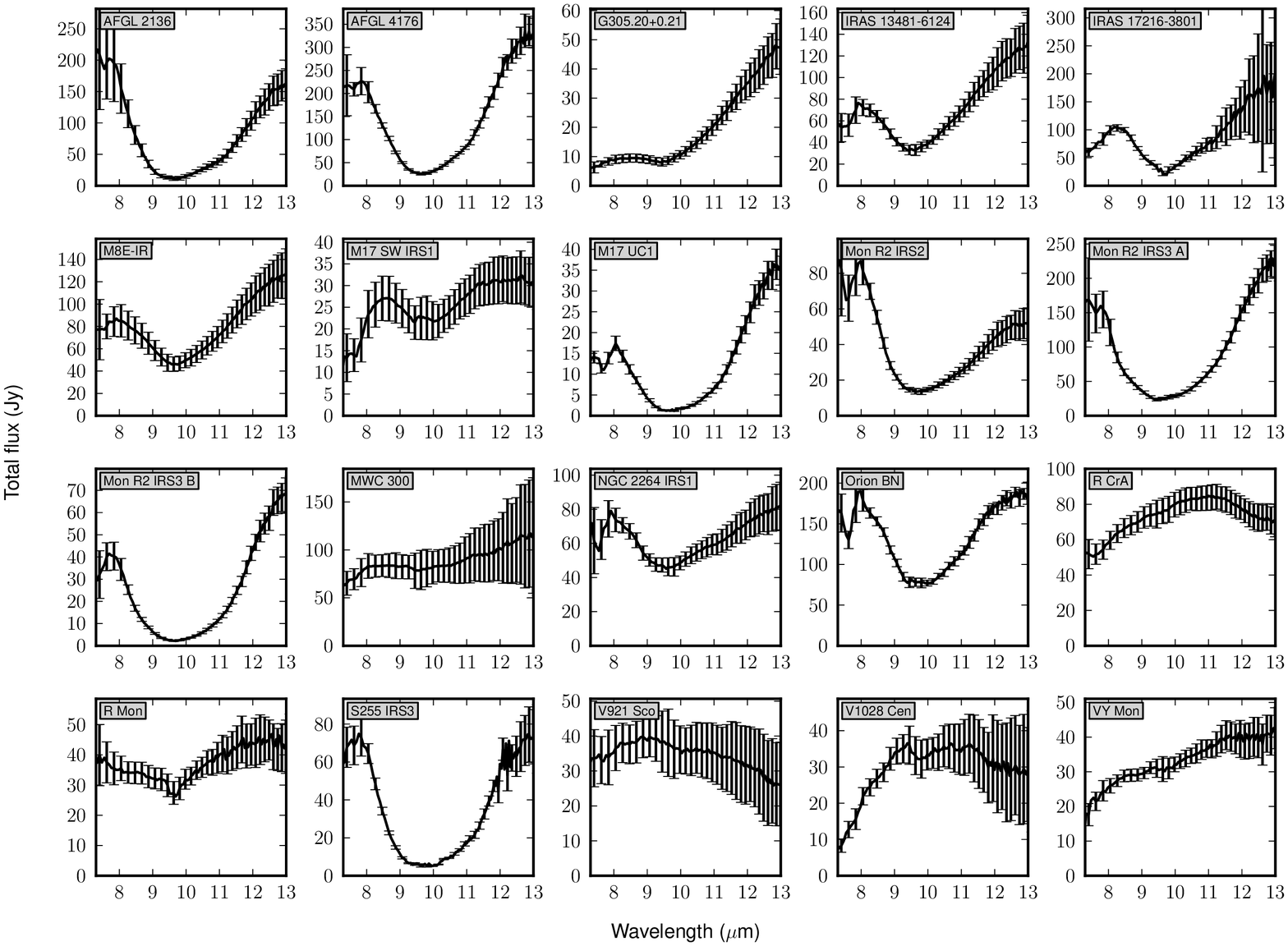}
    \caption{Total $N$-band spectra of the sources measured with
      MIDI.}
    \label{fig_spectra}
  \end{center}
\end{figure*}

\subsection{Long-baseline interferometry with VLTI}

Observations with the MIDI instrument at the VLTI facility of the
European Southern Observatory (ESO) on Cerro Paranal were performed in
the period 2004--2013, using both the 8.2~m Unit Telescopes (UTs) and
the 1.8~m Auxiliary Telescopes (ATs).  MIDI, operating at the VLTI
continuously since 2003, is a two-telescope interferometric
instrument, which is capable of measuring both total and
  correlated (interferometrically combined) flux, spectrally
dispersed over the $N$ band ($\sim8-13$~\micron{}).

The nominal observation procedure with MIDI has already been described
by \citet{Leinert04}, and we refer to that publication for a more
detailed discussion, but provide a brief review of the procedure here.
All data presented in this work used the prism as the dispersive
element ($\lambda/\Delta \lambda \approx 35$), and were carried out in
the HIGHSENS mode, meaning the correlated flux and photometric
measurements are made separately (as opposed to the SCIPHOT mode,
where a beam splitter is used to simultaneously measure the
interferometrically-combined and individual beams from each
telescope).  For the photometric measurements, light is passed, in
sequence, from each of the two telescopes used in the interferometric
measurement through the VLTI optical train to the MIDI instrument,
where it is spectrally dispersed and recorded on the detector.  Each
interferometric and photometric measurement was preceded and/or
followed by a calibration measurement of a bright single star of known
brightness and diameter.  Finally, we note that MIDI can also function
as a simple $N$-band imager, with acquisition images periodically
taken to check the beam position on the chip.  On the UTs, these
images have a pixel scale of 0\farcs084, with a field of view of
$\sim4\arcsec{}$, and can be AO-corrected in the presence of a
suitable optical guide star\footnote{Although regular acquisition
  images are not part of the standard ESO observing procedure with
  MIDI and are not discussed extensively in the present work, some
  results are discussed briefly in Sec.~\ref{sec_ggd27}.}.

We summarize the interferometric observations of the 24 objects in our
sample in Table~\ref{tab_obslog}, where we show the time the fringe
track was started, the telescopes/stations used, as well as the
projected baseline and position angle.  For four of the targets,
despite good weather conditions and no apparent technical problems, we
did not detect fringes; these non-detections are summarized at the end
of Table~\ref{tab_obslog} and discussed in Sec.~\ref{sec_nondetect}.
The $uv$ coverage for the 20 remaining targets is shown in
Fig.~\ref{fig_uvplot}, where black points indicate the MIDI
measurements.  Besides the observations conducted within the framework
of the present survey, we also include the following data previously
published by other investigators: one $uv$ point for AFGL~2136
\citep{deWit11}, and 6 points for R~CrA \citep{Correia08}.  Finally,
we note that observations of select sources, obtained within the
context of the survey presented here, have already been published by
\citet{Linz09,Follert10,Grellmann11,Boley12}.

We reduced all the raw MIDI data (including data published previously)
using version 2.0Beta1 (8 Nov. 2011) of the MIA+EWS package
\citep{Jaffe04}.  This new version of MIA+EWS includes significant
improvements for sources with low correlated flux.  In particular, we
make use of the \texttt{faintpipe} procedures to create masks from the
interference fringe pattern of all the calibrators.  For each night
and baseline configuration, we create a median mask from the
individual calibrator masks, which we use for extracting the
correlated flux and total spectra from the science measurements.
Possible decreases in flux due to slit loss were not taken into
account, as the observed sources are generally much smaller than the
MIDI slit width (2\farcs29 and 0\farcs52 for the ATs and UTs,
respectively).

Calibrated correlated fluxes (in Jy) were computed from the measured
instrumental values using all the calibrator measurements on the same
baseline configuration in a given night.  For each calibrator of known
diameter, taken from the van~Boekel database of mid-infrared
interferometric calibrators \citep{vanBoekelPhD}, we derive a
wavelength-dependent calibration function relating the measured
counts/s of correlated flux on the detector to Jy, and take the mean
over all calibrators to calibrate the science measurements.  For the
estimated uncertainty on the calibrated correlated flux, we include
both the statistical error derived by MIA+EWS (based on the variation
of the correlated flux over the duration of the fringe track) and the
variation of the individual transfer functions derived from the
calibrator measurements.

For each target, we created a median spectrum of the total $N$-band
flux from the individual photometric measurements with MIDI (again,
using all the calibrators on the same baseline configuration in a
given night), which we show in Fig.~\ref{fig_spectra}.  For sources
which were measured with both the UTs and ATs, we used only the
spectra observed with the UTs.  The uncertainty on the total (median)
spectra, shown as error bars in Fig.~\ref{fig_spectra}, is taken as
the standard deviation over the ensemble used in the median.

\subsection{Aperture-masking interferometry with Keck-1}

\begin{figure}
  \begin{center}
    \includegraphics[width=85mm]{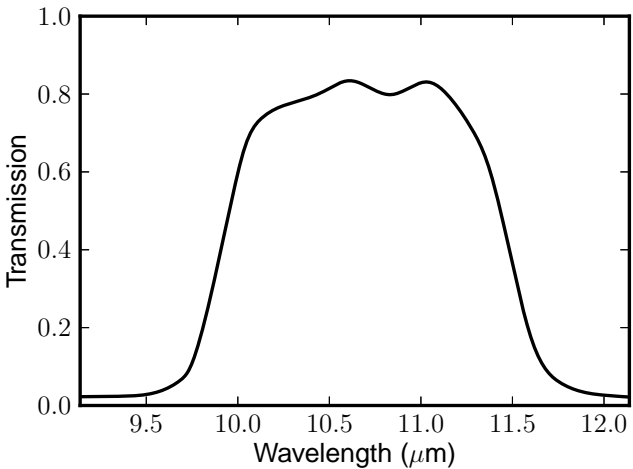}
    \caption{Transmission curve of the Keck 10.6~\micron{} filter.}
    \label{fig_keckfilt}
  \end{center}
\end{figure}

For five of the sources in our sample, we also make use of
previously-published aperture-masking interferometric observations
with the Keck-1 telescope as part of the segment-tilting experiment
by \citet{Monnier09}.  In contrast to the MIDI observations, these
visibility measurements are not spectrally resolved, but rather were
taken with a 10.6~\micron{} filter (transmission curve shown in
Fig.~\ref{fig_keckfilt}) which covers most of the $N$-band.  The $uv$
coverage provided is uniform, with projected baselines of 0.5--9~m.
We show the location of these measurements in $uv$ space as red points
in Fig.~\ref{fig_uvplot}.

\section{Results}

\subsection{Geometric fits to wavelength-averaged visibilities}
\label{sec_geofit}

The MIDI instrument provides spectrally-dispersed interferometric
visibilities over the entire $N$-band, at projected baselines which
range from about 5 to 130~m for the observations presented here.
However, in order to simplify both the analysis and the combination of
the VLTI and Keck observations (available for five of the 20 sources
presented here), we examine the complete data set at a single
wavelength of 10.6~\micron{}, which corresponds to the center of the
filter used for the Keck observations (Fig.~\ref{fig_keckfilt}).  In
order to directly compare the VLTI and Keck data, we define the mean
visibility in the filter as the mean of the correlated flux divided by
the mean of the total flux, where each mean is weighted by the filter
transmission curve.  The uncertainty on the correlated/total flux is
taken as the average of the uncertainty over the filter, weighted by
the correlated/total flux and the filter transmission.

Extensive $uv$ coverage remains difficult to achieve with long-baseline
mid-infrared interferometric observations.  In Fig.~\ref{fig_uvplot},
it can be seen that the coverage ranges anywhere from a single $uv$
point (e.g. M17~UC1) up to several dozen (e.g. IRAS~13481-6124).  For
some objects, radial coverage is quite good, while measurements at
multiple position angles are lacking (e.g. S255~IRS3).  For other
objects (e.g. M8E-IR), position angles are well sampled, but radial
coverage is deficient.

To deal with this variety of $uv$ coverage, for each target we fit up
to a maximum of four geometric models to the visibility levels at
10.6~\micron{}, depending on the distribution of observations in $uv$
space.  In order of increasing complexity, these models are: a
one-dimensional Gaussian (denoted ``1D'', parameterized only by the
FWHM of the Gaussian); a one-dimensional Gaussian plus an overresolved
component (``1DOR'', parameterized by the visibility $V(0)$ of the
overresolved component and the FWHM); a two-dimensional Gaussian plus
an overresolved component (``2DOR'', parameterized by $V(0)$, the FWHM
along the major axis, position angle $\phi$ and inclination angle
$\theta$); and a two-dimensional Gaussian plus a one-dimensional
Gaussian (``2D1D'', parameterized by the ratio of the integrated
fluxes $F_\mathrm{2D}/F_\mathrm{1D}$ of the two components, their
respective sizes FWHM$_\mathrm{2D}$ and FWHM$_\mathrm{1D}$, as well as
$\phi$ and $\theta$).  For the two-dimensional models, we constrain
the two components to be centered, meaning all the models are point
symmetric.  This is consistent with the low differential phases ($\la
10\degr$) measured for most targets/baselines; exceptions to this
(Mon~R2~IRS3~A and Orion~BN) are discussed in
Sec.~\ref{sec_individual}.

For the purposes of the present study, we chose not to consider
potential contributions from unresolved point sources, as at
mid-infrared wavelengths the level of photospheric emission in these
objects is expected to be negligible compared to that of the
circumstellar material.  Additionally, we note that the work of
\citet{Monnier09}, which examined a larger sample of intermediate- and
high-mass YSOs by means of mid-infrared aperture-masking
interferometry, did not find evidence for point-source contributions
in these objects.  Nevertheless, we note that a few objects in our
sample (e.g. V921~Sco, VY~Mon) still show visibilities $\ga10\%$ at
baselines longer than 100~m, meaning that the contribution from an
unresolved component in some sources may be on the order of 10--20\%.

\begin{table*}\tiny
\begin{center}
\caption{Parameters of one-dimensional geometric fits to 10.6~\micron{} visibilities}
\label{tab_fitpars_a}
\begin{tabular}{l l l l l l l}
\hline \hline
Source & \multicolumn{2}{c}{1D Gaussian} & & \multicolumn{3}{c}{1D Gaussian OR} \\
\cline{2-3} \cline {5-7}
 & \multicolumn{1}{c}{FWHM} & \multicolumn{1}{c}{$\chi^2_\mathrm{r}$} & & \multicolumn{1}{c}{$V(0)$} & \multicolumn{1}{c}{FWHM} & \multicolumn{1}{c}{$\chi^2_\mathrm{r}$}  \\
  & \multicolumn{1}{c}{(mas)} & & & & \multicolumn{1}{c}{(mas)} & \\
\hline
AFGL 2136 & $147._{-0.99}^{+1.0}$ &$5.80$& &$0.942_{-0.0064}^{+0.0063}$ & $140._{-1.4}^{+1.3}$ &$4.20$\\\\
AFGL 4176 & $78.9_{-1.2}^{+5.1}$ &$27.0$& &$0.366_{-0.050}^{+0.023}$ & $49.4_{-2.9}^{+1.8}$ &$17.0$\\\\
G305.20+0.21 & $50.5_{-0.90}^{+1.2}$ &$8.05$& &$0.0540_{-0.018}^{+0.033}$ & $25.3_{-5.1}^{+3.8}$ &$2.55$\\\\
IRAS 13481-6124 & $69.3_{-1.9}^{+2.2}$ &$19.5$& &$0.196_{-0.013}^{+0.015}$ & $16.0_{-1.3}^{+1.5}$ &$3.99$\\\\
IRAS 17216-3801 & $94.5_{-6.5}^{+7.6}$ &$7.91$& &$0.260_{-0.056}^{+0.11}$ & $34.5_{-9.5}^{+11.}$ &$4.18$\\\\
M17 SW IRS1 & $38.9_{-0.81}^{+0.97}$ &$2.29$& &$1.02_{-0.17}^{+0.17}$ & $39.1_{-1.5}^{+1.4}$ &$3.05$\\\\
M17 UC1 & $51.1_{-2.0}^{+3.2}$ &---& & & & \\\\
M8E-IR & $30.8_{-0.95}^{+1.1}$ &$8.94$& &$0.234_{-0.037}^{+0.046}$ & $14.1_{-1.9}^{+1.8}$ &$1.37$\\\\
Mon R2 IRS2 & $49.4_{-1.8}^{+1.9}$ &$7.08$& &$0.754_{-0.14}^{+0.29}$ & $43.4_{-5.7}^{+7.3}$ &$9.07$\\\\
Mon R2 IRS3 A & $164._{-0.75}^{+0.75}$ &$60.5$& &$0.786_{-0.0059}^{+0.0061}$ & $135._{-1.3}^{+1.3}$ &$35.0$\\\\
Mon R2 IRS3 B & $37.6_{-0.99}^{+1.2}$ &$0.0281$& & & & \\\\
MWC 300 & $49.5_{-1.6}^{+1.6}$ &$2.28$& &$1.00_{-0.0054}^{+0.0056}$ & $49.4_{-1.7}^{+1.8}$ &$2.32$\\\\
NGC 2264 IRS1 & $37.6_{-1.6}^{+2.0}$ &---& & & & \\\\
Orion BN & $127._{-4.0}^{+4.2}$ &$57.5$& &$0.146_{-0.014}^{+0.023}$ & $26.3_{-2.1}^{+2.9}$ &$9.32$\\\\
R CrA & $87.3_{-1.3}^{+1.3}$ &$15.1$& &$0.940_{-0.0087}^{+0.0080}$ & $75.4_{-3.1}^{+2.6}$ &$14.3$\\\\
R Mon & $105._{-1.8}^{+1.8}$ &$3.84$& &$0.984_{-0.0076}^{+0.0074}$ & $102._{-2.3}^{+2.2}$ &$3.82$\\\\
S255 IRS3 & $75.8_{-4.5}^{+5.0}$ &$10.9$& &$0.258_{-0.044}^{+0.070}$ & $30.0_{-4.3}^{+5.6}$ &$2.76$\\\\
V921 Sco & $16.0_{-1.2}^{+0.32}$ &$4.77$& &$0.744_{-0.082}^{+0.075}$ & $12.9_{-1.1}^{+0.78}$ &$2.37$\\\\
V1028 Cen & $41.8_{-0.81}^{+1.7}$ &$45.5$& &$0.272_{-0.031}^{+0.033}$ & $9.67_{-3.7}^{+4.2}$ &$24.0$\\\\
VY Mon & $19.5_{-0.33}^{+0.53}$ &$56.5$& &$0.319_{-0.031}^{+0.24}$ & $9.46_{-0.87}^{+4.9}$ &$33.1$\\

\hline
\end{tabular}
\end{center}
\end{table*}

\begin{table*}\tiny
\begin{center}
\caption{Parameters of two-dimensional geometric fits to 10.6~\micron{} visibilities}
\label{tab_fitpars_b}
\begin{tabular}{l l l l l l l l l l l l l}
\hline \hline
Source & \multicolumn{5}{c}{2D Gaussian OR} & & \multicolumn{6}{c}{2D Gaussian + 1D Gaussian} \\
\cline{2-6} \cline{8-13}
 & \multicolumn{1}{c}{$V(0)$} & \multicolumn{1}{c}{FWHM} & \multicolumn{1}{c}{$\phi$} & \multicolumn{1}{c}{$\theta$} & \multicolumn{1}{c}{$\chi^2_\mathrm{r}$} & & \multicolumn{1}{c}{$F_\mathrm{2D}/F_\mathrm{1D}$} & \multicolumn{1}{c}{FWHM$_\mathrm{2D}$} & \multicolumn{1}{c}{FWHM$_\mathrm{1D}$} & \multicolumn{1}{c}{$\phi$} & \multicolumn{1}{c}{$\theta$} & \multicolumn{1}{c}{$\chi^2_\mathrm{r}$} \\
  & & \multicolumn{1}{c}{(mas)} & \multicolumn{1}{c}{(deg)} & \multicolumn{1}{c}{(deg)} & & & & \multicolumn{1}{c}{(mas)} & \multicolumn{1}{c}{(mas)} & \multicolumn{1}{c}{(deg)} & \multicolumn{1}{c}{(deg)} & \\
\hline
AFGL 2136 & $0.944_{-0.0064}^{+0.0064}$ & $147._{-1.9}^{+2.1}$ & $53.7_{-6.2}^{+6.5}$ & $25.4_{-2.5}^{+2.4}$ &$3.83$& &$0.707_{-0.043}^{+0.051}$ & $43.9_{-1.5}^{+5.0}$ & $244._{-7.8}^{+8.8}$ & $37.7_{-15.}^{+23.}$ & $46.2_{-8.1}^{+12.}$ & $0.839$\\\\
AFGL 4176 & $0.396_{-0.035}^{+0.018}$ & $60.3_{-1.7}^{+1.3}$ & $110._{-2.3}^{+2.1}$ & $54.7_{-1.9}^{+1.2}$ &$9.30$& &$0.212_{-0.029}^{+0.21}$ & $46.3_{-6.3}^{+5.6}$ & $120._{-6.2}^{+30.}$ & $120._{-8.7}^{+3.0}$ & $63.5_{-8.4}^{+3.0}$ & $6.70$\\\\
G305.20+0.21 & $0.238_{-0.14}^{+0.42}$ & $42.2_{-8.5}^{+7.1}$ & $179._{-25.}^{+15.}$ & $34.2_{-6.0}^{+11.}$ &$3.65$& & & & & & & \\\\
IRAS 13481-6124 & $0.195_{-0.014}^{+0.017}$ & $16.6_{-1.3}^{+2.3}$ & $98.6_{-35.}^{+25.}$ & $27.5_{-6.6}^{+13.}$ &$4.19$& &$0.174_{-0.013}^{+0.017}$ & $13.7_{-1.0}^{+1.4}$ & $104._{-4.8}^{+6.4}$ & $112._{-18.}^{+16.}$ & $43.0_{-9.5}^{+9.5}$ & $1.72$\\\\
IRAS 17216-3801 & $0.522_{-0.18}^{+0.12}$ & $84.8_{-33.}^{+20.}$ & $125._{-9.7}^{+40.}$ & $59.1_{-11.}^{+7.1}$ &$6.69$& & & & & & & \\\\
M17 SW IRS1 & $0.727_{-0.10}^{+0.20}$ & $43.1_{-2.2}^{+13.}$ & $171._{-15.}^{+7.5}$ & $64.7_{-15.}^{+7.7}$ &$1.38$& & & & & & & \\\\
M8E-IR & $0.308_{-0.066}^{+0.10}$ & $21.7_{-4.6}^{+4.9}$ & $118._{-22.}^{+25.}$ & $45.7_{-9.8}^{+9.8}$ &$2.07$& & & & & & & \\\\
Mon R2 IRS3 A & $0.824_{-0.0060}^{+0.0060}$ & $173._{-1.8}^{+1.8}$ & $43.3_{-0.88}^{+0.90}$ & $50.0_{-0.79}^{+0.77}$ &$19.5$& &$1.43_{-0.064}^{+0.088}$ & $128._{-3.3}^{+3.7}$ & $512._{-16.}^{+22.}$ & $40.7_{-1.3}^{+1.3}$ & $55.4_{-1.5}^{+1.4}$ & $6.00$\\\\
MWC 300 & $0.990_{-0.0056}^{+0.0054}$ & $52.0_{-2.0}^{+2.6}$ & $78.4_{-12.}^{+6.5}$ & $60.7_{-2.4}^{+4.9}$ &$1.72$& &$0.130_{-0.020}^{+0.030}$ & $12.7_{-2.4}^{+2.3}$ & $56.0_{-1.6}^{+2.1}$ & $21.5_{-13.}^{+15.}$ & $67.6_{-13.}^{+8.8}$ & $0.971$\\\\
Orion BN & $0.209_{-0.012}^{+0.013}$ & $39.8_{-1.5}^{+1.3}$ & $13.8_{-1.4}^{+1.9}$ & $72.9_{-5.7}^{+4.9}$ &$6.14$& &$0.264_{-0.019}^{+0.021}$ & $39.8_{-3.2}^{+1.3}$ & $745._{-84.}^{+76.}$ & $13.8_{-1.4}^{+1.9}$ & $72.9_{-5.7}^{+4.9}$ & $6.54$\\\\
R CrA & $0.865_{-0.0053}^{+0.0057}$ & $53.5_{-2.9}^{+3.5}$ & $31.9_{-1.1}^{+1.1}$ & $68.8_{-1.9}^{+1.9}$ &$11.9$& &$0.348_{-0.026}^{+0.030}$ & $12.8_{-1.1}^{+1.1}$ & $107._{-2.1}^{+2.3}$ & $176._{-12.}^{+9.4}$ & $44.6_{-7.2}^{+6.1}$ & $2.34$\\\\
R Mon & $0.984_{-0.0075}^{+0.0075}$ & $102._{-1.7}^{+4.4}$ & $142._{-47.}^{+49.}$ & $7.15_{-1.1}^{+14.}$ &$3.97$& &$1.27_{-0.36}^{+0.25}$ & $25.7_{-3.5}^{+1.5}$ & $192._{-26.}^{+22.}$ & $86.3_{-9.0}^{+10.}$ & $63.3_{-11.}^{+8.8}$ & $0.388$\\\\
S255 IRS3 & $0.316_{-0.048}^{+0.085}$ & $92.8_{-24.}^{+22.}$ & $149._{-8.4}^{+7.4}$ & $79.6_{-8.1}^{+2.4}$ &$8.81$& & & & & & & \\\\
V921 Sco & $0.776_{-0.067}^{+0.090}$ & $15.5_{-1.2}^{+1.1}$ & $162._{-9.8}^{+8.1}$ & $48.7_{-8.9}^{+5.2}$ &$0.868$& &$2.94_{-1.3}^{+0.56}$ & $15.2_{-2.5}^{+0.86}$ & $50.5_{-16.}^{+16.}$ & $162._{-14.}^{+14.}$ & $49.1_{-11.}^{+7.7}$ & $0.828$\\\\
V1028 Cen & $0.574_{-0.061}^{+0.061}$ & $39.0_{-2.8}^{+2.4}$ & $172._{-3.4}^{+4.1}$ & $67.0_{-2.7}^{+3.1}$ &$5.71$& &$0.721_{-0.13}^{+0.36}$ & $33.0_{-2.9}^{+4.0}$ & $88.7_{-12.}^{+41.}$ & $174._{-4.5}^{+4.7}$ & $69.7_{-3.4}^{+5.3}$ & $2.74$\\\\
VY Mon & $0.567_{-0.039}^{+0.051}$ & $18.9_{-1.0}^{+1.3}$ & $180._{-5.7}^{+5.5}$ & $59.3_{-2.6}^{+3.2}$ &$6.08$& &$0.843_{-0.22}^{+0.27}$ & $16.8_{-1.8}^{+1.6}$ & $42.8_{-7.4}^{+19.}$ & $4.90_{-7.5}^{+7.5}$ & $61.3_{-3.5}^{+7.5}$ & $0.244$\\

\hline
\end{tabular}
\end{center}
\end{table*}

We show the best-fit parameters for each of these models in
Tables~\ref{tab_fitpars_a} (one-dimensional models) and
\ref{tab_fitpars_b} (two-dimensional models).  The best-fit models are
found using a grid search over parameter space, followed by a
downhill-simplex $\chi^2$-minimization routine.  For each fit, we
report the value of $\chi^2_\mathrm{r}$, i.e. the goodness-of-fit
parameter divided by the number of degrees of freedom of the fit.
Blank rows in Tables~\ref{tab_fitpars_a}-\ref{tab_fitpars_b} indicate
cases where either the number of model parameters exceeds the number
of observed $uv$ points, or the fit did not converge (e.g. due to way
the observations are distributed in $uv$ space).

\begin{figure*}
  \begin{center}
    \includegraphics[width=180mm]{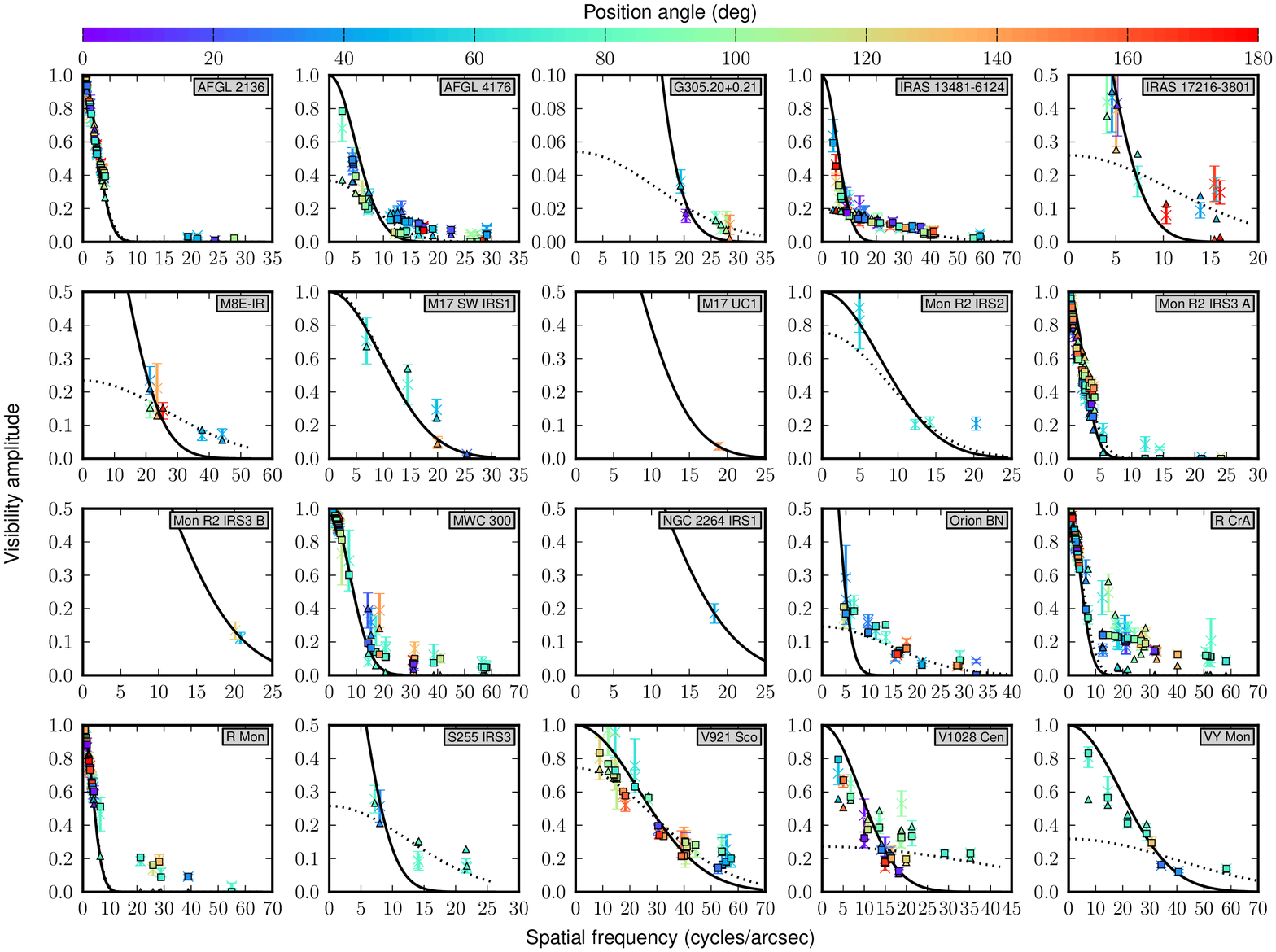}
    \caption{Best-fit geometric models to the 10.6~\micron{}
      visibilities.  The observations are shown as crosses with error
      bars; the color corresponds to the position angle of the
      measurement, shown in the color bar at the top of the plot.  The
      1D model is shown as a solid line, and the 1DOR model is shown
      as a dotted line.  The triangles show the 2DOR model, while the
      squares show the 2D1D model.  For a description of the different
      models, see Sec.~\ref{sec_geofit}.}
    \label{fig_visfit}
  \end{center}
\end{figure*}

\begin{figure*}
  \begin{center}
    \includegraphics[width=180mm]{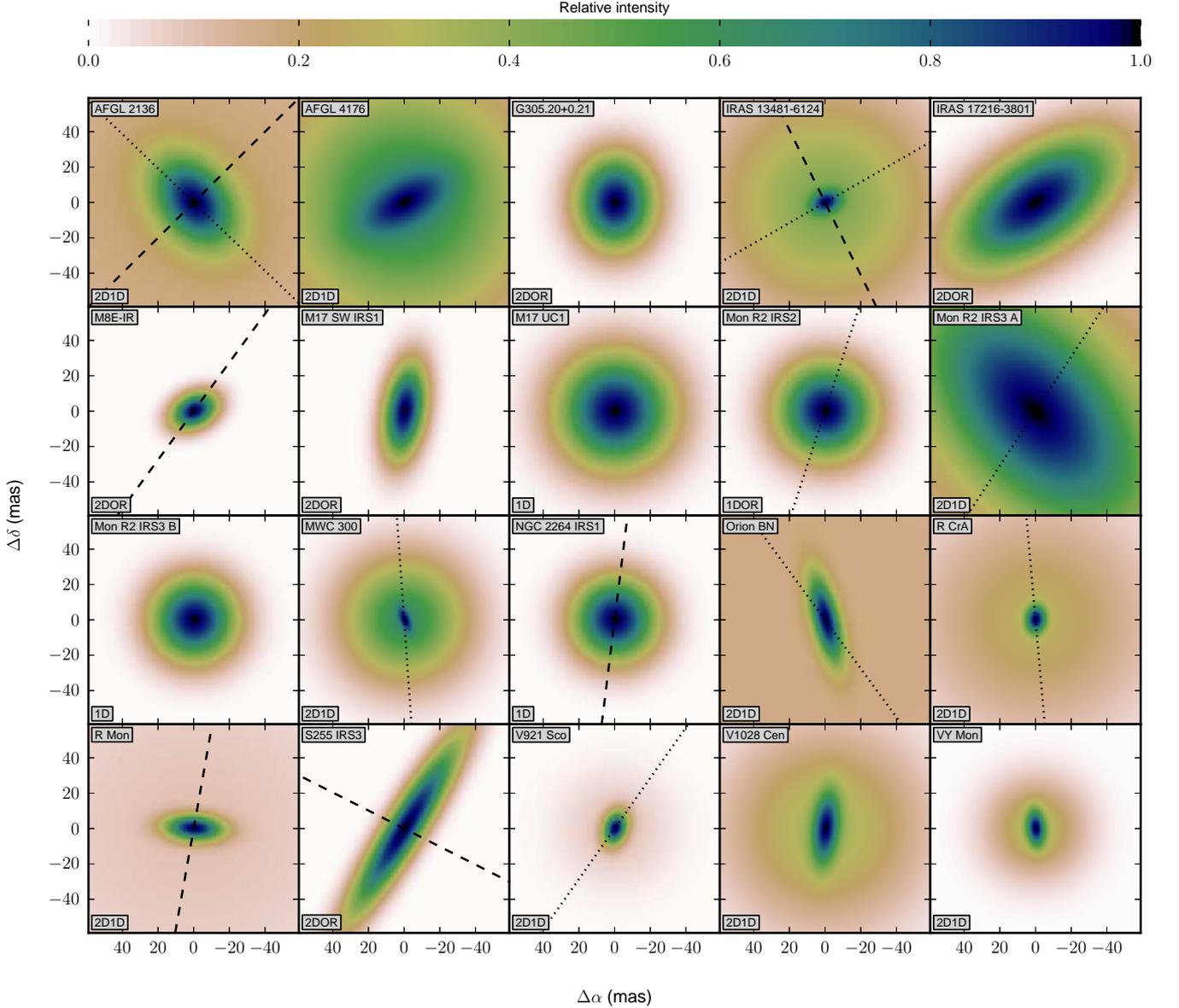}
    \caption{The intensity distributions of the best-fit models
      described in Sec.~\ref{sec_geofit}.  The model parameters are
      shown in Tables~\ref{tab_fitpars_a} and \ref{tab_fitpars_b}.
      Where available, we have indicated outflow and/or disk
      orientations as dashed and dotted lines, respectively (see
      Table~\ref{tab_sourceprops} for disk/outflow directions and
      references).}
    \label{fig_images}
  \end{center}
\end{figure*}

\begin{figure*}
  \begin{center}
    \includegraphics[width=180mm]{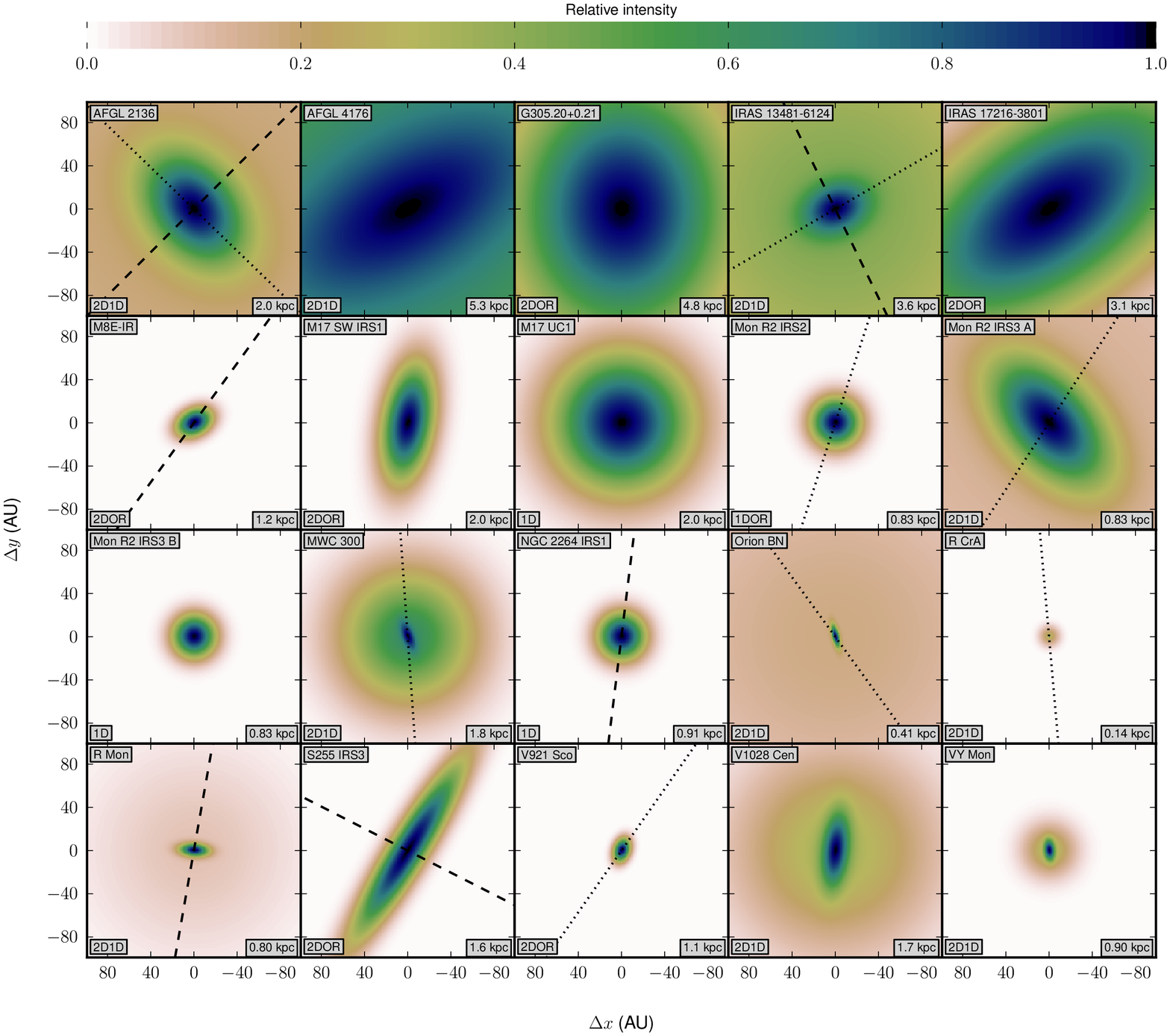}
    \caption{Same as Fig.~\ref{fig_images}, but the images have been
      scaled to linear distances in AU using the distances from
      Table~\ref{tab_sourceprops}.}
    \label{fig_images_au}
  \end{center}
\end{figure*}

The uncertainties on the fitting parameters are derived using a
Monte-Carlo approach, where we generate synthetic observations,
normally distributed about the mean observed visibility level with a
standard deviation equal to the measurement uncertainty, and repeat
the optimization process.  In Fig.~\ref{fig_visfit}, we show the
observed 10.6~\micron{} visibility levels as a function of spatial
frequency for each source, together with each best-fit model.  The
image intensity distributions of the best-fit models are shown in
Fig.~\ref{fig_images} (spatial scale in mas) and
Fig.~\ref{fig_images_au} (spatial scale in AU, using the distances
from Table~\ref{tab_sourceprops}).

\subsection{Characterization of the spectrally-dispersed correlated
  flux levels}
\label{sec_cflux}

\begin{figure*}
  \begin{center}
    \includegraphics[scale=0.85,bb=0 386 570 793]{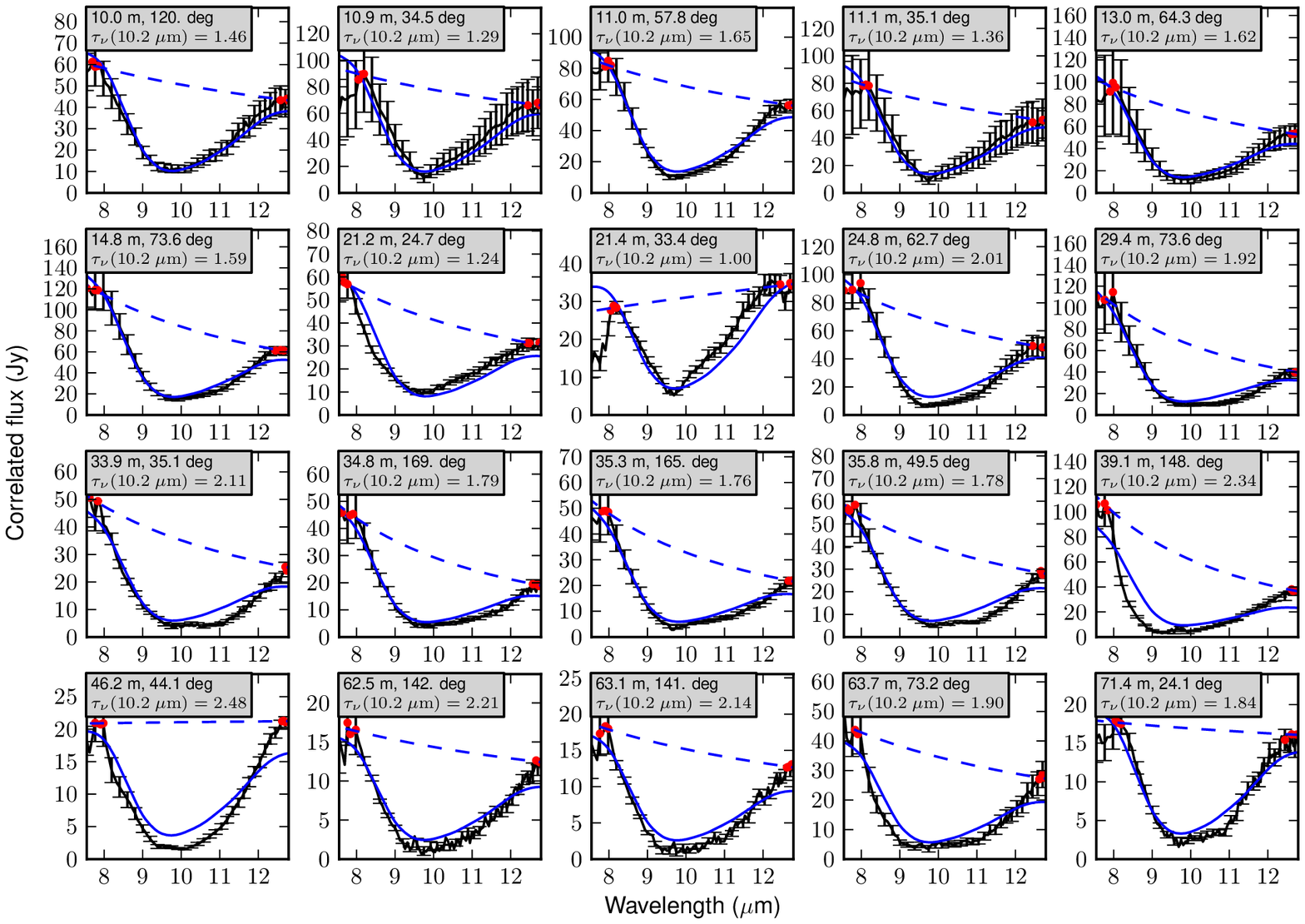}
    \caption{\cfluxcapfit{Orion BN}}
    \label{fig_cflux_orionbn}
  \end{center}
\end{figure*}

Besides the wavelength-averaged visibility levels, the MIDI
measurements can also be considered in terms of their spectral
behavior.  In this section, rather than focusing on the
visibility amplitude (i.e., the ratio of the correlated flux to the
total flux), we consider the correlated flux measured in each
observation.

We show the correlated flux as a function of wavelength for each point
in $uv$ space for one source, Orion BN, in
Fig.~\ref{fig_cflux_orionbn}.  The remaining objects are presented in
Figs.~\ref{fig_cflux_afgl2136}--\ref{fig_cflux_vymon} of the online
version of the journal.  In general, we note that sources which show
silicate absorption in their total $N$-band spectra
(Fig.~\ref{fig_spectra}) also show absorption in the correlated flux.
Qualitatively, this is to be expected, as foreground material (which
may include an extended envelope, in addition to the diffuse ISM) will
indiscriminately absorb emission from all spatial scales.  However,
departures from this may be observed if the absorption occurs on
scales resolved by MIDI.  Additionally, the contribution of emission
to the total and (spatially-filtered) correlated flux levels can, in
general, depend on both the spatial distribution and spectral
characteristics of the emitting material.

\begin{table}
  \begin{center}
    \caption{Optical depth of the silicate absorption feature at a
      wavelength of 10.2~\micron{}}
    \label{tab_optdepth}
    \begin{tabular}{l c c c}
      \hline \hline
      Source & $\tau_\nu(10.2~\mu\mathrm{m})$ & $<\tau_\nu(10.2~\mu\mathrm{m})>$ & $\sigma_{\tau_\nu}$ \\
      \hline
      AFGL 2136 & 2.19 & 2.33 & 0.20 \\
      AFGL 4176 & 1.94 & 1.82 & 0.29 \\
      G305.20+0.21 & 0.50 & 1.71 & 0.76 \\
      IRAS 13481-6124 & 0.81 & 1.30 & 0.13 \\
      IRAS 17216-3801 & 1.21 & 0.99 & 0.20 \\
      M8E-IR & 0.67 & 0.74 & 0.09 \\
      M17 UC1 & 2.53 & 1.81 & --- \\
      Mon R2 IRS2 & 1.40 & 1.27 & 0.05 \\
      Mon R2 IRS3 A & 1.77 & 2.65 & 0.17 \\
      Mon R2 IRS3 B & 2.59 & 2.69 & 0.06 \\
      NGC 2264 IRS1 & 0.42 & 0.35 & --- \\
      Orion BN & 0.85 & 1.77 & 0.37 \\
      R Mon & 0.24 & 0.92 & 0.25 \\
      S255 IRS3 & 2.31 & 2.33 & 0.34 \\
      \hline
    \end{tabular}
    \tablefoot{The second column shows the depth measured in the
      spectrum of the total flux (Fig.~\ref{fig_spectra}), while the
      depth in the correlated flux, averaged over all measurements, is
      shown in the third column.  The standard deviation of the depth
      measured in the correlated flux spectra is shown in the fourth
      column.}
  \end{center}
\end{table}

In order to evaluate the depth of the silicate feature in each
individual spectrum, we fit a continuum in the form of a power law,
shown as a blue line in Figs.~\ref{fig_cflux_orionbn} and
\ref{fig_cflux_afgl2136}--\ref{fig_cflux_vymon}, to the maxima of the
correlated flux in the ranges 7.5--8.2~\micron{} and
12.4--12.8~\micron{} (shown as red points in the figures).  We
evaluate the optical depth of the silicate feature as
$\tau_\nu(\lambda) =
\ln{(F_\nu^\mathrm{(c)}(\lambda)/F_\nu(\lambda))}$, where $F_\nu$ is
the measured correlated flux, and $F_\nu^\mathrm{(c)}$ is the
continuum fit.

In Table~\ref{tab_optdepth}, for the sources which show silicate
absorption, we show the derived optical depth at a wavelength of
10.2~\micron{}\footnote{Although the central wavelength of the
  silicate feature is nominally $\sim9.7$~\micron{}, the
  9.3--10.1~\micron{} region of the $N$-band is affected by telluric
  ozone absorption.}.  We show both the optical depth
$\tau_\nu(10.2~\mu\mathrm{m})$ of the feature in the total spectrum,
as well as the mean optical depth ${<\tau_\nu(10.2~\mu\mathrm{m})>}$
of the feature in the correlated flux spectra, where we have taken the
mean value and standard deviation $\sigma_{\tau_\nu}$ of each target
over all the observed baselines and position angles.

\subsection{Fitting the correlated flux absorption spectra}
\label{sec_absfit}

A common approach used to model mid-infrared spectra, much simpler
than full radiative transfer modeling, is to fit the observed spectrum
with a blackbody or power-law spectrum, modified by experimental
and/or laboratory dust opacities.  For objects which show little or no
silicate absorption, this approach has been used to model the
mid-infrared emission features seen in T~Tauri stars
\citep[e.g.][]{Olofsson10} and Herbig Ae/Be stars
\citep[e.g.][]{Juhasz10}.

Here, we apply a similar approach to the \emph{absorption} spectra of
13 of the objects in our sample, which have spectra dominated by
silicate absorption\footnote{We exclude Mon~R2~IRS3~A from this
  analysis, which has a much wider absorption feature than can be fit
  by the dust species used here (see Fig.~\ref{fig_cflux_monr2irs3a}).}.  In
our model, we let the correlated flux $F_\nu$ be described by a
power-law continuum extincted by foreground material, i.e.
\begin{equation}
F_\nu(\lambda) = C \lambda^p e^{-\tau_\nu(\lambda)},
\end{equation}
where the constant $C$ and the exponent $p$, both baseline-dependent,
define the continuum.  The absorption optical depth
$\tau_\nu(\lambda)$ is 1) given by the product of the column density
and extinction cross-section (defined by the dust composition) of the
absorbing material, and 2) defined to be the same for all
baselines/position angles.  In this context, this ``foreground''
absorbing material then includes components from both the source
envelope, in addition to the diffuse ISM along the line of sight, with
the former most likely being the dominant contributor.

\begin{figure}
  \begin{center}
    \includegraphics[width=85mm]{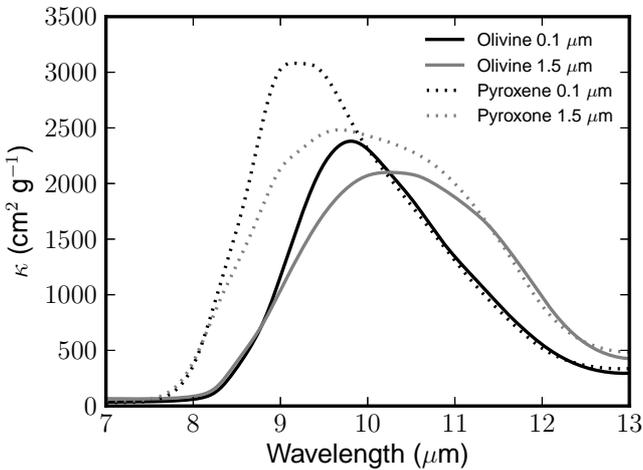}
    \caption{Mass absorption coefficients for amorphous MgFeSiO$_4$
      (solid lines) and amorphous Mg$_{0.5}$Fe$_{0.5}$SiO$_3$ (dotted
      lines).  Black lines are for a grain radius of 0.1~\micron{},
      and the gray lines are for a grain radius of 1.5~\micron{}.}
    \label{fig_kappas}
  \end{center}
\end{figure}

For the dust components, we limit ourselves to two sizes each (0.1 and
1.5~\micron) of spherical grains of amorphous material, in the form of
olivine (Mg$_{2x}$Fe$_{2-2x}$SiO$_4$) and pyroxene
(Mg$_x$Fe$_{1-x}$SiO$_3$) glasses.  We find that the opacities of pure
Mg$_2$SiO$_4$/MgSiO$_3$ grains presented by \citet{Jaeger03} are
unable to reproduce the observed absorption spectra, while the
inclusion of iron in the silicates leads to satisfactory fits.  We
therefore adopt $x = 0.5$ (equal Mg/Fe content) and calculate mass
absorption coefficients for each grain type and size, shown in
Fig.~\ref{fig_kappas}.  These absorption coefficients were calculated
using Mie theory and the optical properties of the grains from
\citet{Dorschner95}.  We note, however, that \citet{Min05} have shown
that pure Mg-rich silicates ($x=1.0$) yield 10~\micron{} silicate
features of nearly identical shape to the iron-rich silicates that we
use, if the mass absorption coefficients are calculated for a
statistical distribution of grain sizes and
shapes\footnote{\citet{Min05} use a distribution of hollow spheres
  (DHS).}. Hence, our choice of silicate composition should be seen as
a practical way of matching the observed silicate features, although
it does not provide evidence for the presence of iron in the observed
silicates.

This simplified model makes two important assumptions about the
correlated flux spectra which may not hold in all cases.  Namely, that
the underlying emission is well-described by a power law, and that the
foreground absorbing screen, which implicitly includes contributions
from both an envelope and the diffuse ISM, is independent of spatial
scale.  Deviations from a power law for the emission can occur not
only in the form of spectral features in the emitting material, but
can also arise from signatures of the spatial structure in the
correlated flux (e.g. from a binary signal or zero crossings in the
visibility amplitude).  That the foreground absorbing screen is
independent of spatial scale means that this absorbing material is
essentially uniform on the spatial scales probed by the observations.
This may not be true, for example, if significant absorption is
occurring at scales already resolved by MIDI.  These two assumptions
are difficult to verify \emph{a priori}, but for now we merely take
them as postulates, and discuss their validity for individual sources
in Sec.~\ref{sec_individual}.

\begin{table*}
  \begin{center}
    \caption{Derived column densities and relative contributions
      (by mass) of absorbing material}
    \label{tab_dustcomp}
    \begin{tabular}{l c c c c c c c c}
      \hline \hline
      & & \multicolumn{2}{c}{MgFeSiO$_4$} & & \multicolumn{2}{c}{Mg$_{0.5}$Fe$_{0.5}$SiO$_3$} \\
      \cline {3-4} \cline{6-7}
      Source & Mass col. density & $r=0.1$~\micron & $r=1.5$~\micron & & $r=0.1$~\micron & $r=1.5$~\micron \\
      & ($10^{-4}$ g~cm$^{-2}$) & (\%) & (\%) & & (\%) & (\%) \\
      \hline
AFGL 2136 & $12.7_{-1.1}^{+0.78}$ & $ 2.5_{-2.3}^{+30.2}$ & $ 2.1_{-2.0}^{+22.1}$ & & $75.5_{-39.8}^{+13.8}$ & $19.9_{-10.0}^{+50.3}$ \\\\
AFGL 4176 & $8.37_{-0.26}^{+0.16}$ & $49.5_{-8.9}^{+7.4}$ & $ 0.6_{-0.5}^{+5.2}$ & & $32.9_{-5.0}^{+6.9}$ & $16.9_{-9.2}^{+7.2}$ \\\\
G305.20+0.21 & $7.75_{-0.37}^{+0.37}$ & $63.9_{-30.6}^{+24.1}$ & $ 2.9_{-2.6}^{+34.0}$ & & $29.9_{-13.2}^{+19.1}$ & $ 3.3_{-3.0}^{+44.9}$ \\\\
IRAS 13481-6124 & $5.70_{-0.090}^{+0.20}$ & $63.3_{-6.3}^{+3.7}$ & $ 0.4_{-0.3}^{+3.6}$ & & $36.0_{-3.0}^{+3.9}$ & $ 0.4_{-0.3}^{+3.4}$ \\\\
IRAS 17216-3801 & $4.75_{-0.17}^{+0.098}$ & $80.5_{-16.7}^{+4.0}$ & $ 1.0_{-1.0}^{+13.8}$ & & $17.9_{-4.3}^{+7.7}$ & $ 0.6_{-0.6}^{+8.0}$ \\\\
M17 UC1 & $8.77_{-0.35}^{+0.49}$ & $89.6_{-26.1}^{+34.2}$ & $ 3.5_{-3.2}^{+43.9}$ & & $ 2.5_{-2.3}^{+30.7}$ & $ 4.4_{-4.2}^{+61.5}$ \\\\
M8E-IR & $3.68_{-0.048}^{+0.034}$ & $70.6_{-9.5}^{+5.0}$ & $28.2_{-3.9}^{+8.9}$ & & $ 0.3_{-0.3}^{+3.6}$ & $ 0.9_{-0.8}^{+6.9}$ \\\\
Mon R2 IRS2 & $6.64_{-0.14}^{+0.077}$ & $75.9_{-22.9}^{+3.0}$ & $ 1.5_{-1.3}^{+17.9}$ & & $21.6_{-3.1}^{+4.4}$ & $ 0.9_{-0.9}^{+6.9}$ \\\\
Mon R2 IRS3 B & $12.9_{-0.58}^{+1.9}$ & $53.0_{-31.3}^{+31.1}$ & $ 2.6_{-2.5}^{+39.6}$ & & $41.9_{-24.4}^{+5.2}$ & $ 2.5_{-2.5}^{+36.7}$ \\\\
NGC 2264 IRS1 & $1.68_{-0.055}^{+0.27}$ & $42.9_{-25.2}^{+28.0}$ & $ 2.8_{-2.7}^{+41.4}$ & & $51.8_{-20.8}^{+17.7}$ & $ 2.4_{-2.3}^{+20.0}$ \\\\
Orion BN & $7.22_{-0.29}^{+0.19}$ & $71.9_{-21.7}^{+7.9}$ & $ 1.6_{-1.4}^{+16.2}$ & & $25.3_{-7.8}^{+4.5}$ & $ 1.2_{-1.0}^{+11.9}$ \\\\
R Mon & $3.22_{-0.065}^{+0.067}$ & $78.2_{-19.2}^{+6.6}$ & $ 2.1_{-2.0}^{+18.6}$ & & $17.7_{-7.0}^{+8.0}$ & $ 2.1_{-1.9}^{+16.0}$ \\\\
S255 IRS3 & $11.2_{-0.70}^{+0.77}$ & $64.0_{-30.2}^{+62.8}$ & $ 4.7_{-4.5}^{+70.6}$ & & $26.1_{-14.3}^{+40.1}$ & $ 5.2_{-5.0}^{+72.5}$ \\
      \hline
    \end{tabular}
  \end{center}
\end{table*}

In total, for a source with $n$ correlated flux spectra (i.e., $n$
points in $uv$ space, where each point represents a spectrum with
$\sim100$ wavelength channels), there are $2n+4$ free parameters in
this model ($2n$ parameters for the power-law continua, and 4
parameters describing the column densities of each dust type/size).
We use a downhill-simplex $\chi^2$-minimization routine to optimize
all $2n+4$ parameters simultaneously.  For an initial solution, we fit
a continuum as described in Sec.~\ref{sec_cflux}, take an equal mix of
the four grain types/sizes, and let the total optical depth at
10.2~\micron{} be equal to unity.  We derive uncertainties on the fit
parameters using a Monte-Carlo approach, where we generate synthetic
correlated flux values, normally spread about the observed values,
with a standard deviation equal to the noise (i.e., random) error on
the measurement as derived by the MIA+EWS package\footnote{Note that
  the error bars shown in Figs.~\ref{fig_cflux_orionbn} and
  \ref{fig_cflux_afgl2136}--\ref{fig_cflux_vymon} represent the
  combination of this random error, plus the uncertainty on the
  absolute calibration.  The absolute flux calibration introduces a
  factor which varies only slowly with wavelength, and thus is not
  expected to change the shape of the silicate feature.  The random
  error, which characterizes the wavelength-to-wavelength scatter in
  the correlated flux of an individual spectrum, is about an order of
  magnitude smaller than the calibration error.}  and repeat the fit
procedure 100 times; the optimum model parameters and their
uncertainties are given by the mean and standard deviation,
respectively, of the parameters in this ensemble.  We show the
best-fit models using this procedure as a solid blue line in
Figs.~\ref{fig_cflux_orionbn} and
\ref{fig_cflux_afgl2136}--\ref{fig_cflux_vymon}, and list the total
column densities of the absorbing material and the relative mass
fraction of each dust component in Table~\ref{tab_dustcomp}.

\subsection{Non-detections}
\label{sec_nondetect}

For four of the 24 targets which we attempted to observe with MIDI, we
were unable to find interferometric fringes despite good technical and
weather conditions.  Thus, for the observations listed at the end of
Table~\ref{tab_obslog}, these sources were completely overresolved by
the interferometer.  In this section, we discuss each of these
non-detections in turn.

\subsubsection{GGD~27-ILL}
\label{sec_ggd27}

\begin{figure}
  \begin{center}
    \includegraphics[width=85mm]{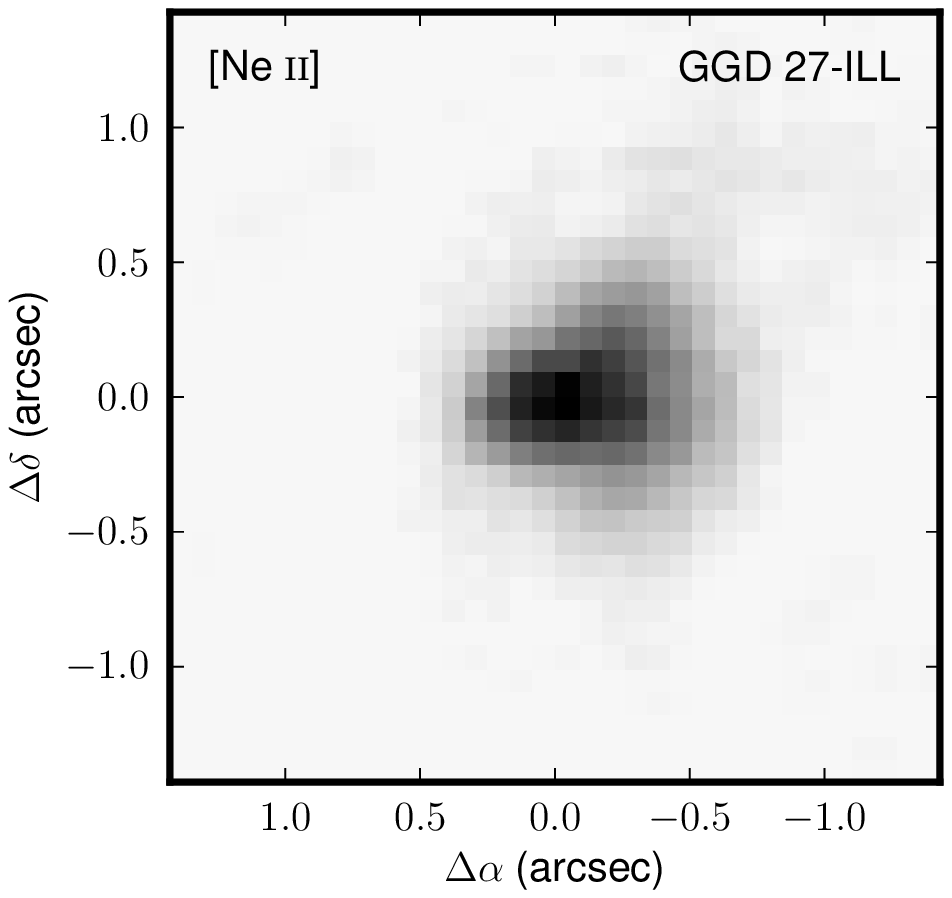}
    \caption{Acquisition image of GGD~27-ILL obtained with UT3 on
      2004-06-03 using the [\ion{Ne}{II}] narrow-band filter (center
      12.80~\micron{}, width 0.21~\micron{}).  North is up, east is
      left.}
    \label{fig_ggd27acq}
  \end{center}
\end{figure}

The star-forming region GGD~27 received early attention when
\citet{Gyulbudaghian78} revealed the presence of two strong
Herbig-Haro objects in the vicinity.  These were later cataloged as
HH~80/81, which is the name under which this region is known in the
radio astronomy community. The central cm radio source is associated
with the infrared source IRAS 18162-2048 \citep{Rodriguez89}, and
drives a thermal ionized jet with bipolar components spanning a total
extent of 5.3~pc \citep{Marti93}.  Recent (sub-)mm interferometry
indicates that this jet-driving source is associated with a rotating
molecular disk \citep[e.g.,][]{Fernandez11,Carrasco12}. The central
source is not directly visible at wavelengths $<3$~\micron{}, but its
location was inferred by means of near-infrared polarimetry of the
surrounding reflection nebula by \citet{Tamura91}. Subsequently
referred to as GGD~27-ILL, the central source was later detected at
10~\micron{} \citep{Stecklum97}, 4.7~\micron{} \citep{Aspin94}, and
even 3.8~\micron{} \citep{Linz03}.

We attempted to observe GGD~27-ILL in June 2004.  A fringe search was
performed on the UT2-UT3 baseline, with a projected length of 44~m and
a position angle of 22\degr{}, which is well-aligned with the ionized
jet.  No fringes could be found, and no further attempts were possible
on the following days due to technical problems during this MIDI
commissioning run. However, during this period MIDI could be used as a
simple mid-infrared imager.  An acquisition image with a total
exposure time of 125~s using the narrow-band [\ion{Ne}{II}] filter,
shown in Fig.~\ref{fig_ggd27acq}, reveals that the source appears
slightly extended at mid-infrared wavelengths already with an
8~m-class telescope.  Another attempt to observe this source on June
30, 2010 was foiled by poor atmospheric conditions during that night,
when even a valid acquisition\footnote{The acquisition was attempted
  directly in the mid-infrared in this case, since GGD~27-ILL is too
  faint at 2.2~\micron{} to be centered by the now-common VLTI-IRIS
  beam stabilizer.} of the target ($\sim5$~Jy at 11.7~\micron{}) could
not be achieved with the UTs.

\subsubsection{IRAS~16164-5046}

This source was identified as a YSO by \citet{Braz87} based on
infrared photometry and SED arguments. It is associated with an
UC\ion{H}{II} region \citep[e.g.,][]{Urquhart07}. \citet{Bik06}
revealed the presence of CO overtone emission at 2.3~\micron{} in a
$K$-band spectroscopic survey, where this source is referred to as
16164nr3636\footnote{Note that there is a coordinate offset of almost
  10\arcsec{} in their target list for this source, which has a
  position $\alpha=16^\mathrm{h}$20$^\mathrm{m}$11\fs14,\break
  $\delta=-50$\degr53\arcmin15\farcs9 (J2000) according to 2MASS.}  in
their nomenclature. The same data were used by \citet{Bik04} for more
detailed modeling of the CO line profiles. They propose that the CO
overtone emission from this source arises in a circumstellar disk with
an inclination of $\sim30$\degr.  Combined with the SED and radio data
listed by \citet{Bik06}, and adopting a distance of 3.8--4.0~kpc
\citep{Urquhart12,Moises11}, this indicates the presence of a
circumstellar disk around a mid- to late O-star. \citet{Wheelwright10}
refined the 2.3~\micron{} CO line spectroscopy for this source
(G332.8256-00.5498 in their nomenclature) by applying
spectroastrometric techniques. While no significant spectroastrometric
signal could be revealed, the disk interpretation for the CO emission
was re-enforced, and an updated disk inclination of
42.3$^{+4.4}_{-11.7}$~deg was derived.

We attempted to observe this target with three intermediate AT
baselines between 30 and 65~m. In all cases, the fringe search was not
successful.  In combination with the moderate inclination of 42\degr{}
for the potential circumstellar disk, this may indicate the existence
of a substantial envelope around this object, and hence the related
mid-infrared emission might be overresolved by MIDI. In addition, the
UC\ion{H}{II} region associated with this source may also give rise to
extended mid-infrared emission.

\subsubsection{IRAS~17136-3617}

\begin{figure*}
  \begin{center}
    \includegraphics[width=180mm]{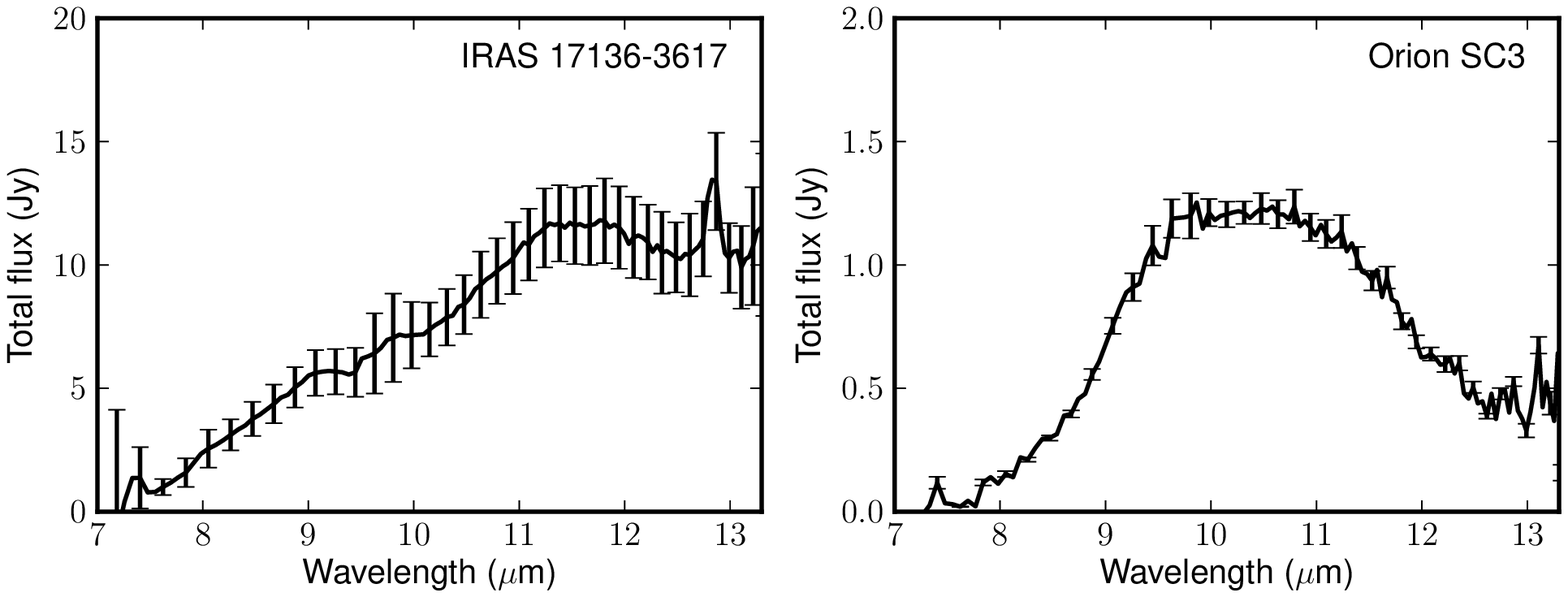}
    \caption{Total spectrum of IRAS~17136-3617 (left) and Orion~SC3
      (right) measured with the 8.2~m UT telescopes on 2006-05-18 and
      2005-02-26, respectively.  The error bars represent the
      uncertainty in the absolute flux calibration.}
    \label{fig_sc3gm24spect}
  \end{center}
\end{figure*}

We attempted to observe IRAS~17136-3617 (aka GM~24) two times with
MIDI, using projected baselines of 25 and 46~m.  However, despite
successful acquisition and good weather conditions, no fringes were
found, indicating the source is completely overresolved at these
baselines.  We show a spectrum of the total flux, measured with the
8.2~m UT telescopes, in the left panel of Fig.~\ref{fig_sc3gm24spect}.
We note the distinct absence of any silicate absorption feature,
despite the apparently large visual extinction of the near-infrared
source IRS~3A \citep[$A_V \ga 50$~mag;][]{Tapia09}, with which the
mid-infrared source is probably associated.  This source also
represents the only detection of the [\ion{Ne}{II}] nebular emission
line at 12.8~\micron{} in our sample.

\subsubsection{Orion~SC3}

Thermal infrared imaging of the Orion Trapezium region, including the
famous Ney-Allen nebula, revealed the presence of arc-shaped
structures pointing towards the most luminous (multiple) star in the
Trapezium, $\theta^1$~Ori~C \citep{Hayward94}. Among these newly-found
structures was the peculiar object SC3, which appears circularly
symmetric at 8.8 and 11.7~\micron{}, despite a very small projected
distance (1\farcs8) from the O6.5 star
$\theta^1$~Ori~C. \citet{Robberto02} hypothesize SC3 to be a face-on
circumstellar disk in the background of $\theta^1$~Ori~C, which by
chance is projected close to the O~star.

We performed one fringe search utilizing the shortest UT baseline
UT2-UT3 in February 2005.  We did the initial centering on the
mid-infrared peak; the neighboring O star, which also shows weak
mid-infrared emission \citep[cf.][]{Robberto05}, is outside the field
of view for these MIDI UT observations.  No fringes were detected.
Hence, given the low total flux of SC3 of $<2$~Jy, a source with a
large fraction of extended emission will give rise to very low
correlated flux levels, which may not be detectable by
MIDI. Mid-infrared imaging with T-ReCS at Gemini South from the same
year \citep{Smith05} gives some credence to this suggestion: their
single-dish resolution is sufficient to spatially resolve SC3 into a
compact peak surrounded by a more extended halo with a size of
$\sim1\farcs5$.  The spectrum of the total flux, shown in the right
panel of Fig.~\ref{fig_sc3gm24spect}, shows a clear silicate emission
feature, which supports the interpretation of a circumstellar disk
with low inclination.  We note the absence of the [\ion{Ne}{II}]
12.8~\micron{} line, which we find consistent with the idea that SC3
is not directly subjected to the intense radiation from the Trapezium
O~star $\theta^1$~OriC.

\section{Analysis and discussion}
\label{sec_analysis}

\subsection{$N$-band geometry}
\label{sec_analysis_geom}

The results of the geometric modeling described in
Sec.~\ref{sec_geofit} show a range of linear sizes and inclinations.
In general, the mid-infrared intensity distributions of the sources
are not circularly symmetric: when our $uv$ coverage is sufficient to
fit a two-dimensional model, we find inclinations ranging from
30\degr{} to 80\degr{} (corresponding to axial ratios of about 1.15 to
6), indicating the presence of non-spherical structures on scales of
tens of milliarcseconds.

For 12 of the targets observed in our sample, independent information
about disks and/or outflows is available in the literature, which can
provide insights into the physical nature of the asymmetries present
in the MIDI data.  In Figs.~\ref{fig_images} and \ref{fig_images_au},
we show, where available, orientations collected from the literature
(see Table~\ref{tab_sourceprops} for values and references) for
circumstellar disks (dotted lines) and/or outflows (dashed lines).

In general, we see a tendency for the compact, mid-infrared emission
to trace disk structures, albeit with exceptions.  As extreme cases,
the compact, mid-infrared emission of AFGL~2136 is both elongated
along the direction of the circumstellar disk and perpendicular to the
outflow, while the situation in the case of M8E-IR appears to be
reversed: the MIDI visibilities seem to trace an outflow component (no
disk detection for this object has been reported in the literature).
For six sources with known disks and/or outflow components, and where
we fit a two-dimensional geometric model (IRAS~13481-6124, MWC~300,
R~CrA, R~Mon, S255~IRS3, and V921~Sco), the structures resolved by
MIDI are roughly parallel to the disk/perpendicular to the outflow.
In the case of Mon~R2~IRS3~A, the mid-infrared emission is roughly
perpendicular to the polarization disk reported by \citet{Yao97}.
This and other sources are discussed in more detail in
Sec.~\ref{sec_individual}.

Finally, we note that the geometric analysis of the
wavelength-averaged visibilities presented here represents only a
first attempt at understanding the physical structures responsible for
emission in these sources.  In particular, analyses of the
\emph{wavelength dependence} of the visibilities have the potential to
unlock a wealth of additional information about these objects, and we
encourage such endeavors.

\subsection{Mid-infrared size vs. luminosity}
\label{sec_size_vs_L}

At near-infrared wavelengths, using a model consisting of a thin ring
representing the inner radius of a circumstellar disk,
\citet{Monnier02} showed a tight correlation between luminosity and
ring radius for YSOs with luminosities in the range
$\sim10^0$--$10^5$~\Lsun{}.  This correlation arises both because the
dust sublimation radius varies with $\sqrt{L_*}$
\citep[e.g.][]{Monnier02}, and because the near-infrared emission from
grains near the dust evaporation temperature ($\sim1500$~K) is
confined to a small region of the disk.  This relation seems to break
down at higher luminosities ($L \ga 10^4$~\Lsun{}), which may indicate
the presence of optically-thick circumstellar gas interior to the dust
evaporation radius \citep{Monnier05}.

The relation between size and luminosity has been examined at
mid-infrared wavelengths, as well.  Using the FWHM of a Gaussian fit
to the mid-infrared visibilities, \citet{Monnier09} and
\citet{Grellmann11} found that although the mid-infrared size of YSOs
is correlated with luminosity, there is considerably more scatter in
the relation than is seen at near-infrared wavelengths.  Indeed, this
is to be expected; as noted in both these works, at mid-infrared
wavelengths the notion of ``size'' becomes more complex than a simple
ring radius: the mid-infrared intensity distribution is complicated by
a combination of geometrical effects (e.g. flaring, shadowing),
temperature/wavelength effects (the extent of the near-infrared
emission is limited by the Wien cut-off at the dust sublimation
temperature) and the fact that the mid-infrared radiation is expected
to arise from a variety of structures (e.g. disk wall, disk surface,
envelope).

\begin{figure}
  \begin{center}
    \includegraphics[width=85mm]{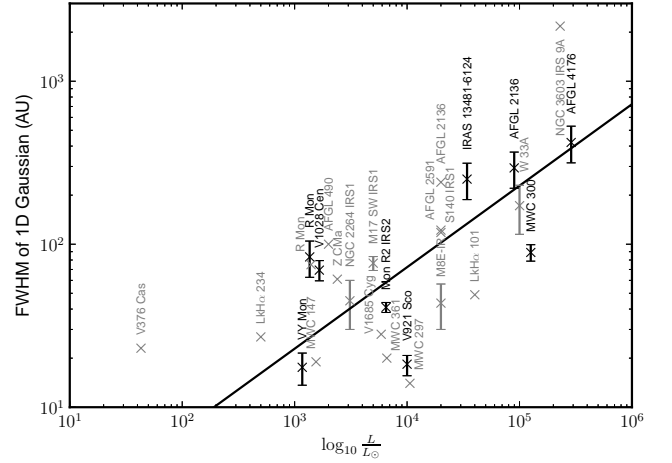}
    \caption{Linear size vs. luminosity.  The black crosses show the
      sizes derived in this work from the one-dimensional Gaussian
      models for selected sources (see Sec.~\ref{sec_size_vs_L}),
      while the gray crosses show the results compiled by
      \citet{Grellmann11}.  The solid line shows a fit to the black
      points for FWHM\,$\propto\sqrt{L}$.}
    \label{fig_size_vs_L}
  \end{center}
\end{figure}

For sources for which the luminosity is known, we show the FWHM size
we derive from the one-dimensional Gaussian geometric model
vs. luminosity as black crosses in Fig.~\ref{fig_size_vs_L}, together
with the sizes and luminosities compiled by \citet{Grellmann11} as
gray crosses.  In our own sample, in order to restrict ourselves to
sources where we do not completely overresolve the extended structure
with the interferometer, we limit the selection to those sources for
which we measure $V \ga 0.5$ at the shortest baselines.  Besides the
uncertainty on the derived angular size, in the derivation of the
linear size we also account for the uncertainty in the distance to
each source.  If no value for the uncertainty in distance is available
in Table~\ref{tab_sourceprops}, we adopt a value of 25\%, which is
appropriate for kinematic and photometric distances.  As both
\citet{Monnier09} and \citet{Grellmann11} found, we see considerable
scatter in the mid-infrared size vs. luminosity relation.  A scaling
relation of FWHM~$\propto\sqrt{L}$, shown as a black line in
Fig.~\ref{fig_size_vs_L}, is \emph{consistent} with the mid-infrared
sizes, but is by no means as well-defined as at near-infrared
wavelengths.

\subsection{Absorption spectra and dust composition}

As Table~\ref{tab_optdepth} shows, the optical depth of the silicate
absorption feature is sometimes substantially lower in the total
spectrum than it is in the correlated flux spectra.  This is probably
due to the fact that the total spectrum contains contributions of
emission from all scales, while the spatial filtering of the
interferometer limits this contribution to a single scale in each
correlated flux spectrum.  In this context, the true shape of the
absorption feature is best represented in the short-baseline
interferometric measurements, where contributions from hot emitting
material (which is presumably more dominant at smaller spatial scales)
are minimized.

In general, we find that the power-law emission/foreground absorption
model described in Sec.~\ref{sec_absfit} for the 13 sources in
Table~\ref{tab_dustcomp} adequately reproduces the observed correlated
flux spectra in almost all cases (the fits for the sources from
Table~\ref{tab_dustcomp} are shown as blue lines in
Figs.~\ref{fig_cflux_orionbn} and
\ref{fig_cflux_afgl2136}--\ref{fig_cflux_vymon}).  As the column
density and composition of the absorbing material is constrained to be
the same for all baselines, this means that the absorption (due to
silicates) is both uniform over the spatial scales probed by MIDI, and
physically detached from those scales.  Furthermore, the underlying
emission also seems to be well-approximated by a power law.

There are, however, some cases where this simple model is unable to
reproduce the observed correlated flux levels.  For example, for
AFGL~4176 (Fig.~\ref{fig_cflux_afgl4176}) the fits at short baselines
are quite good, but become progressively worse at long baselines.  At
the longest spatial scales, the imprint of the spatial structure on
the correlated flux levels becomes apparent \citep{Boley12}, meaning
that the assumption of the underlying emission having a power-law
spectrum is incorrect.  Such spatial effects may also be present in
G305.20+0.21 (Fig.~\ref{fig_cflux_g305}) and Mon~R2~IRS3~A
(Fig.~\ref{fig_cflux_monr2irs3a}).

Besides effects from the source geometry, the true emission spectrum
behind the foreground absorption screen may, in the general case,
contain spectral features (from, for example, silicates, crystalline
grain species, etc.).  Such features may become more prominent at
longer baselines due to potential emission arising from a
circumstellar disk \citep[e.g][]{vanBoekel04}.  However, accounting
for emission features in the spectra of deeply-embedded sources is a
complicated process, and beyond the scope of the present work.

As shown in Table~\ref{tab_dustcomp}, for most sources the majority of
the absorbing material can be attributed to amorphous (glass) olivine.
AFGL~2136 stands alone in being clearly dominated by amorphous
pyroxene, while AFGL~4176 and NGC~2264~IRS1 have roughly equal
contributions (by mass) of amorphous pyroxene and olivine dust.  Three
sources (AFGL~2136, AFGL~4176 and M8E-IR) show a significant
population of large grains (1.5~\micron{}), but for the other ten
sources fit the amount of mass contained in large grains is limited to
a few percent.  In general, however, we find that the grain size
distribution, and to some extent the grain composition, is highly
variable throughout our sample.  Although this is expected, given the
qualitatively different shapes of the silicate absorption feature in
both the total and correlated flux spectra, it should be taken as a
reminder when modeling these and other similar objects that the dust
composition may differ significantly from that of the diffuse
interstellar medium.

Finally, we note that while the exact proportions derived for the
grain species and listed in Table~\ref{tab_dustcomp} are dependent on
the opacity templates used (see Sec.~\ref{sec_absfit}), the general
trends derived from fitting the dust features of mid-infrared spectra
(size distribution, pyroxene/olivine ratio) have been shown not to
depend strongly on the specifics of the grains
\citep[e.g.][]{Juhasz09}.  In this sense, regardless of the specifics
of the iron content or grain structure (shape, porosity, etc.) of the
amorphous silicates used, larger grains tend to give a wider
10~\micron{} feature relative to smaller grains, and the center of
this feature tends to be at shorter wavelengths for grains with a
pyroxene stoichiometry than for grains with an olivine stoichiometry.
Thus, the \emph{general conclusions} about the size distributions and
olivine/pyroxene ratios derived in Sec.~\ref{sec_absfit} can be
considered robust.  For further discussion of this topic, we direct
the interested reader to the works by \citet{Bouwman08} and
\citet{Juhasz10}, and references therein.

\subsection{Notes on selected sources}
\label{sec_individual}

\subsubsection{AFGL~2136}

AFGL~2136~IRS1 is from the classic list of BN-type objects
\citep{Henning84}.  It is the illuminating infrared source of the
so-called Juggler reflection nebula
\citep{Kastner92}. \citet{Menten04} found faint centimeter emission
(their component RS4) closely associated with this target. The nature
of this emission, which is optically thick up to frequencies of
$\sim43$~GHz, is not clarified yet, although an ionized jet is the
most probable explanation.  As mentioned in
Sec.~\ref{sec_analysis_geom}, the sizes derived from the MIDI
visibilities for AFGL~2136 show significant elongation parallel to the
major axis of a circumstellar disk, which was revealed in
near-infrared polarimetric images by \citet{Minchin91}, and
perpendicular to the molecular outflow found by \citet{Kastner94}.
The FWHM size of the two-dimensional Gaussian in our geometric model
is 44~mas (88~AU at 2~kpc), and the correlated flux levels are well
reproduced by the foreground-absorption model described in
Sec.~\ref{sec_absfit} (see Fig.~\ref{fig_cflux_afgl2136}).

\subsubsection{AFGL~4176}

AFGL~4176 is a southern high-luminosity MYSO
\citep{Henning84,Persi86}.  The position angle and elongation derived
from the 2D1D model here are consistent with the more complicated
temperature-gradient disk model presented by \citet{Boley12}.  The
correlated flux (Fig.~\ref{fig_cflux_afgl4176}) is reproduced well by
our absorption model at shorter baselines, with the fits becoming
progressively worse towards longer baselines, which we find consistent
with the expectation that emission at these angular scales is
dominated by hot dust \citep{Boley12}.

\subsubsection{G305.20+0.21}

The surveys presented by \citet{Walsh99,Walsh01} established the
presence of a bright compact infrared source within the star-forming
complex G305. There is an \ion{H}{II} region some 10\arcsec{} to the
southwest \citep{Walsh07}, but the infrared source itself is not
associated with centimeter free-free emission
\citep{Phillips98,Walsh07}. The extrapolated IRAS luminosity is $\ga
10^5$~\Lsun{} \citep{Walsh01}, which together with the cm
non-detection supports our view that this object is a massive YSO in a
pre-UC\ion{H}{II}-region stage. Further interest is triggered by the
presence of a group of aligned class-II methanol masers coincident
with this infrared object \citep[component G305.202+0.207 in the work
  by][]{Phillips98}. Soon after the finding of a number of such
sources (including G305), it was speculated that such aligned masers
might trace edge-on circumstellar disks \citep{Norris98}.  However,
there is also evidence that methanol maser distributions may instead
trace outflows \citep[e.g.][]{deBuizer03,deBuizer12}.

The geometric fits for this object indicate a large (200~AU at
4.8~kpc) structure, elongated along a position angle of
$\sim34\degr{}$.  This is roughly consistent with the orientation of
the linear maser distribution of $\sim57\degr{}$ found by
\citet{Norris98}; however, as noted above, it is not clear that the
maser orientation can be readily attributed to a disk or an outflow.
The correlated flux levels (Fig.~\ref{fig_cflux_g305}) show features
which cannot be reproduced by our absorption model.

\subsubsection{IRAS~13481-6124}

Though it was identified as a MYSO in the 1980s \citep{Persson87}, and
was found to feature an ionized wind \citep{Beck91}, there were not
many more observational studies available for this southern object
until recently.  The discoveries of both a disk and a jet/outflow
system, however, have been recently reported around this MYSO
\citep{Kraus10,Stecklum10}.

The $uv$ coverage of the MIDI observations of this source is
extensive, and the orientation derived from the 2D1D geometric model
(position angle $\sim112\degr$) is consistent with the
near-infrared disk \citep{Kraus10}.  The correlated flux levels
(Fig.~\ref{fig_cflux_iras13481}) are well reproduced by our absorption
model at all baselines.

\subsubsection{IRAS~17216-3801}

After its initial identification as a MYSO by \citet{Persson87}, this
interesting source has been somewhat neglected. This may have been
caused by its location on the southern sky, and more than 1\degr{}
away from the Galactic plane. It is very bright in the $K$ band, which
enabled early speckle observations by
\citet{Leinert97,Leinert01}. However, these revealed just a faint
(scattering) halo, but no more complex, spatially-resolved structures,
and no binarity. The object bears some resemblance to IRAS 13481-6124
in terms of infrared appearance and luminosity, but the presence of outflow
or jet activity has not yet been investigated.

The 2DOR geometric model shows that the two-dimensional component
needs to be quite large in linear scale (FWHM $\sim260$~AU at 3.1~kpc)
in order to fit the data.  The correlated flux spectra shown in
Fig.~\ref{fig_cflux_iras17216}, as well as the total $N$-band
spectrum, have an unusual triangular appearance which is only
partially reproduced by our simple absorption model.

\subsubsection{M8E-IR}

The study by \citet{Simon84} was the first work to disentangle the
M8~East region into an UC\ion{H}{II} region with no near-infrared
counterpart\footnote{Infrared emission from the UC\ion{H}{II} region
  was revealed much later, and only in the mid-infrared at 24.5 $\mu$m
  \citep{Linz09, deWit09}.} and a strong infrared source, subsequently
called M8E-IR, which is around 7\arcsec{} away. \citet{Simon84}
emphasized the similarities between M8E-IR and the BN object in Orion.
M8E-IR was later confirmed not to be associated with free-free
emission from ionized gas at wavelengths of 2 and 6~cm \citep[cf.][
  and their source region Mol~37]{Molinari98}.  \citet{Simon85}
speculated on the existence of a small circumstellar disk around
M8E-IR based on thermal infrared lunar occultation data. Their
indications for a small elongated structure came mainly from their
3.8~\micron{} data, while their 10~\micron{} data are in agreement
with a more symmetric intensity distribution.

Our geometric fits of this object show that the compact mid-infrared
emission is elongated along the CO outflow reported by
\citet{Mitchell91}.  This is consistent with the work of
\citet{Linz09}, who concluded that the role of a circumstellar disk in
this system is probably negligible.

\subsubsection{M17~SW~IRS1}

This object is in the list of BN-type objects by \citet{Henning84}. It
is also known as the Kleinmann-Wright (KW) object \citep{Kleinmann73},
and is situated in the dusty south-western part of the M17
star-forming complex.  \citet{Porter98} analyzed the $H$- and $K$-band
spectra of this object, and concluded that they resemble the spectrum
coming from a concealed Herbig~Be star. \citet{Chini04} revealed
multiplicity within a few thousand AU in the near-infrared. However,
the actual KW object is the dominating YSO in that core region, and is
estimated to have a luminosity equivalent to that of an early B-type
star.  \citet{Follert10} originally published the MIDI data included
here, and used them to select appropriate models from the
\citet{Robitaille07} model grid of YSOs.  Our geometrical analysis
agrees with their assessment that a strongly-inclined configuration
($\theta \ga 60\degr$) is necessary to explain the observations.

\subsubsection{M17~UC1}

This object is a deeply-embedded young massive star in the M17
complex.  It is known to be associated with a hyper-compact
\ion{H}{II} region \citep{Felli80,Felli84} with an electron density of
greater than $10^5$~cm$^{-3}$.  Additionally, broad ($\ga35$~\kms)
radio recombination lines have been found \citep{Johnson98}. One of
the interpretations for such line widths, which are in excess of the
usual Doppler broadening, is the infall of material onto, and/or the
rotation of, an accretion disk \citep[e.g.][]{Sewilo04}. In the
infrared, \citet{Chini00} found that UC1 is weak at 1--2.2~\micron{}
wavelengths. However, there is another infrared source $\sim9000$~AU
away (IRS~5S), which is a stronger source in the near-infrared (but
similarly bright in the $N$-band) which might form a bound high-mass
system together with UC1.

This object was not in our original target list, but was observed as a
backup target during a different MIDI program on our
request. Therefore, we have only a single visibility
measurement. Hence, we can just derive a one-dimensional Gaussian
size.  Nevertheless, these data show that it is at least possible to
obtain fringes for hyper-compact \ion{H}{II} regions. The mid-infrared
emission is not overresolved as a whole, as might be the tendency for
more-evolved ultra-compact \ion{H}{II} regions (see our example of
IRAS~17136-3617 in Sec.~\ref{sec_nondetect}).

\subsubsection{Mon~R2~IRS2}

Mon~R2 is a massive star forming region, with an UC\ion{H}{II} region
lying just off the center of a molecular core.  There are several
well-known compact infrared sources around this region, among them
Mon~R2~IRS2.  IRS2 is situated in the geometrical center of a
ring-like near-infrared structure, and has been shown to be the main
source of illumination \citep{Aspin90}. Furthermore, a large-scale
bipolar CO outflow exists in the region
\citep[e.g.][]{Giannakopoulou97}, but its association with IRS2 is not
established \citep[e.g.,][]{Jaffe03}. IRS2 has a strong infrared
excess, but is not a source of free-free emission. It is unresolved in
$K$-band speckle observations with UKIRT \citep{Alvarez04}.  The
extent of the mid-infrared emission we find is quite compact
($\la$40~AU at 830~pc), which might be an imprint of the lower
luminosity of Mon~R2~IRS2.

\subsubsection{Mon~R2~IRS3~A}

The IRS3 region is another activity center within the Mon~R2
star-forming complex. Several near-infrared speckle studies have
addressed the multiplicity on arcsecond scales
\citep{Koresko93,Preibisch02,Alvarez04}. The dominating source at
2~\micron{}, as well as 10~\micron{}, is IRS3~A \citep[using the same
  nomenclature as][]{Preibisch02}. This source is surrounded by a
bipolar near-infrared nebula (position angle $\sim 30\degr$),
suggesting it to be embedded in a thick disk or torus, with pronounced
polar cavities.

The MIDI visibilities are low, and most of the correlated flux spectra
are box-like and cannot be reproduced by our absorbing dust models
(Sec.~\ref{sec_absfit}).  The fit with the 2D1D model, using the
combined MIDI and Keck data, yields a position angle of $41\degr$,
i.e. similar to the orientation of the near-infrared bipolar nebula,
and consistent with the fit to the Keck data by \citet{Monnier09}.
Finally, we note that the MIDI data show differential phases
significantly deviating from zero, which is indicative of deviations
from point symmetry.

\subsubsection{Mon~R2~IRS3~B}

A second YSO is visible in the mid-infrared in the vicinity of
Mon~R2~IRS3~A, located 0\farcs83 to the north-east
\citep[cf.][]{Monnier09}, termed IRS3~B by \citet{Preibisch02}. A
high-velocity outflow was found by \citet{Giannakopoulou97} roughly
perpendicular to the dominating molecular outflow in Mon~R2~IRS3,
which itself has a position angle of $\sim 135\degr$.
\citet{Preibisch02} speculate that IRS3~B might be the driving source
for this high-velocity flow, based on the near-infrared appearance of
shocked H$_2$ blobs along a position angle of 50\degr{}, which seem to
emanate from this source.  This suggests a collimating circumstellar
structure around IRS3~B, oriented roughly perpendicular to the shock
features at 50\degr{}.

With MIDI, IRS3~B shows higher visibility levels than IRS3~A, but is
clearly weaker regarding the total $N$-band flux.  Therefore, MIDI
observations were only possible using the UTs.  The two measurements,
with nearly perpendicular baseline angles (but similar projected
baseline lengths) show the highest optical depth in the silicate
absorption feature of the correlated fluxes among all of our targets
(see Table~\ref{tab_optdepth} and Fig.~\ref{fig_cflux_monr2irs3b}).
The two correlated flux spectra show some differences in shape, but
the general flux levels are not drastically different. Given that the
two MIDI baseline angles are roughly parallel and perpendicular to the
potential disk, this suggests that this disk is not seen edge-on, but
instead at a moderate inclination.  Contrary to IRS3~A, the absorption
of the correlated flux from IRS3~B is reproduced well by our
absorption model.

\subsubsection{Orion~BN}

This very strong near- and mid-infrared source was recognized by
\citet{Becklin67} as a distinct brightness peak within the Orion KL
region. Although it quickly achieved the status of a prototype object
for a sample of high-mass protostars \citep[e.g.,][]{Henning84}, many
details about the nature of Orion~BN are still unclear. The peculiar
proper motion of BN is well-established \citep{Plambeck95}, however,
the actual origin of this object is still debated. It might stem from
the dissolution of a multiple system in Orion~KL itself
\citep[e.g.][]{Rodriguez05,Gomez08}, while another group proposes
Orion BN to be a runaway source, ejected from the Orion Trapezium
\citep{Tan04,Chatterjee12}. The mid-infrared peak of BN is surrounded
by extended emission, with a pronounced extension to the southwest,
and the compact emission shows a barely-resolved elongation along the
northwest-southeast direction in the Keck 12.5~\micron{} imaging by
\citet{Shuping04}.  Earlier, the presence of an ultra-dense ionized
circumstellar wind was revealed \citep[e.g.,][]{Scoville83}, and
recent radio recombination line interferometry revealed a velocity
gradient of 10~\kms{} in the northeast-southwest direction
\citep{Rodriguez09}, which might indicate motion in an ionized
jet. Finally, near-infrared imaging polarimetry revealed polarization
patterns consistent with a circumstellar disk, with an elongation
along a direction of 126\degr{} \citep{Jiang05}.

Our MIDI data were collected primarily with the ATs, and show the
presence of a source which is clearly resolved. The derived angular
sizes of the two-dimensional component in arcseconds translates into a
relatively small linear size ($\sim10$--20~AU).  We see clear
indications for deviations from spherical symmetry, and the 2D1D
geometric model can only fit some of the data points reasonably
well. The position angle of the two-dimensional component is around
14\degr{}, i.e. roughly consistent with the axis of the disk of
36\degr{} proposed by \citet{Jiang05}. However, the quality of the
2D1D model fit is rather poor ($\chi^2_\mathrm{r}=6.54$) and this
value may not be well determined.  In particular, the data points with
position angles of around 70\degr{} pose larger problems for the 2D1D
model, indicating the presence of more complex structures than can be
reproduced with this model.  Finally, we note that BN shows
significant differential phase signals in the MIDI data
\citep[cf.][]{Linz08}, indicating deviations from point symmetry.

\section{Summary and conclusions}

In this work, we presented the results of a nine-year program to
observe massive young stellar objects with the two-telescope
mid-infrared interferometric instrument MIDI on the Very Large
Telescope Interferometer of the European Southern Observatory on Cerro
Paranal, Chile.  We measured spectrally-resolved, long-baseline
interferometric visibilities in the 8--13~\micron{} wavelength range
for 20 massive YSOs, with projected baselines ranging from 5 to 130~m.
Where available, we combined these data with previously-published
measurements of the sources in our sample, both from MIDI, and the
Keck segment tilting experiment of \citet{Monnier09}.

Using simple geometric models of the visibilities at a wavelength of
10.6~\micron{}, we derived emission scales and showed significant
asymmetries in the majority of the objects.  Although there is a
general trend of increasing mid-infrared size with luminosity, we
confirm a much looser correlation than that reported for YSOs at
near-infrared wavelengths, as has been found also by \citet{Monnier09}
and \citet{Grellmann11}.  A comparison with literature information
about disks and/or outflows shows that, in general, neither one of
these components is solely responsible for the compact $N$-band
emission.  Consequently, caution should be used when attributing
mid-infrared emission to a particular physical structure for objects
which may have both an outflow and a disk.  For some of the objects in
our sample, the long-baseline MIDI measurements represent the first
detection of any asymmetries; these results can thus be used in
planning future observations to directly detect disk and/or outflow
components in these sources.

For 13 sources which show a deep silicate absorption feature, we fit a
simple model to the correlated flux, consisting of a smooth
(power-law) spectrum, extincted by a foreground screen of material
with a column density and dust composition (a mix of large and small
grains of magnesium-iron silicates, fit for each source) independent
of projected baseline or position angle.  In almost all cases, this
model adequately reproduces the observed correlated fluxes, implying
that the absorbing material is both largely uniform, and detached from
the spatial scales probed by MIDI.  Furthermore, both the grain size
distribution and composition vary throughout the sample.

We make the MIDI observations presented here available in reduced,
electronic form via the CDS in OIFITS format \citep[for a description
  of the OIFITS standard, see the work by][]{Pauls05}, with the hope
that they will prove useful for more detailed studies of these massive
YSOs.

\begin{acknowledgements}

We thank J.~Monnier for providing the reduced Keck data, J.-U.~Pott
for observing the object M17~UC1 upon our request, G.~Meeus for
providing the R~CrA data, and J.~Bouwman for useful discussions
regarding the dust opacities.  It is our pleasure to thank the
anonymous referee, who carefully reviewed this manuscript and provided
helpful critique and suggestions for its improvement.

\end{acknowledgements}

\bibliography{refs}

\Online

\begin{appendix}
\section{Online material}

\begin{longtable}{l l l c c c c}
\caption{Observation log}
\label{tab_obslog}
\\ \hline \hline
Source & Date/time (UTC) & Telescopes & Proj. baseline & Position angle & Avg. visibility & ESO Program ID\\
& & & (m) & (deg.) & \\
\hline
\endfirsthead
\\ \hline \hline
Source & Date/time (UTC) & Telescopes & Proj. baseline & Position angle & Avg. visibility & ESO Program ID\\
& & & (m) & (deg.) & \\
\hline
\endhead
\hline
\endfoot
\hline
\multicolumn{6}{l}{\textbf{Notes.} $\dagger$ Cannot be observed with the ATs due to performance limitations of the STRAP
/AO}\\
\multicolumn{6}{l}{guiding system \citep[cf.][]{Linz11}.}
\endlastfoot
AFGL 2136$^\dagger$ & 2005-06-24~~03:49 & UT3-UT4 & $61.3$ & $107.0$ & $0.03$ & 75.C-0755(A)\\
 & 2005-06-26~~03:02 & UT1-UT2 & $53.5$ & $14.0$ & $0.01$ & 75.C-0755(B)\\
 & 2006-05-18~~08:44 & UT2-UT3 & $46.4$ & $45.9$ & $0.04$ & 77.C-0440(A)\\
 & 2008-06-23~~08:04 & UT2-UT3 & $42.6$ & $47.4$ & $0.03$ & 381.C-0607(A)\\
\\
AFGL 4176 & 2005-03-02~~08:08 & UT3-UT4 & $61.6$ & $112.1$ & $0.02$ & 74.C-0389(B)\\
 & 2005-06-24~~04:54 & UT3-UT4 & $61.8$ & $169.3$ & $0.02$ & 75.C-0755(A)\\
 & 2005-06-25~~22:50 & UT1-UT2 & $49.3$ & $19.3$ & $0.07$ & 75.C-0755(B)\\
 & 2005-06-26~~02:05 & UT1-UT2 & $41.9$ & $49.8$ & $0.08$ & 75.C-0755(B)\\
 & 2005-06-26~~03:27 & UT1-UT2 & $36.3$ & $62.4$ & $0.04$ & 75.C-0755(B)\\
 & 2005-06-27~~03:31 & UT2-UT3 & $30.3$ & $85.6$ & $0.08$ & 75.C-0755(B)\\
 & 2006-02-22~~05:46 & E0-G0 & $16.0$ & $41.9$ & $0.30$ & 76.C-0757(A)\\
 & 2006-04-18~~08:01 & D0-G0 & $26.5$ & $121.4$ & $0.06$ & 77.C-0440(C)\\
 & 2006-04-24~~07:30 & E0-G0 & $13.4$ & $119.4$ & $0.31$ & 77.C-0440(B)\\
 & 2007-02-11~~06:24 & E0-G0 & $16.0$ & $40.5$ & $0.23$ & 77.C-0440(B)\\
 & 2007-02-12~~06:18 & G0-H0 & $31.9$ & $40.0$ & $0.13$ & 77.C-0440(C)\\
 & 2007-03-15~~07:08 & G0-H0 & $31.2$ & $77.7$ & $0.11$ & 78.C-0712(B)\\
 & 2007-03-16~~09:13 & G0-H0 & $28.3$ & $106.8$ & $0.07$ & 78.C-0712(B)\\
 & 2007-04-18~~04:55 & E0-G0 & $15.6$ & $77.8$ & $0.20$ & 78.C-0712(A)\\
 & 2007-05-24~~01:51 & E0-G0 & $15.8$ & $68.7$ & $0.27$ & 78.C-0712(A)\\
 & 2007-06-19~~02:35 & G0-H0 & $28.9$ & $101.0$ & $0.06$ & 78.C-0712(B)\\
 & 2007-07-03~~01:31 & E0-G0 & $14.6$ & $98.9$ & $0.28$ & 78.C-0712(A)\\
 & 2011-05-09~~02:15 & B2-C1 & $9.8$ & $21.7$ & $0.47$ & 84.C-1072(B)\\
 & 2011-05-09~~02:49 & B2-C1 & $9.6$ & $27.0$ & $0.46$ & 84.C-1072(B)\\
 & 2011-05-09~~03:14 & B2-C1 & $9.5$ & $31.0$ & $0.49$ & 84.C-1072(B)\\
 & 2011-05-09~~03:29 & B2-C1 & $9.4$ & $33.4$ & $0.52$ & 84.C-1072(B)\\
 & 2011-05-09~~04:54 & B2-D0 & $25.9$ & $46.4$ & $0.18$ & 84.C-1072(C)\\
 & 2011-05-09~~05:27 & B2-D0 & $24.7$ & $51.6$ & $0.16$ & 84.C-1072(C)\\
 & 2011-05-09~~08:28 & B2-C1 & $5.3$ & $80.9$ & $0.61$ & 84.C-1072(B)\\
 & 2011-05-10~~02:23 & B2-D0 & $29.3$ & $23.5$ & $0.20$ & 84.C-1072(C)\\
 & 2011-05-10~~03:46 & B2-D0 & $27.7$ & $36.7$ & $0.19$ & 84.C-1072(C)\\
 & 2012-02-01~~09:04 & A1-B2 & $10.7$ & $103.5$ & $0.51$ & 84.C-1072(C)\\
 & 2012-02-24~~05:20 & D0-H0 & $63.8$ & $37.3$ & $0.06$ & 84.C-1072(D)\\
 & 2012-02-24~~05:30 & D0-H0 & $63.8$ & $39.8$ & $0.06$ & 84.C-1072(D)\\
 & 2012-02-24~~09:29 & D0-H0 & $59.9$ & $92.0$ & $0.05$ & 84.C-1072(D)\\
 & 2012-03-26~~09:05 & I1-K0 & $32.4$ & $38.6$ & $0.11$ & 84.C-1072(A)\\
 & 2012-03-26~~09:41 & I1-K0 & $30.8$ & $43.1$ & $0.10$ & 84.C-1072(A)\\
 & 2012-03-27~~05:21 & I1-K0 & $38.0$ & $9.1$ & $0.10$ & 84.C-1072(A)\\
 & 2012-03-27~~05:26 & I1-K0 & $37.9$ & $9.7$ & $0.09$ & 84.C-1072(A)\\
 & 2012-03-27~~05:34 & I1-K0 & $37.8$ & $10.9$ & $0.09$ & 84.C-1072(A)\\
 & 2012-03-27~~07:09 & I1-K0 & $36.2$ & $23.8$ & $0.10$ & 84.C-1072(A)\\
 & 2012-03-29~~04:08 & I1-K0 & $38.2$ & $179.9$ & $0.09$ & 84.C-1072(A)\\
 & 2012-04-18~~06:39 & D0-H0 & $57.8$ & $101.7$ & $0.03$ & 84.C-1072(D)\\
 & 2012-04-18~~06:49 & D0-H0 & $57.2$ & $104.0$ & $0.03$ & 84.C-1072(D)\\
\\
G305.20+0.21$^\dagger$ & 2005-03-02~~07:13 & UT3-UT4 & $61.0$ & $106.9$ & $0.03$ & 74.C-0389(B)\\
 & 2005-02-27~~03:35 & UT2-UT3 & $44.7$ & $4.8$ & $0.03$ & 74.C-0389(A)\\
 & 2005-03-02~~05:26 & UT3-UT4 & $56.7$ & $83.4$ & $0.02$ & 74.C-0389(B)\\
 & 2005-03-02~~06:10 & UT3-UT4 & $58.9$ & $93.4$ & $0.03$ & 74.C-0389(B)\\
 & 2005-06-24~~02:46 & UT3-UT4 & $62.3$ & $146.7$ & $0.01$ & 75.C-0755(A)\\
 & 2005-06-26~~01:20 & UT1-UT2 & $42.5$ & $48.1$ & $0.03$ & 75.C-0755(B)\\
\\
IRAS 13481-6124 & 2009-02-15~~04:55 & UT2-UT3 & $44.9$ & $3.4$ & $0.14$ & 82.C-0899(A)\\
 & 2009-05-02~~04:04 & D0-H0 & $62.4$ & $77.3$ & $0.11$ & 83.C-0387(D)\\
 & 2009-05-03~~04:58 & D0-G1 & $67.6$ & $141.1$ & $0.13$ & 83.C-0387(F)\\
 & 2009-07-02~~00:41 & G1-K0 & $73.1$ & $38.9$ & $0.11$ & 83.C-0387(I)\\
 & 2009-07-04~~00:58 & E0-H0 & $45.1$ & $90.7$ & $0.21$ & 83.C-0012(A)\\
 & 2009-07-04~~23:23 & E0-G0 & $15.8$ & $71.0$ & $0.39$ & 83.C-0387(A)\\
 & 2009-07-05~~23:43 & E0-H0 & $46.9$ & $75.9$ & $0.13$ & 83.C-0387(B)\\
 & 2009-07-06~~02:49 & E0-H0 & $40.0$ & $118.7$ & $0.17$ & 83.C-0387(B)\\
 & 2009-07-06~~23:23 & G0-H0 & $31.5$ & $72.5$ & $0.23$ & 83.C-0387(C)\\
 & 2009-07-07~~02:42 & G0-H0 & $26.8$ & $117.9$ & $0.23$ & 83.C-0387(C)\\
 & 2009-07-07~~02:53 & G0-H0 & $26.4$ & $120.7$ & $0.22$ & 83.C-0387(C)\\
 & 2010-05-04~~01:45 & A0-K0 & $128.0$ & $48.6$ & $0.07$ & 85.C-0406(A)\\
 & 2010-05-04~~04:32 & A0-K0 & $122.5$ & $84.8$ & $0.06$ & 85.C-0406(A)\\
 & 2010-05-05~~02:03 & G1-K0 & $79.8$ & $16.0$ & $0.13$ & 85.C-0403(A)\\
 & 2010-05-05~~03:54 & A0-G1 & $87.9$ & $114.1$ & $0.11$ & 85.C-0403(A)\\
 & 2010-05-05~~04:06 & A0-G1 & $88.3$ & $116.4$ & $0.10$ & 85.C-0403(A)\\
 & 2010-05-05~~05:49 & A0-G1 & $90.4$ & $138.2$ & $0.08$ & 85.C-0403(A)\\
 & 2010-05-05~~07:24 & A0-G1 & $90.4$ & $159.3$ & $0.08$ & 85.C-0403(A)\\
 & 2012-05-01~~05:01 & B2-C1 & $9.0$ & $41.6$ & $0.71$ & 89.C-0968(A)\\
 & 2012-05-01~~07:10 & B2-D0 & $22.1$ & $61.2$ & $0.33$ & 89.C-0968(A)\\
 & 2012-05-02~~05:15 & A1-B2 & $11.2$ & $128.9$ & $0.46$ & 89.C-0968(A)\\
 & 2012-05-02~~06:50 & A1-C1 & $13.5$ & $115.3$ & $0.33$ & 89.C-0968(A)\\
 & 2012-05-02~~07:50 & A1-D0 & $22.0$ & $98.8$ & $0.20$ & 89.C-0968(A)\\
 & 2012-05-02~~08:40 & A1-B2 & $11.3$ & $174.5$ & $0.49$ & 89.C-0968(A)\\
 & 2012-05-02~~09:08 & B2-D0 & $15.4$ & $81.4$ & $0.33$ & 89.C-0968(A)\\
 & 2013-02-12~~06:38 & B2-D0 & $30.3$ & $8.6$ & $0.33$ & 90.C-0717(A)\\
 & 2013-02-12~~08:16 & C1-D0 & $19.6$ & $24.2$ & $0.38$ & 90.C-0717(A)\\
 & 2013-02-13~~04:03 & B2-D0 & $30.0$ & $164.2$ & $0.17$ & 90.C-0717(A)\\
 & 2013-02-13~~04:46 & C1-D0 & $20.2$ & $171.2$ & $0.29$ & 90.C-0717(A)\\
 & 2013-02-13~~05:50 & C1-D0 & $20.3$ & $1.5$ & $0.27$ & 90.C-0717(A)\\
 & 2013-02-13~~06:22 & A1-D0 & $34.6$ & $21.5$ & $0.29$ & 90.C-0717(A)\\
 & 2013-02-13~~07:13 & A1-D0 & $34.2$ & $31.1$ & $0.29$ & 90.C-0717(A)\\
 & 2013-02-13~~08:22 & B2-D0 & $29.2$ & $25.7$ & $0.27$ & 90.C-0717(A)\\
 & 2013-02-16~~04:54 & G1-H0 & $55.2$ & $158.5$ & $0.14$ & 90.C-0717(B)\\
 & 2013-02-16~~09:00 & G1-H0 & $57.2$ & $11.2$ & $0.17$ & 90.C-0717(B)\\
 & 2013-02-16~~09:32 & D0-I1 & $82.3$ & $108.0$ & $0.09$ & 90.C-0717(B)\\
\\
IRAS 17216-3801 & 2012-05-01~~05:48 & B2-C1 & $11.3$ & $13.9$ & $0.37$ & 89.C-0968(A)\\
 & 2012-05-01~~09:23 & A1-B2 & $11.0$ & $134.1$ & $0.34$ & 89.C-0968(A)\\
 & 2012-05-01~~09:46 & B2-C1 & $10.0$ & $40.2$ & $0.41$ & 89.C-0968(A)\\
 & 2012-05-02~~04:34 & A1-B2 & $8.9$ & $91.5$ & $0.41$ & 89.C-0968(A)\\
 & 2012-05-02~~07:11 & A1-C1 & $16.0$ & $69.6$ & $0.22$ & 89.C-0968(A)\\
 & 2012-05-02~~08:14 & A1-D0 & $34.2$ & $51.5$ & $0.15$ & 89.C-0968(A)\\
 & 2012-05-02~~09:33 & B2-D0 & $30.5$ & $39.5$ & $0.11$ & 89.C-0968(A)\\
 & 2013-02-12~~07:33 & B2-D0 & $33.8$ & $165.1$ & $0.16$ & 90.C-0717(A)\\
 & 2013-02-12~~08:46 & C1-D0 & $22.6$ & $175.5$ & $0.12$ & 90.C-0717(A)\\
 & 2013-02-13~~07:48 & A1-D0 & $35.1$ & $176.9$ & $0.18$ & 90.C-0717(A)\\
\\
M8E-IR$^\dagger$ & 2004-06-05~~08:06 & UT1-UT3 & $96.8$ & $42.7$ & $0.09$ & 60.A-9224(A)\\
 & 2004-06-05~~09:53 & UT1-UT3 & $82.8$ & $44.6$ & $0.08$ & 60.A-9224(A)\\
 & 2004-08-01~~01:50 & UT2-UT3 & $46.6$ & $38.4$ & $0.26$ & 273.C-5044(A)\\
 & 2005-03-02~~08:58 & UT3-UT4 & $46.8$ & $94.1$ & $0.17$ & 74.C-0389(B)\\
 & 2005-06-24~~07:36 & UT3-UT4 & $51.6$ & $137.8$ & $0.24$ & 75.C-0755(A)\\
 & 2005-06-26~~00:26 & UT1-UT2 & $55.7$ & $173.4$ & $0.15$ & 75.C-0755(B)\\
\\
M17 SW IRS1 & 2005-06-24~~08:23 & UT3-UT4 & $43.9$ & $140.7$ & $0.09$ & 75.C-0755(A)\\
 & 2005-06-26~~04:25 & UT1-UT2 & $55.8$ & $24.3$ & $0.03$ & 75.C-0755(B)\\
 & 2006-05-18~~09:59 & UT2-UT3 & $43.5$ & $48.4$ & $0.30$ & 77.C-0440(A)\\
 & 2006-08-10~~01:27 & D0-G0 & $31.7$ & $69.6$ & $0.45$ & 77.C-0440(C)\\
 & 2006-08-13~~00:22 & E0-G0 & $15.1$ & $64.7$ & $0.69$ & 77.C-0440(B)\\
\\
M17 UC1$^\dagger$ & 2006-06-12~~09:33 & UT3-UT4 & $41.3$ & $146.9$ & $0.03$ & 77.D-0709(B)\\
\\
Mon R2 IRS2 & 2009-02-15~~00:48 & UT2-UT3 & $44.6$ & $39.9$ & $0.19$ & 82.C-0899(A)\\
 & 2009-11-14~~04:11 & E0-G0 & $10.8$ & $55.8$ & $0.78$ & 84.C-1072(A)\\
 & 2009-11-15~~04:06 & E0-G0 & $10.7$ & $55.4$ & $0.84$ & 84.C-1072(A)\\
 & 2009-11-15~~05:16 & G0-H0 & $26.9$ & $64.9$ & $0.19$ & 84.C-1072(B)\\
 & 2009-11-15~~06:39 & G0-H0 & $31.1$ & $70.6$ & $0.20$ & 84.C-1072(B)\\
\\
Mon R2 IRS3 A & 2005-02-27~~02:26 & UT2-UT3 & $46.2$ & $46.3$ & $0.04$ & 74.C-0389(A)\\
 & 2006-03-13~~01:19 & UT3-UT4 & $52.9$ & $117.0$ & $0.02$ & 76.C-0757(E)\\
 & 2009-10-02~~08:04 & G0-H0 & $26.6$ & $64.4$ & $0.11$ & 84.C-1072(B)\\
 & 2009-11-15~~04:36 & E0-G0 & $12.0$ & $60.3$ & $0.24$ & 84.C-1072(A)\\
 & 2009-11-15~~06:59 & G0-H0 & $31.6$ & $71.5$ & $0.09$ & 84.C-1072(B)\\
\\
Mon R2 IRS3 B & 2005-02-27~~02:49 & UT2-UT3 & $45.6$ & $46.2$ & $0.16$ & 74.C-0389(A)\\
 & 2006-03-13~~02:20 & UT3-UT4 & $44.1$ & $125.0$ & $0.17$ & 76.C-0757(E)\\
\\
MWC 300 & 2009-05-03~~08:34 & D0-G1 & $69.2$ & $133.6$ & $0.15$ & 83.C-0387(F)\\
 & 2009-05-04~~06:21 & G1-H0 & $67.6$ & $174.3$ & $0.09$ & 83.C-0387(G)\\
 & 2009-05-04~~08:52 & G1-H0 & $68.4$ & $11.2$ & $0.06$ & 83.C-0387(G)\\
 & 2009-06-30~~05:24 & A0-K0 & $127.4$ & $73.2$ & $0.05$ & 83.C-0387(E)\\
 & 2009-06-30~~05:35 & A0-K0 & $126.8$ & $73.4$ & $0.05$ & 83.C-0387(E)\\
 & 2009-06-30~~06:06 & A0-K0 & $123.5$ & $73.8$ & $0.06$ & 83.C-0387(E)\\
 & 2009-06-30~~08:26 & A0-K0 & $84.6$ & $70.8$ & $0.16$ & 83.C-0387(E)\\
 & 2009-07-01~~04:25 & A0-G1 & $90.0$ & $114.5$ & $0.10$ & 83.C-0387(H)\\
 & 2009-07-04~~05:56 & E0-H0 & $46.0$ & $73.8$ & $0.15$ & 83.C-0012(B)\\
 & 2009-07-04~~07:26 & E0-H0 & $37.7$ & $72.9$ & $0.23$ & 83.C-0012(B)\\
 & 2009-07-05~~04:28 & E0-G0 & $16.0$ & $72.3$ & $0.67$ & 83.C-0387(A)\\
 & 2009-07-06~~05:57 & E0-H0 & $45.4$ & $73.9$ & $0.14$ & 83.C-0387(B)\\
 & 2009-07-07~~05:01 & G0-H0 & $31.8$ & $73.3$ & $0.07$ & 83.C-0387(C)\\
 & 2010-05-04~~09:34 & A0-K0 & $125.6$ & $73.6$ & $0.04$ & 85.C-0406(A)\\
 & 2010-05-20~~05:49 & H0-I1 & $40.7$ & $142.8$ & $0.35$ & 85.C-0406(C)\\
 & 2012-05-01~~07:34 & B2-D0 & $31.3$ & $21.0$ & $0.36$ & 89.C-0968(A)\\
 & 2012-05-02~~05:59 & A1-B2 & $10.2$ & $113.2$ & $0.70$ & 89.C-0968(A)\\
 & 2012-05-02~~09:55 & B2-D0 & $33.8$ & $33.3$ & $0.29$ & 89.C-0968(A)\\
\\
NGC 2264 IRS1$^\dagger$ & 2009-02-15~~01:12 & UT2-UT3 & $40.2$ & $43.9$ & $0.17$ & 82.C-0899(A)\\
\\
Orion BN & 2005-02-27~~00:29 & UT2-UT3 & $46.2$ & $44.1$ & $0.07$ & 74.C-0389(A)\\
 & 2006-09-14~~08:36 & E0-G0 & $13.0$ & $64.3$ & $0.29$ & 78.C-0712(A)\\
 & 2006-10-20~~06:00 & D0-G0 & $24.8$ & $62.7$ & $0.25$ & 78.C-0712(B)\\
 & 2006-12-15~~01:44 & E0-G0 & $11.0$ & $57.8$ & $0.28$ & 78.C-0712(A)\\
 & 2006-12-15~~06:41 & E0-G0 & $14.8$ & $73.6$ & $0.35$ & 78.C-0712(A)\\
 & 2007-02-09~~03:04 & G0-H0 & $29.4$ & $73.6$ & $0.25$ & 78.C-0712(B)\\
 & 2013-02-12~~02:14 & A1-B2 & $10.0$ & $119.9$ & $0.22$ & 90.C-0717(A)\\
 & 2013-02-12~~04:15 & B2-C1 & $11.1$ & $35.1$ & $0.29$ & 90.C-0717(A)\\
 & 2013-02-12~~04:40 & B2-C1 & $10.9$ & $34.5$ & $0.35$ & 90.C-0717(A)\\
 & 2013-02-13~~00:17 & C1-D0 & $21.2$ & $24.7$ & $0.19$ & 90.C-0717(A)\\
 & 2013-02-13~~02:04 & A1-D0 & $35.8$ & $49.5$ & $0.14$ & 90.C-0717(A)\\
 & 2013-02-13~~03:05 & B2-D0 & $33.9$ & $35.1$ & $0.12$ & 90.C-0717(A)\\
 & 2013-02-13~~05:07 & C1-D0 & $21.4$ & $33.4$ & $0.17$ & 90.C-0717(A)\\
 & 2013-02-16~~00:32 & H0-I1 & $39.1$ & $148.3$ & $0.19$ & 90.C-0717(B)\\
 & 2013-02-16~~01:19 & D0-H0 & $63.7$ & $73.2$ & $0.12$ & 90.C-0717(B)\\
 & 2013-02-16~~01:57 & D0-G1 & $63.1$ & $140.8$ & $0.05$ & 90.C-0717(B)\\
 & 2013-02-16~~02:04 & D0-G1 & $62.5$ & $141.7$ & $0.05$ & 90.C-0717(B)\\
 & 2013-02-16~~02:43 & H0-I1 & $35.3$ & $164.7$ & $0.12$ & 90.C-0717(B)\\
 & 2013-02-16~~03:06 & H0-I1 & $34.8$ & $168.6$ & $0.11$ & 90.C-0717(B)\\
 & 2013-02-16~~04:01 & G1-H0 & $71.4$ & $24.1$ & $0.07$ & 90.C-0717(B)\\
\\
R CrA & 2004-07-09~~09:10 & UT2-UT3 & $27.4$ & $63.5$ & $0.48$ & 73.C-0828(B)\\
 & 2004-07-28~~02:04 & UT2-UT3 & $46.6$ & $30.5$ & $0.20$ & 73.C-0828(B)\\
 & 2004-07-29~~23:49 & UT2-UT3 & $46.2$ & $9.5$ & $0.21$ & 73.C-0828(B)\\
 & 2004-07-30~~03:32 & UT2-UT3 & $45.6$ & $43.3$ & $0.34$ & 73.C-0828(B)\\
 & 2004-07-30~~05:25 & UT2-UT3 & $40.6$ & $54.9$ & $0.26$ & 73.C-0828(B)\\
 & 2004-07-30~~05:33 & UT2-UT3 & $40.1$ & $55.5$ & $0.30$ & 73.C-0828(B)\\
 & 2004-07-30~~05:40 & UT2-UT3 & $39.6$ & $56.1$ & $0.27$ & 73.C-0828(B)\\
 & 2005-06-23~~07:33 & UT1-UT4 & $115.3$ & $76.1$ & $0.26$ & 75.C-0801(B)\\
 & 2005-06-25~~07:29 & UT1-UT4 & $114.6$ & $76.6$ & $0.09$ & 75.C-0801(B)\\
 & 2005-06-25~~07:39 & UT1-UT4 & $112.5$ & $77.8$ & $0.10$ & 75.C-0801(B)\\
 & 2005-06-26~~05:57 & UT1-UT4 & $127.5$ & $65.3$ & $0.10$ & 75.C-0801(B)\\
 & 2005-06-28~~04:40 & UT3-UT4 & $61.2$ & $105.0$ & $0.22$ & 75.C-0801(A)\\
 & 2005-06-28~~07:40 & UT3-UT4 & $58.8$ & $134.5$ & $0.24$ & 75.C-0801(A)\\
 & 2005-07-21~~01:29 & UT3-UT4 & $53.3$ & $90.8$ & $0.32$ & 75.C-0801(A)\\
 & 2005-07-21~~02:26 & UT3-UT4 & $58.7$ & $98.9$ & $0.22$ & 75.C-0801(A)\\
 & 2005-07-21~~04:35 & UT3-UT4 & $62.3$ & $117.6$ & $0.33$ & 75.C-0801(A)\\
 & 2005-09-18~~02:10 & UT1-UT4 & $111.2$ & $78.5$ & $0.18$ & 75.C-0801(B)\\
 & 2009-05-03~~09:33 & D0-G1 & $71.4$ & $138.5$ & $0.20$ & 83.C-0387(F)\\
 & 2009-05-04~~07:06 & G1-H0 & $70.1$ & $175.9$ & $0.18$ & 83.C-0387(G)\\
 & 2009-05-04~~09:33 & G1-H0 & $69.5$ & $12.0$ & $0.17$ & 83.C-0387(G)\\
 & 2009-07-01~~06:55 & A0-G1 & $88.3$ & $133.1$ & $0.15$ & 83.C-0387(H)\\
 & 2009-07-04~~08:36 & E0-H0 & $32.3$ & $106.1$ & $0.54$ & 83.C-0012(D)\\
 & 2009-07-05~~01:21 & E0-G0 & $14.0$ & $32.5$ & $0.67$ & 83.C-0387(A)\\
 & 2009-07-05~~05:21 & E0-G0 & $15.8$ & $76.0$ & $0.37$ & 83.C-0387(A)\\
 & 2009-07-06~~04:43 & E0-H0 & $47.9$ & $70.9$ & $0.34$ & 83.C-0387(B)\\
 & 2009-07-06~~07:38 & E0-H0 & $37.8$ & $96.8$ & $0.36$ & 83.C-0387(B)\\
 & 2009-07-07~~00:56 & G0-H0 & $27.5$ & $28.4$ & $0.17$ & 83.C-0387(C)\\
 & 2009-07-07~~01:10 & G0-H0 & $27.9$ & $31.8$ & $0.16$ & 83.C-0387(C)\\
\\
R Mon & 2010-01-13~~04:29 & D0-G1 & $61.9$ & $130.9$ & $0.26$ & 84.C-0183(F)\\
 & 2010-01-14~~05:05 & D0-H0 & $63.3$ & $71.1$ & $0.17$ & 84.C-0183(G)\\
 & 2010-01-16~~02:24 & E0-G0 & $14.1$ & $75.1$ & $0.56$ & 84.C-0183(A)\\
 & 2010-01-16~~05:19 & E0-H0 & $46.7$ & $69.9$ & $0.22$ & 84.C-0183(C)\\
 & 2010-01-19~~04:43 & G1-K0 & $85.3$ & $34.7$ & $0.13$ & 84.C-0183(H)\\
 & 2010-01-21~~02:40 & A0-K0 & $120.6$ & $74.8$ & $0.06$ & 84.C-0183(E)\\
 & 2010-01-21~~06:02 & A0-G1 & $56.8$ & $116.9$ & $0.22$ & 84.C-0183(E)\\
\\
S255 IRS3 & 2010-01-15~~03:31 & E0-G0 & $15.8$ & $74.8$ & $0.28$ & 84.C-0183(D)\\
 & 2010-01-16~~03:46 & E0-H0 & $47.8$ & $73.5$ & $0.07$ & 84.C-0183(C)\\
 & 2010-01-17~~02:54 & G0-H0 & $30.8$ & $76.7$ & $0.10$ & 84.C-0183(B)\\
 & 2010-01-17~~03:05 & G0-H0 & $31.1$ & $76.0$ & $0.10$ & 84.C-0183(B)\\
 & 2010-01-17~~04:40 & E0-H0 & $47.5$ & $68.9$ & $0.08$ & 84.C-0183(B)\\
 & 2013-02-13~~01:09 & C1-D0 & $17.6$ & $32.6$ & $0.30$ & 90.C-0717(A)\\
\\
V921 Sco & 2009-05-02~~08:33 & D0-H0 & $59.0$ & $87.2$ & $0.58$ & 83.C-0387(D)\\
 & 2009-05-03~~07:49 & D0-G1 & $71.2$ & $141.0$ & $0.42$ & 83.C-0387(F)\\
 & 2009-05-04~~04:03 & G1-H0 & $67.7$ & $169.0$ & $0.38$ & 83.C-0387(G)\\
 & 2009-05-04~~08:08 & G1-H0 & $66.8$ & $16.2$ & $0.44$ & 83.C-0387(G)\\
 & 2009-06-30~~00:01 & A0-K0 & $118.7$ & $36.4$ & $0.25$ & 83.C-0387(E)\\
 & 2009-06-30~~00:36 & A0-K0 & $121.9$ & $44.2$ & $0.29$ & 83.C-0387(E)\\
 & 2009-06-30~~06:51 & A0-K0 & $88.5$ & $112.4$ & $0.25$ & 83.C-0387(E)\\
 & 2009-07-01~~03:20 & A0-G1 & $90.2$ & $117.5$ & $0.31$ & 83.C-0387(H)\\
 & 2009-07-01~~04:46 & A0-G1 & $89.6$ & $132.6$ & $0.25$ & 83.C-0387(H)\\
 & 2009-07-02~~01:11 & G1-K0 & $89.3$ & $12.9$ & $0.31$ & 83.C-0387(I)\\
 & 2009-07-04~~02:08 & E0-H0 & $48.0$ & $64.6$ & $0.81$ & 83.C-0012(C)\\
 & 2009-07-04~~06:34 & E0-H0 & $33.3$ & $112.1$ & $0.66$ & 83.C-0012(B)\\
 & 2009-07-04~~06:49 & E0-H0 & $31.8$ & $115.9$ & $0.79$ & 83.C-0012(B)\\
 & 2009-07-06~~02:03 & E0-H0 & $48.0$ & $65.2$ & $0.73$ & 83.C-0387(B)\\
 & 2009-07-06~~06:50 & E0-H0 & $30.9$ & $118.2$ & $0.72$ & 83.C-0387(B)\\
 & 2009-07-07~~02:16 & G0-H0 & $32.0$ & $68.0$ & $1.03$ & 83.C-0387(C)\\
 & 2009-07-07~~05:18 & G0-H0 & $26.3$ & $98.5$ & $1.03$ & 83.C-0387(C)\\
 & 2009-07-07~~07:03 & G0-H0 & $19.4$ & $123.2$ & $0.81$ & 83.C-0387(C)\\
 & 2010-05-04~~03:08 & A0-K0 & $115.4$ & $27.4$ & $0.15$ & 85.C-0406(A)\\
 & 2010-05-04~~05:13 & A0-K0 & $125.8$ & $54.5$ & $0.19$ & 85.C-0406(A)\\
 & 2010-05-04~~08:23 & A0-K0 & $118.5$ & $86.7$ & $0.31$ & 85.C-0406(A)\\
 & 2010-05-04~~10:04 & A0-K0 & $97.1$ & $105.2$ & $0.29$ & 85.C-0406(A)\\
 & 2010-05-05~~06:34 & A0-G1 & $89.1$ & $112.7$ & $0.26$ & 85.C-0403(A)\\
 & 2010-05-05~~08:35 & A0-G1 & $89.5$ & $133.3$ & $0.32$ & 85.C-0403(A)\\
 & 2010-05-05~~09:22 & A0-G1 & $87.7$ & $142.7$ & $0.33$ & 85.C-0403(A)\\
 & 2010-05-05~~10:08 & A0-G1 & $85.7$ & $152.6$ & $0.30$ & 85.C-0403(A)\\
 & 2010-05-20~~05:11 & H0-I1 & $38.5$ & $143.9$ & $0.61$ & 85.C-0406(C)\\
 & 2010-05-20~~07:01 & H0-I1 & $40.2$ & $158.1$ & $0.56$ & 85.C-0406(C)\\
\\
V1028 Cen & 2009-05-02~~02:41 & D0-H0 & $63.8$ & $68.8$ & $0.28$ & 83.C-0387(D)\\
 & 2009-05-02~~07:09 & D0-H0 & $43.8$ & $124.3$ & $0.22$ & 83.C-0387(D)\\
 & 2009-07-03~~23:28 & E0-H0 & $46.7$ & $78.8$ & $0.43$ & 83.C-0012(A)\\
 & 2009-07-05~~00:18 & E0-G0 & $14.9$ & $88.6$ & $0.72$ & 83.C-0387(A)\\
 & 2009-07-06~~01:12 & E0-H0 & $41.0$ & $100.0$ & $0.33$ & 83.C-0387(B)\\
 & 2009-07-07~~00:08 & G0-H0 & $29.8$ & $88.2$ & $0.48$ & 83.C-0387(C)\\
 & 2010-05-20~~06:28 & H0-I1 & $40.0$ & $9.3$ & $0.15$ & 85.C-0406(C)\\
 & 2010-05-21~~04:08 & E0-H0 & $41.4$ & $98.7$ & $0.63$ & 85.C-0403(C)\\
 & 2012-05-01~~06:18 & B2-C1 & $8.4$ & $50.1$ & $0.70$ & 89.C-0968(A)\\
 & 2012-05-02~~06:15 & A1-B2 & $11.0$ & $152.1$ & $0.69$ & 89.C-0968(A)\\
 & 2013-02-12~~05:27 & C1-D0 & $22.0$ & $4.8$ & $0.51$ & 90.C-0717(A)\\
 & 2013-02-12~~06:09 & B2-D0 & $32.8$ & $11.0$ & $0.44$ & 90.C-0717(A)\\
 & 2013-02-13~~03:29 & B2-D0 & $32.8$ & $168.0$ & $0.38$ & 90.C-0717(A)\\
 & 2013-02-13~~05:26 & C1-D0 & $22.0$ & $5.2$ & $0.42$ & 90.C-0717(A)\\
 & 2013-02-13~~06:40 & A1-D0 & $35.5$ & $32.7$ & $0.40$ & 90.C-0717(A)\\
 & 2013-02-13~~08:39 & B2-D0 & $31.0$ & $32.1$ & $0.53$ & 90.C-0717(A)\\
 & 2013-02-16~~03:24 & H0-I1 & $23.8$ & $116.8$ & $0.44$ & 90.C-0717(B)\\
 & 2013-02-16~~06:18 & D0-I1 & $77.2$ & $80.4$ & $0.29$ & 90.C-0717(B)\\
 & 2013-02-16~~06:45 & H0-I1 & $36.0$ & $138.1$ & $0.29$ & 90.C-0717(B)\\
\\
VY Mon & 2010-01-13~~03:23 & D0-G1 & $67.2$ & $128.4$ & $0.33$ & 84.C-0183(F)\\
 & 2010-01-14~~03:45 & D0-H0 & $63.2$ & $74.0$ & $0.39$ & 84.C-0183(G)\\
 & 2010-01-15~~04:19 & E0-G0 & $16.0$ & $72.5$ & $0.82$ & 84.C-0183(D)\\
 & 2010-01-16~~04:37 & E0-H0 & $47.8$ & $71.5$ & $0.43$ & 84.C-0183(C)\\
 & 2010-01-17~~03:43 & G0-H0 & $31.8$ & $73.7$ & $0.65$ & 84.C-0183(B)\\
 & 2010-01-19~~02:56 & G1-K0 & $75.1$ & $28.7$ & $0.17$ & 84.C-0183(H)\\
 & 2010-01-19~~05:51 & G1-K0 & $89.2$ & $35.8$ & $0.13$ & 84.C-0183(H)\\
 & 2010-01-22~~03:58 & A0-K0 & $127.9$ & $72.2$ & $0.16$ & 84.C-0183(I)\\
\multicolumn{6}{l}{\it Sources without fringe detections:}\\
GGD 27 & 2004-06-02 04:00 & UT2-UT3 & 43.9 & 21.6 & --- & 60.A-9224(A)\\
\\
IRAS 16164-5046 & 2009-05-02 06:16 & D0-H0 & 63.5 & 251.7 & --- & 83.C-0387(D)\\
 & 2009-05-03 05:33 & D0-G1 & 65.4 & 303.8 & --- & 83.C-0387(F)\\
 & 2009-07-05 03:16 & G0-H0 & 30.3 & 265.4 & --- & 83.C-0387(A)\\
\\
IRAS 17136-3617 & 2006-05-18 03:39 & UT2-UT3 & 46.3 & 36.2 & --- & 77.C-0440(A)\\
 & 2006-05-26 08:34 & D0-G0 & 25.2 & 76.5 & --- & 77.C-0440(C)\\
\\
Orion SC3 & 2005-02-27 01:38 & UT2-UT3 & 46.5 & 46.1 & --- & 74.C-0389(A)\\

\end{longtable}

\begin{figure*}
  \begin{center}
    \includegraphics[scale=0.85,bb=0 670 470 793]{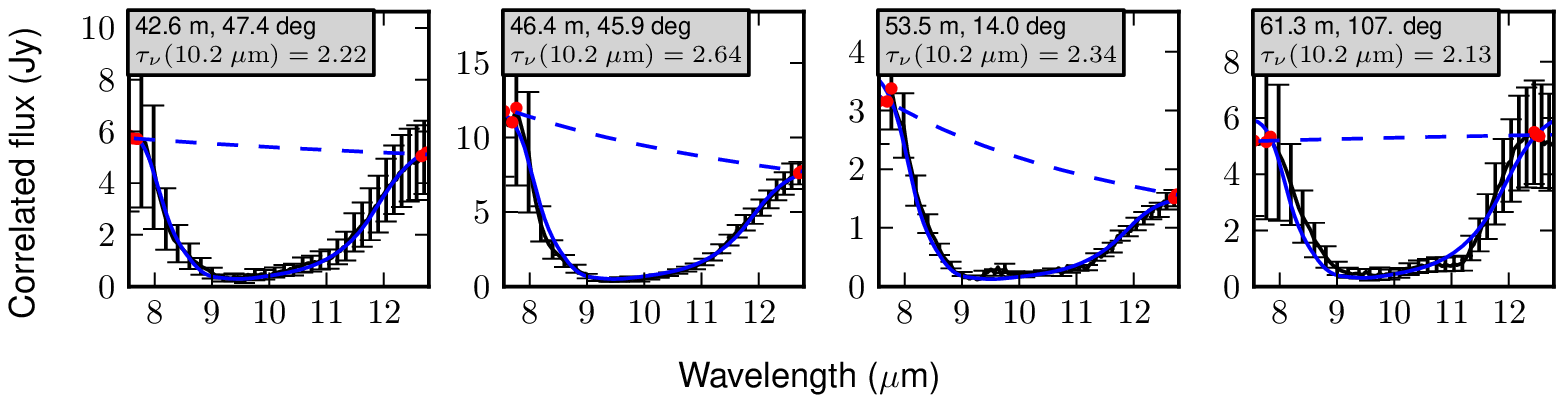}
    \caption{\cfluxcapfit{AFGL~2136}}
    \label{fig_cflux_afgl2136}
  \end{center}
\end{figure*}

\begin{figure*}
  \begin{center}
    \includegraphics[scale=0.85,bb=0 0 570 793]{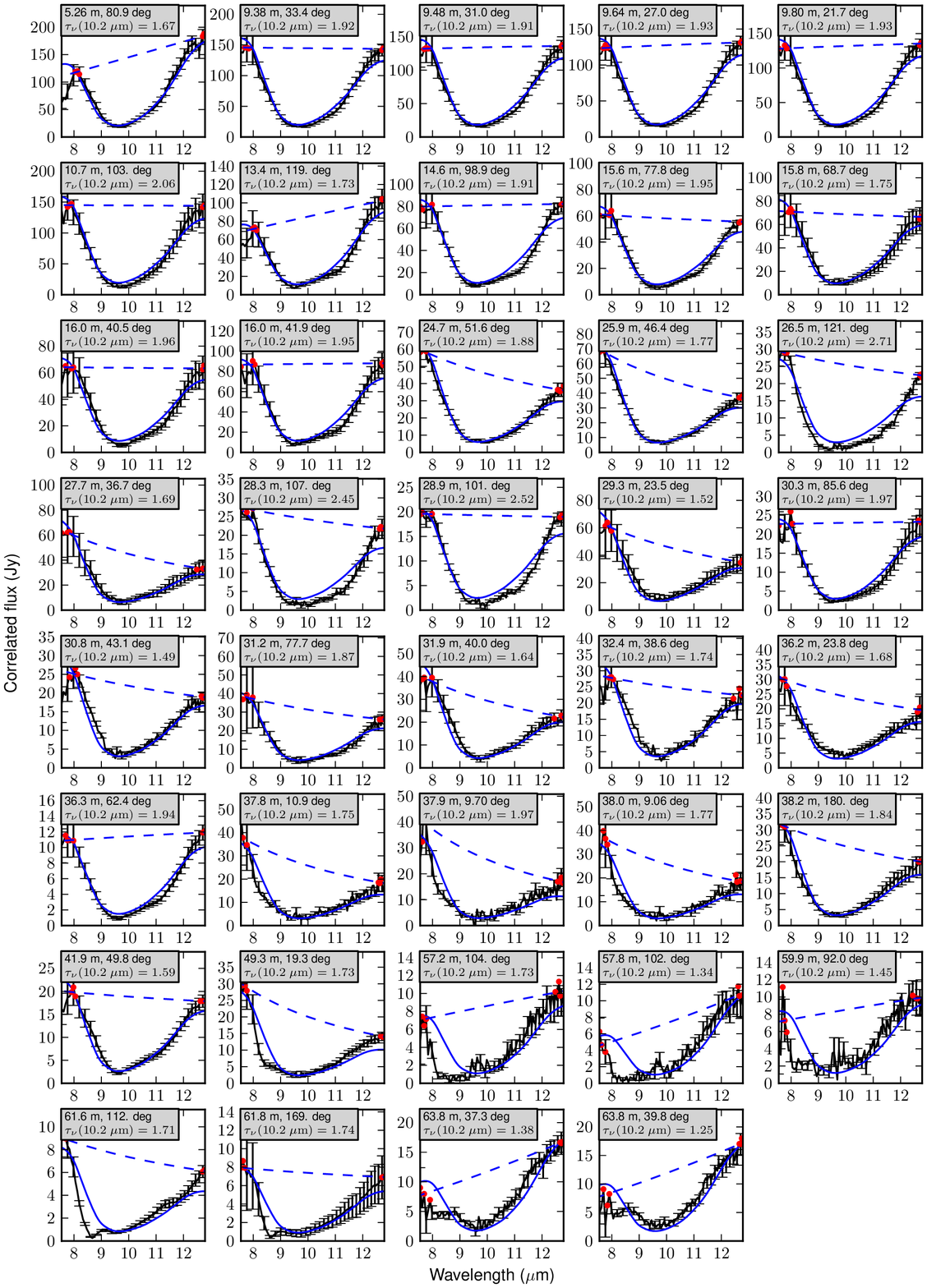}
    \caption{\cfluxcapfit{AFGL~4176}}
    \label{fig_cflux_afgl4176}
  \end{center}
\end{figure*}

\begin{figure*}
  \begin{center}
    \includegraphics[scale=0.85,bb=0 570 367 793]{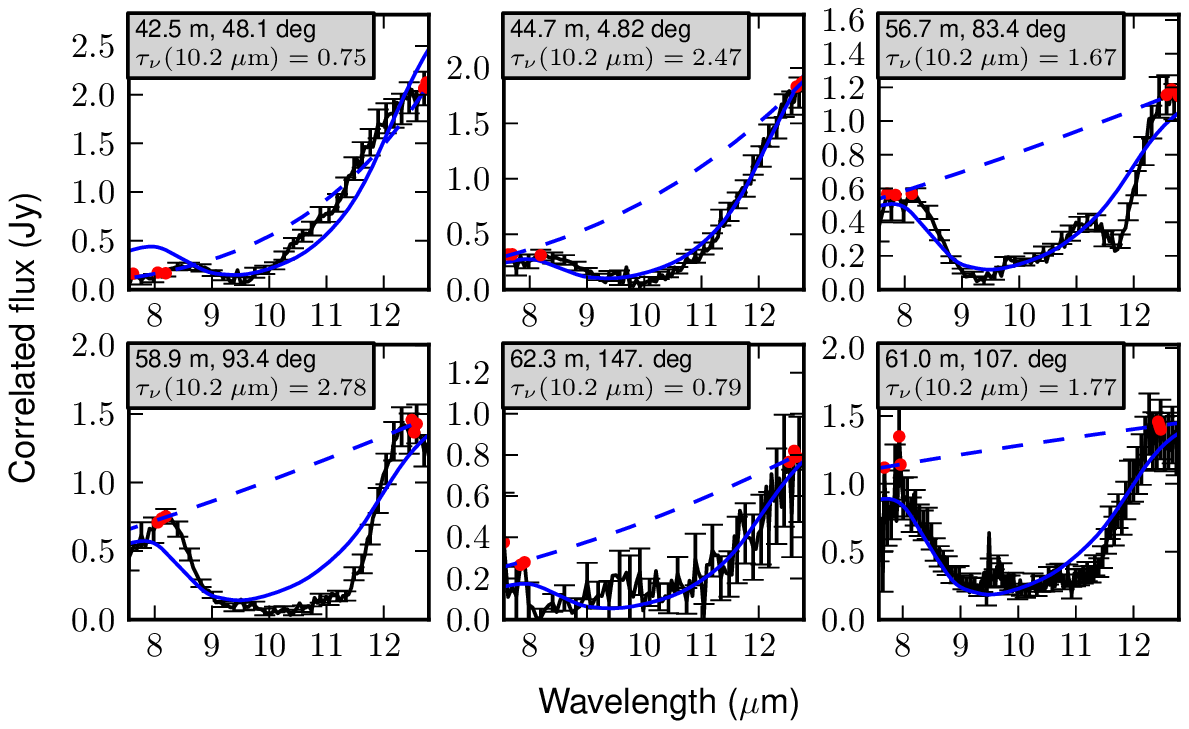}
    \caption{\cfluxcapfit{G305.20+0.21}}
    \label{fig_cflux_g305}
  \end{center}
\end{figure*}

\begin{figure*}
  \begin{center}
    \includegraphics[scale=0.85,bb=0 0 570 793]{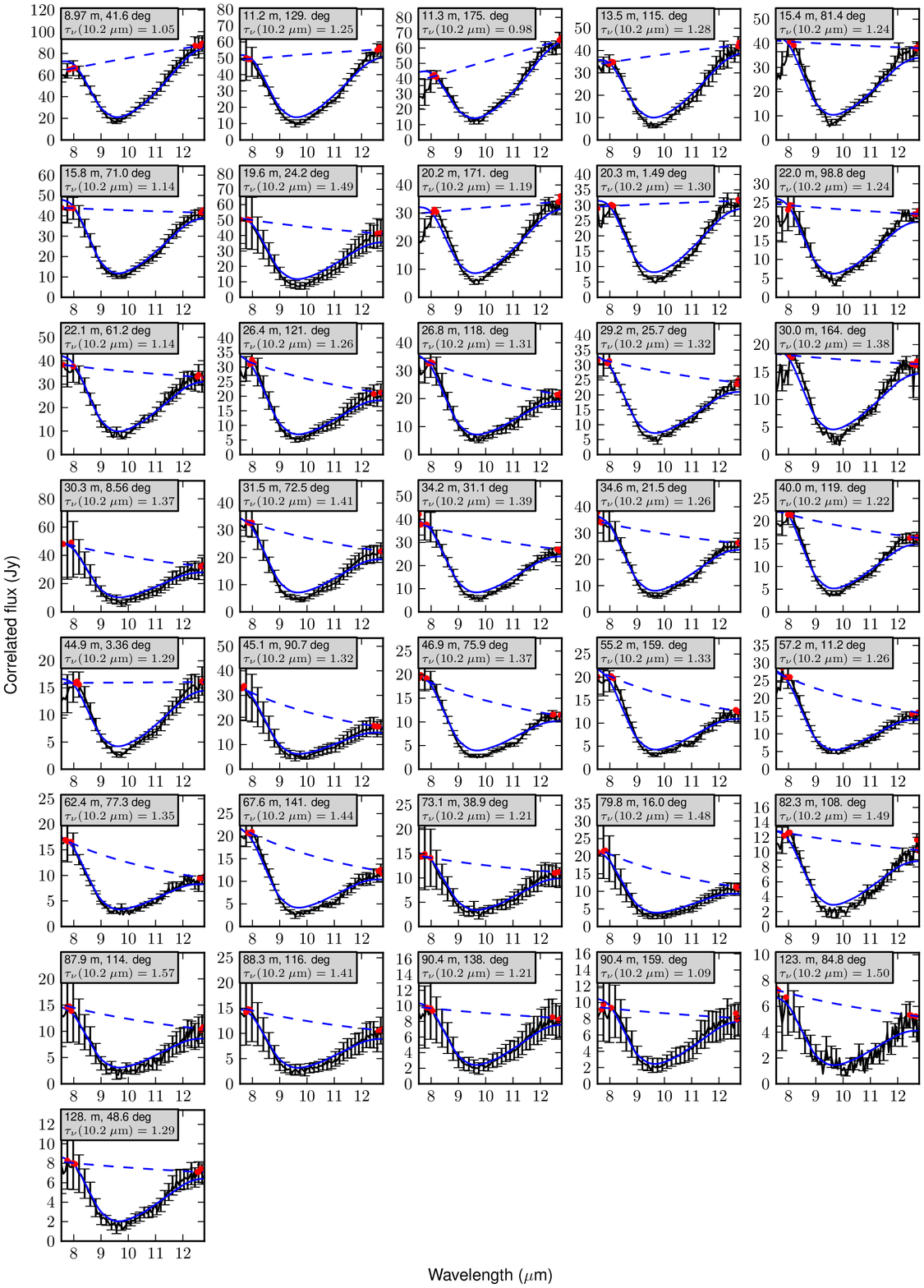}
    \caption{\cfluxcapfit{IRAS~13481-6124}}
    \label{fig_cflux_iras13481}
  \end{center}
\end{figure*}

\begin{figure*}
  \begin{center}
    \includegraphics[scale=0.85,bb=0 570 570 793]{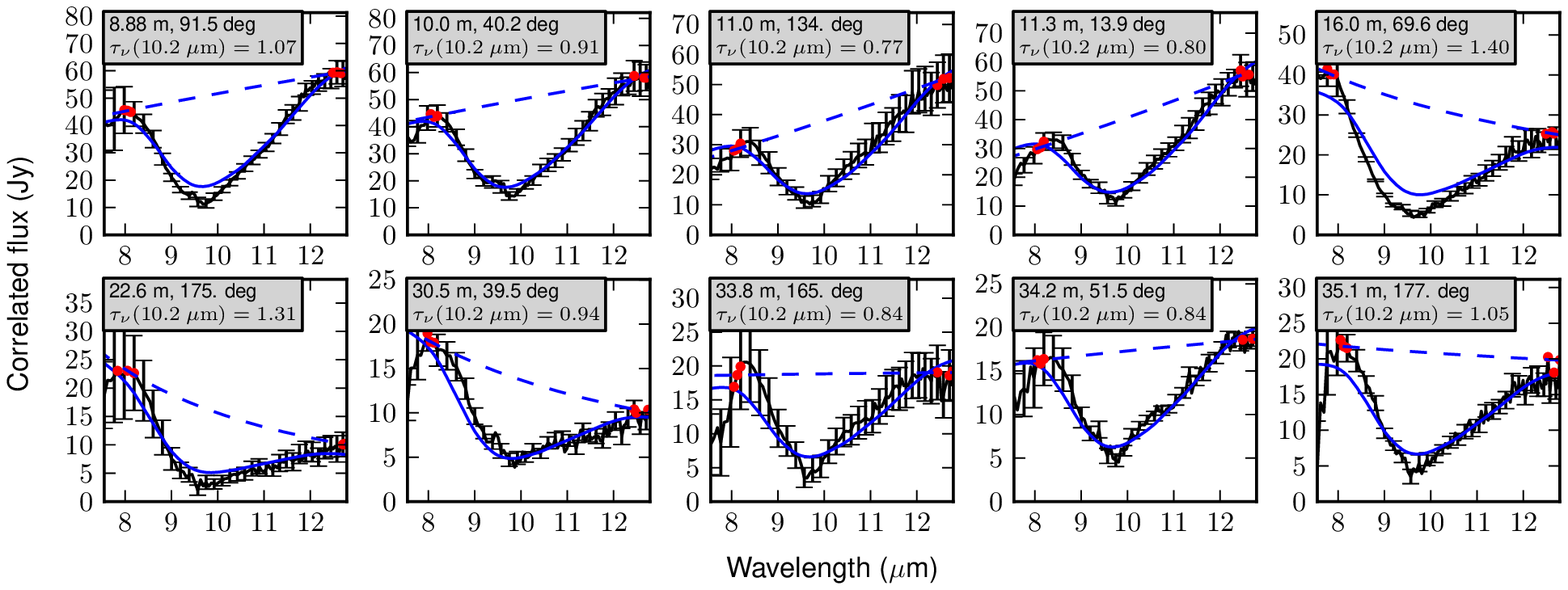}
    \caption{\cfluxcapfit{IRAS~17216-3801}}
    \label{fig_cflux_iras17216}
  \end{center}
\end{figure*}

\begin{figure*}
  \begin{center}
    \includegraphics[scale=0.85,bb=0 570 367 793]{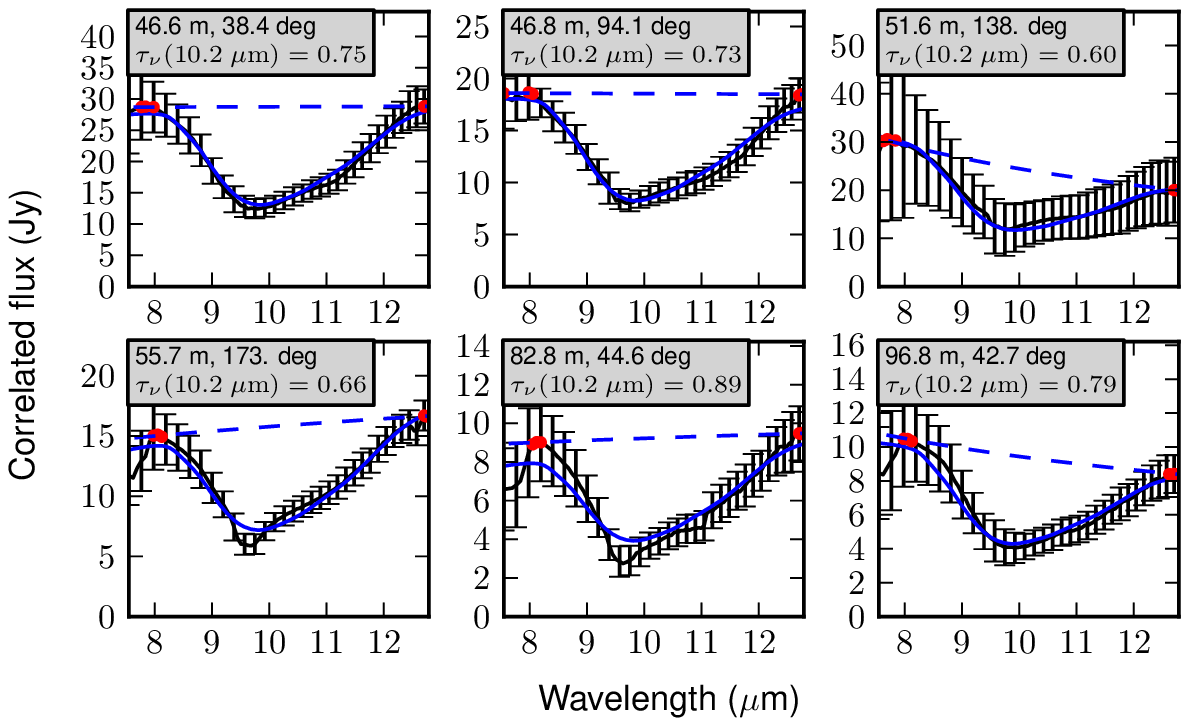}
    \caption{\cfluxcapfit{M8E-IR}}
    \label{fig_cflux_m8eir}
  \end{center}
\end{figure*}

\begin{figure*}
  \begin{center}
    \includegraphics[scale=0.85,bb=0 670 570 793]{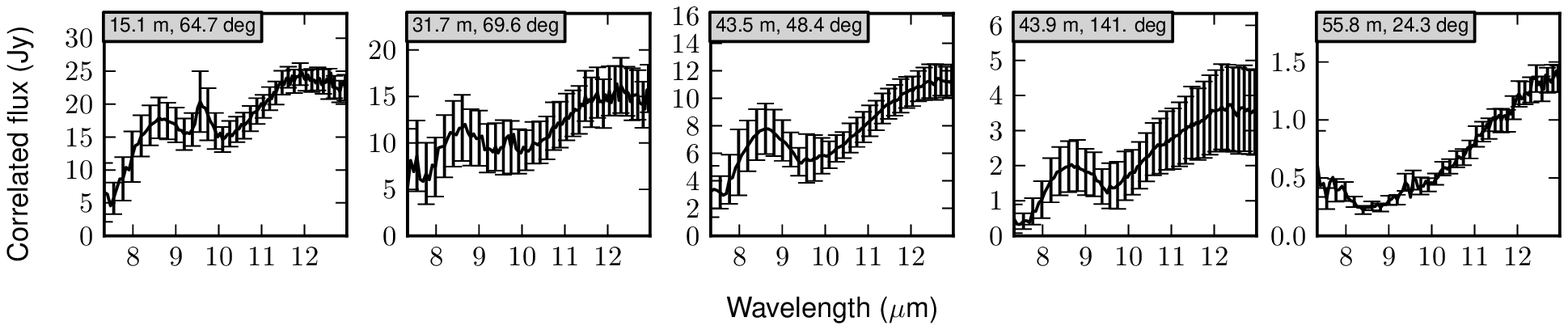}
    \caption{\cfluxcap{M17~SW~IRS1}}
    \label{fig_cflux_m17swirs1}
  \end{center}
\end{figure*}

\begin{figure*}
  \begin{center}
    \includegraphics[scale=0.85,bb=0 670 143 793]{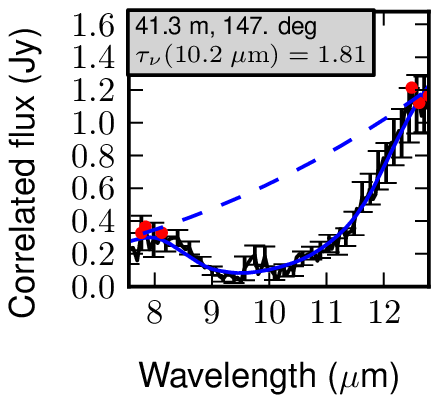}
    \caption{\cfluxcapfit{M17~UC1}}
    \label{fig_cflux_m17uc1}
  \end{center}
\end{figure*}

\begin{figure*}
  \begin{center}
    \includegraphics[scale=0.85,bb=0 670 570 793]{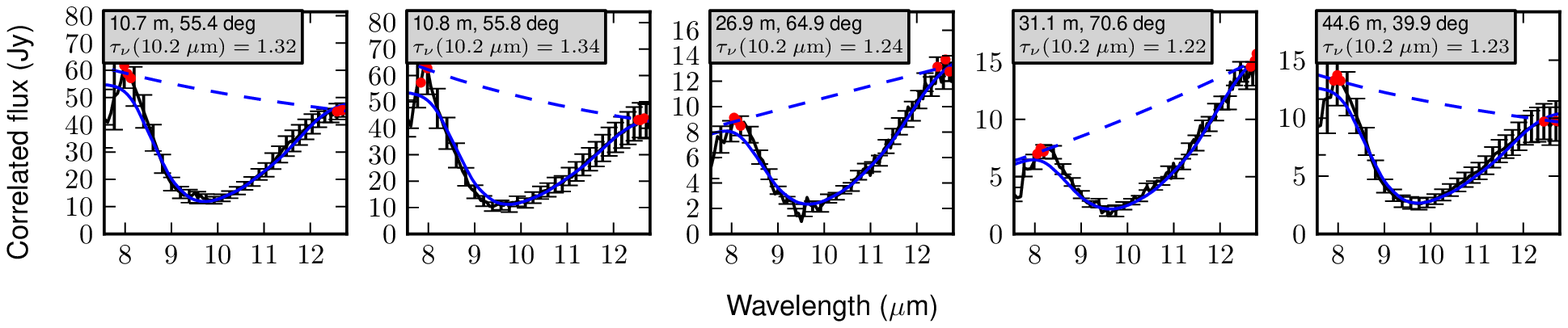}
    \caption{\cfluxcapfit{Mon~R2~IRS2}}
    \label{fig_cflux_monr2irs2}
  \end{center}
\end{figure*}

\begin{figure*}
  \begin{center}
    \includegraphics[scale=0.85,bb=0 670 570 793]{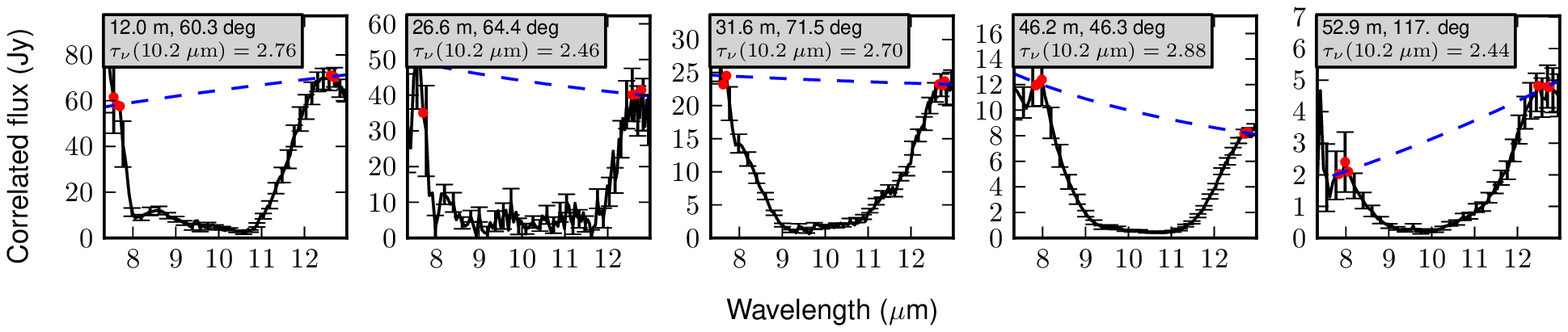}
    \caption{\cfluxcapcont{Mon~R2~IRS3~A}}
    \label{fig_cflux_monr2irs3a}
  \end{center}
\end{figure*}

\begin{figure*}
  \begin{center}
    \includegraphics[scale=0.85,bb=0 670 252 793]{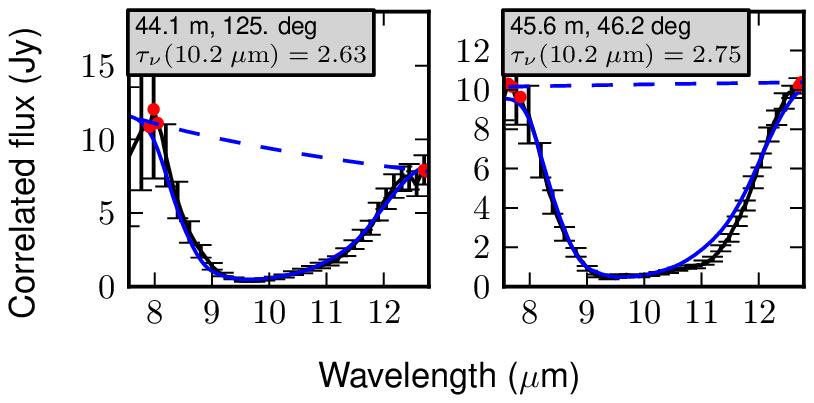}
    \caption{\cfluxcapfit{Mon~R2~IRS3~B}}
    \label{fig_cflux_monr2irs3b}
  \end{center}
\end{figure*}

\begin{figure*}
  \begin{center}
    \includegraphics[scale=0.85,bb=0 386 570 793]{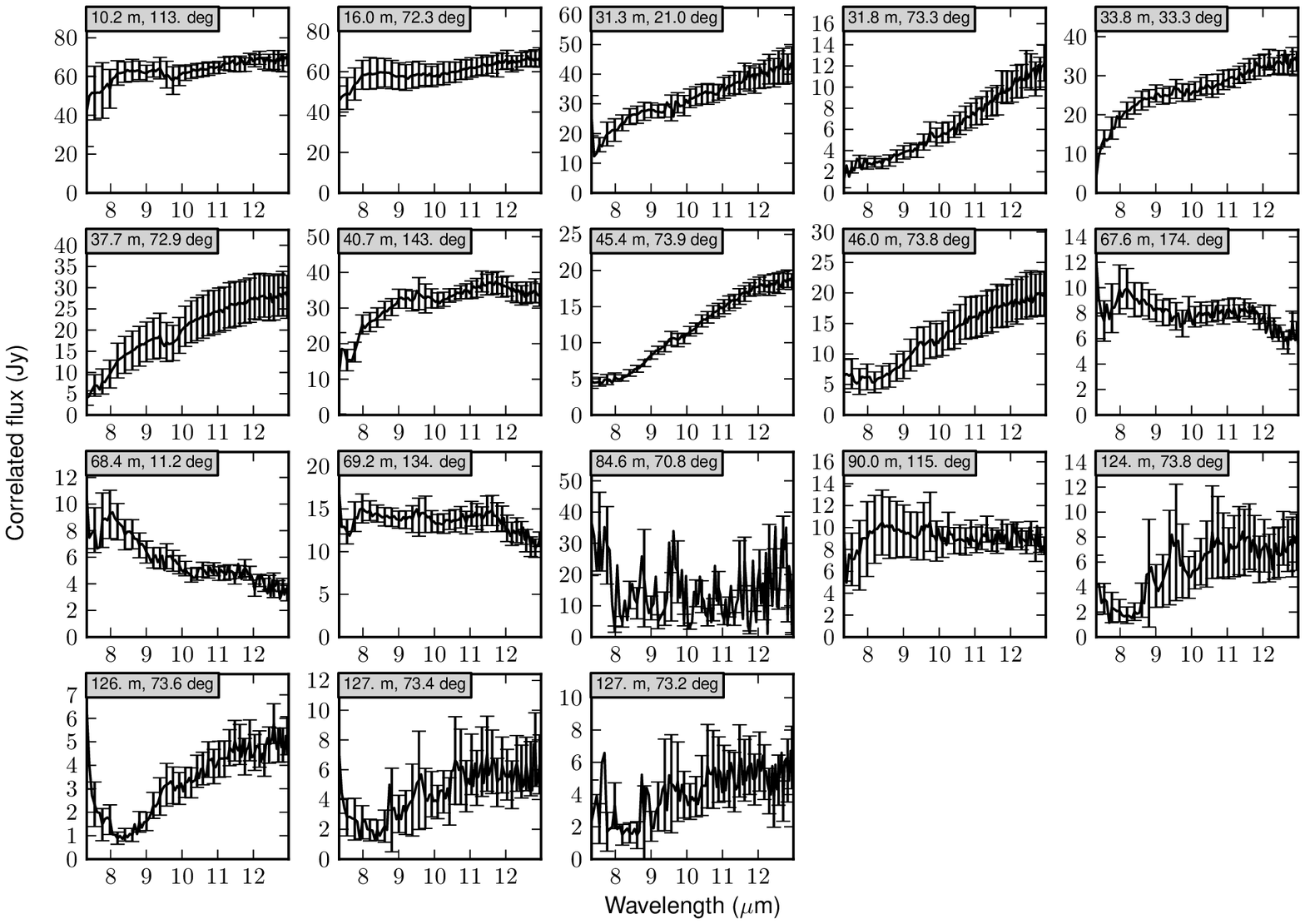}
    \caption{\cfluxcap{MWC~300}}
    \label{fig_cflux_hip90617}
  \end{center}
\end{figure*}

\begin{figure*}
  \begin{center}
    \includegraphics[scale=0.85,bb=0 670 143 793]{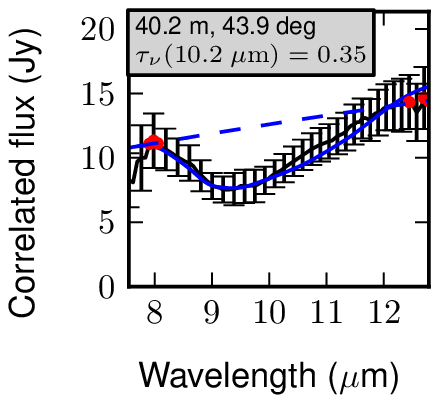}
    \caption{\cfluxcapfit{NGC~2264~IRS1}}
    \label{fig_cflux_ngc2264irs1}
  \end{center}
\end{figure*}

\begin{figure*}
  \begin{center}
    \includegraphics[scale=0.85,bb=0 200 570 793]{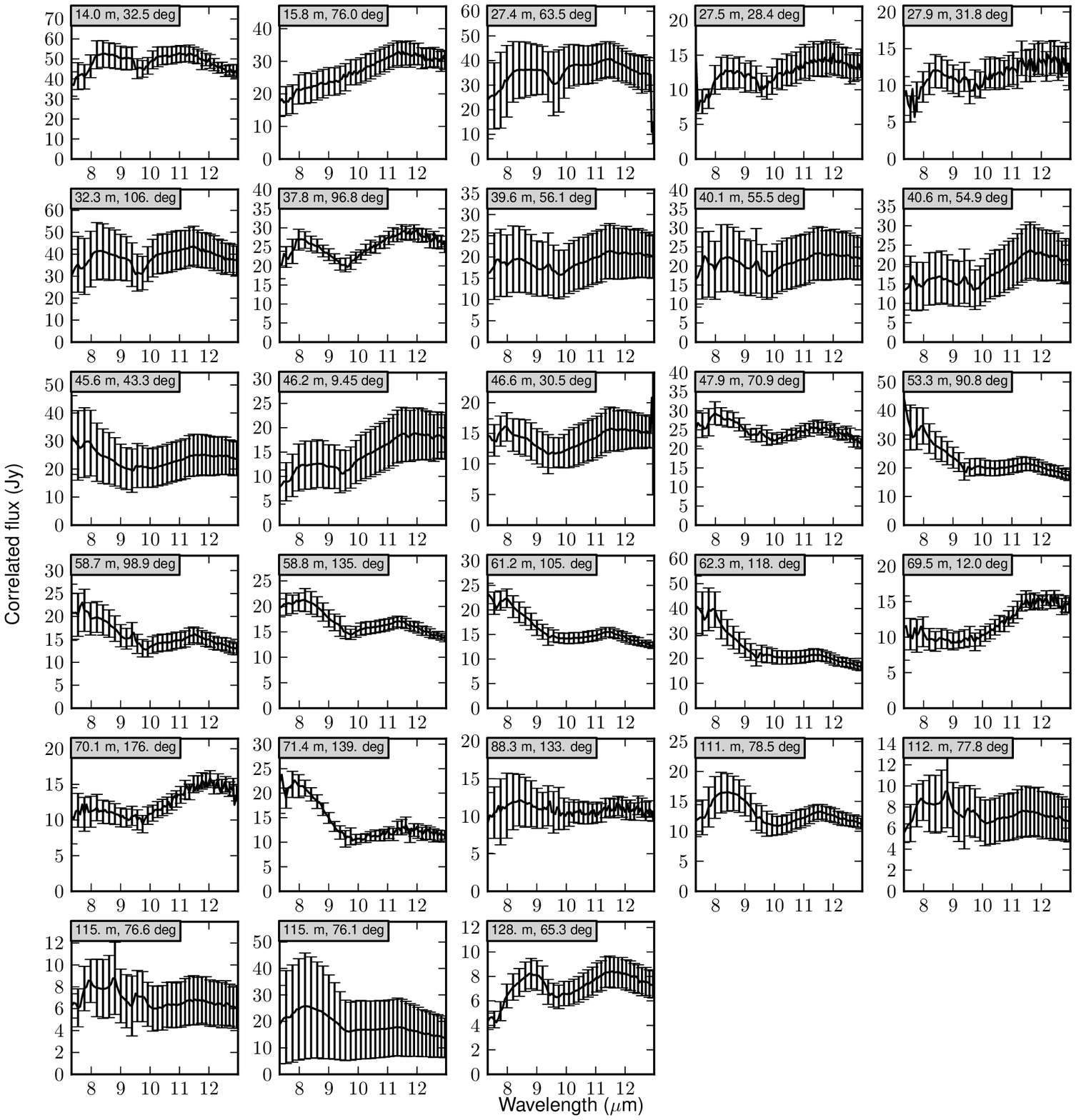}
    \caption{\cfluxcap{R~CrA}}
    \label{fig_cflux_rcra}
  \end{center}
\end{figure*}

\begin{figure*}
  \begin{center}
    \includegraphics[scale=0.85,bb=0 570 470 793]{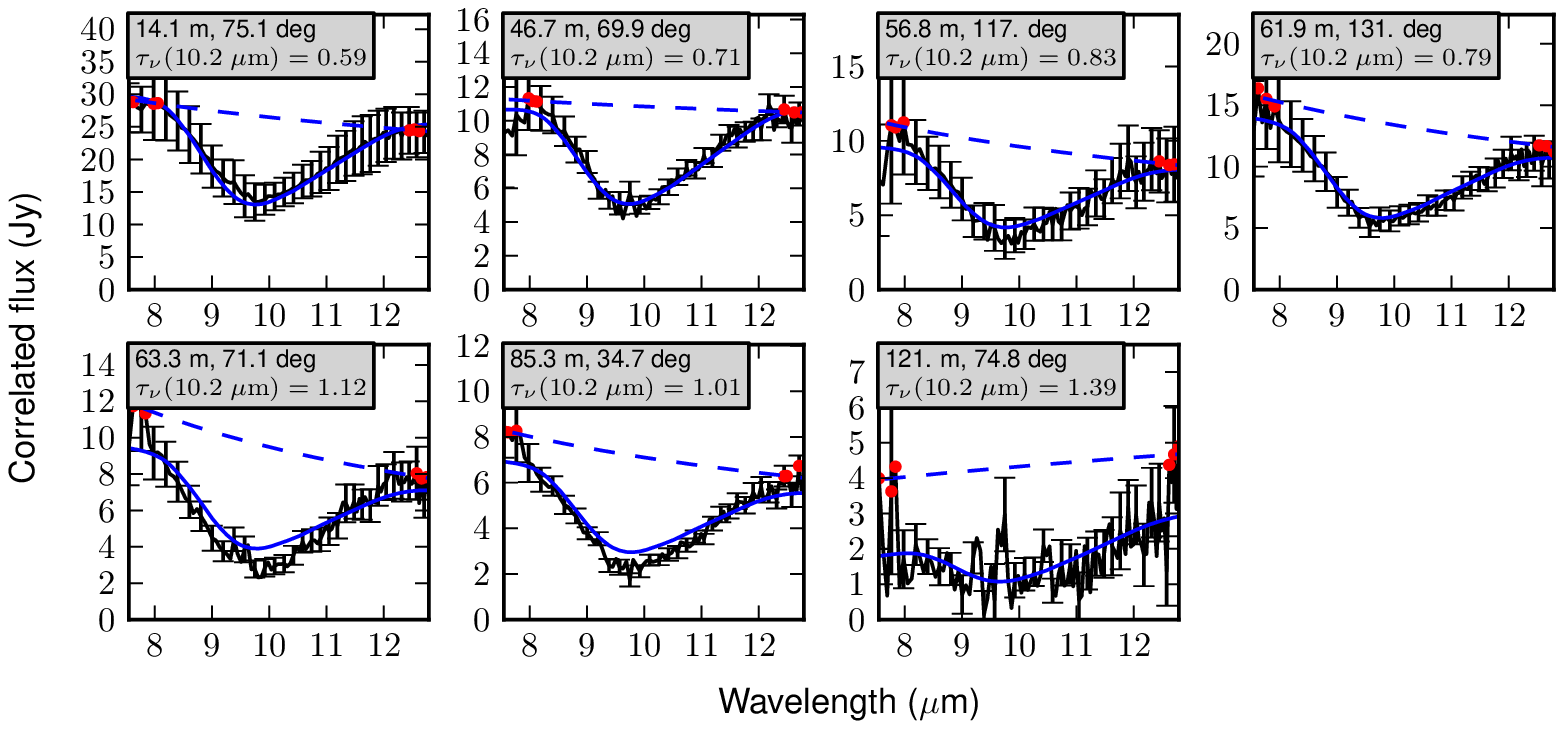}
    \caption{\cfluxcapfit{R~Mon}}
    \label{fig_cflux_rmon}
  \end{center}
\end{figure*}

\begin{figure*}
  \begin{center}
    \includegraphics[scale=0.85,bb=0 570 367 793]{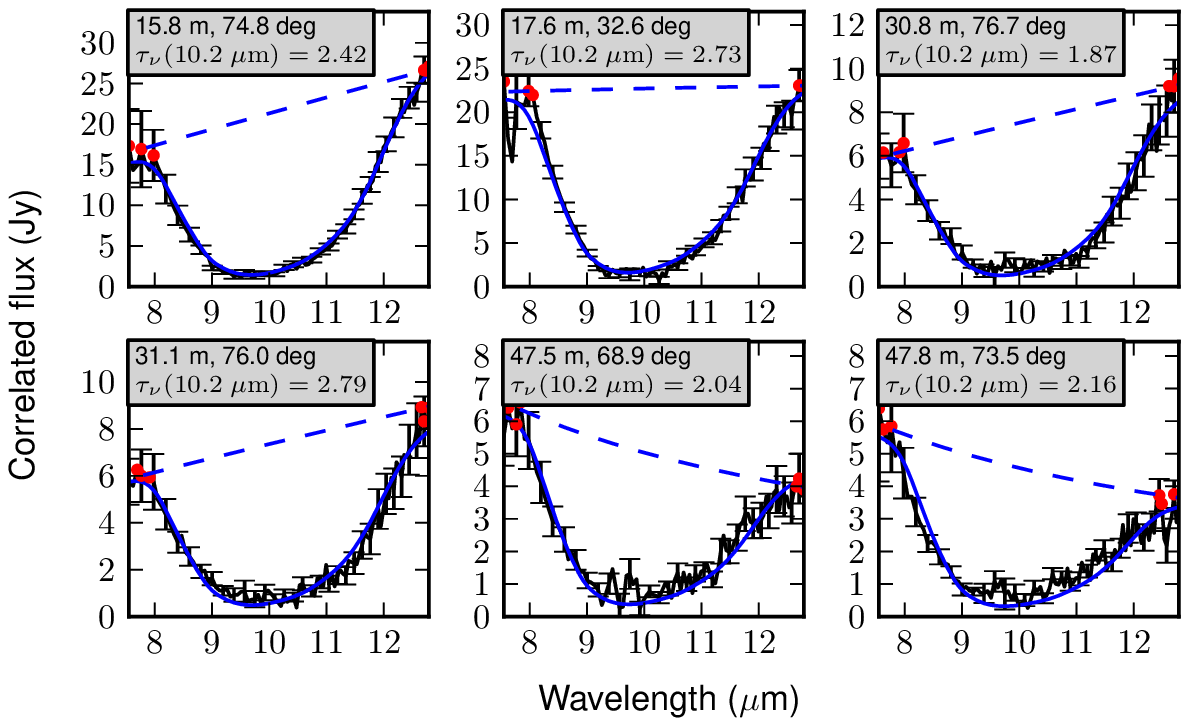}
    \caption{\cfluxcapfit{S255~IRS3}}
    \label{fig_cflux_s255irs3}
  \end{center}
\end{figure*}

\begin{figure*}
  \begin{center}
    \includegraphics[scale=0.85,bb=0 200 570 793]{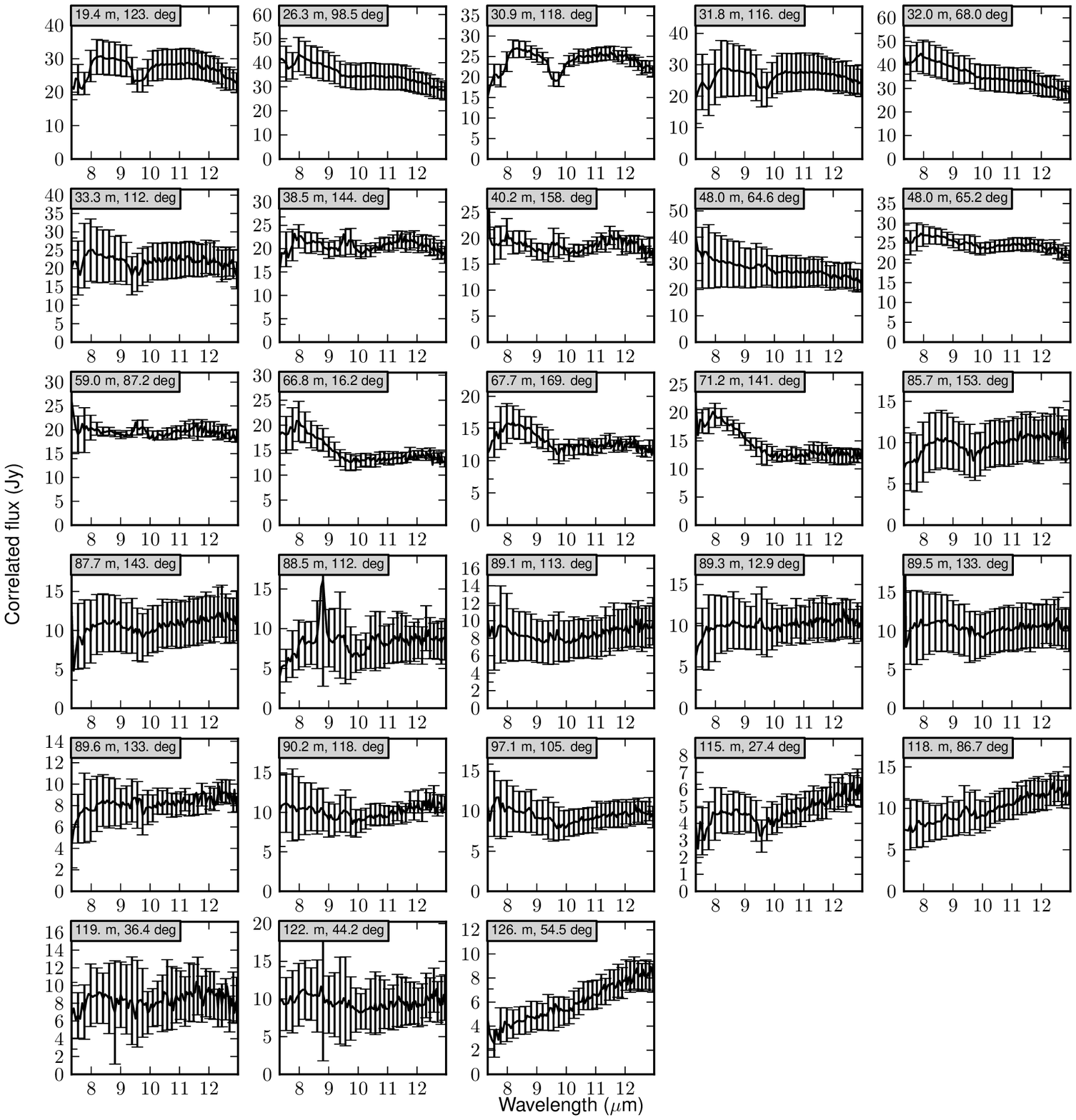}
    \caption{\cfluxcap{V921~Sco}}
    \label{fig_cflux_v921sco}
  \end{center}
\end{figure*}

\begin{figure*}
  \begin{center}
    \includegraphics[scale=0.85,bb=0 386 570 793]{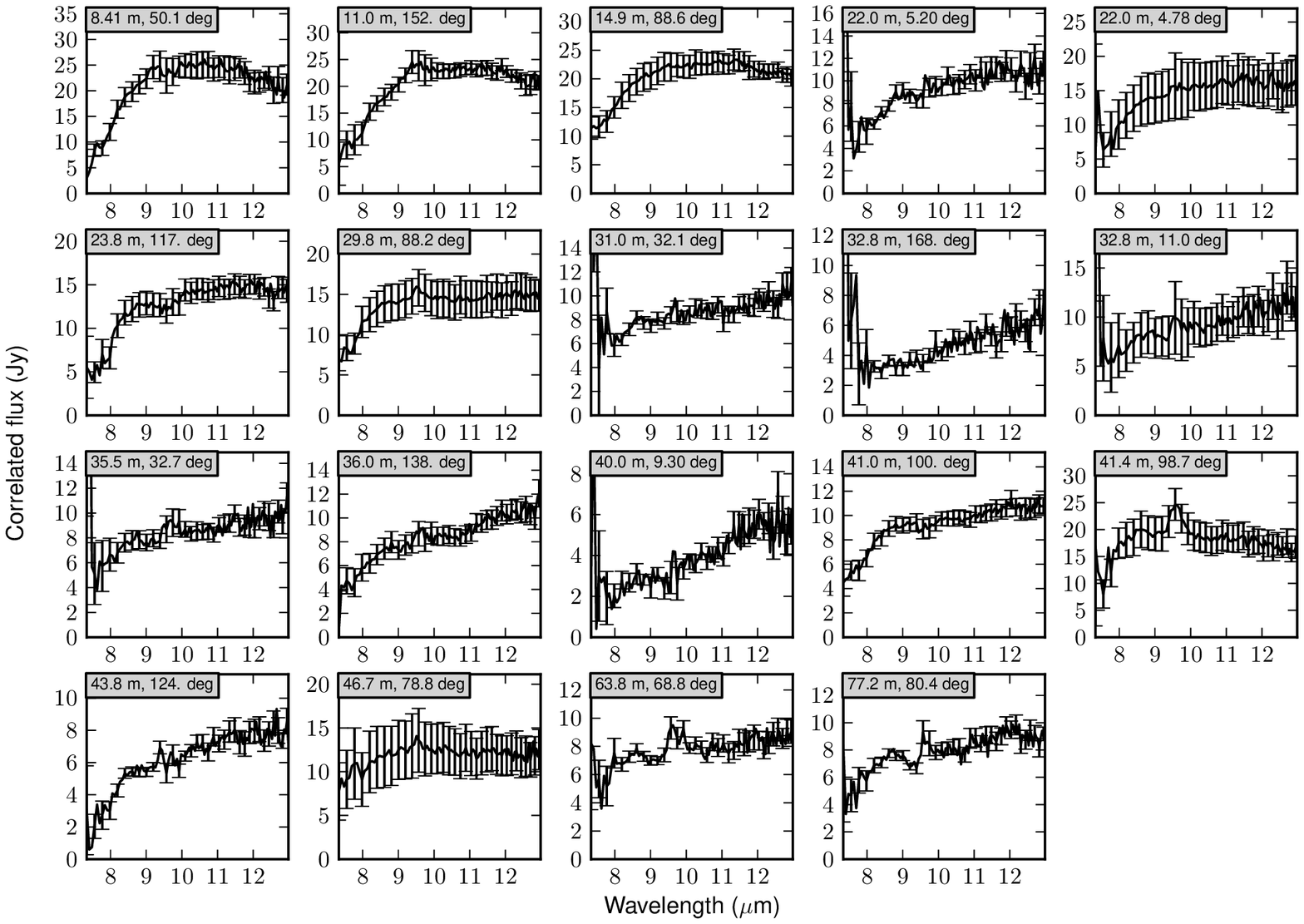}
    \caption{\cfluxcap{V1028~Cen}}
    \label{fig_cflux_v1028cen}
  \end{center}
\end{figure*}

\begin{figure*}
  \begin{center}
    \includegraphics[scale=0.85,bb=0 570 470 793]{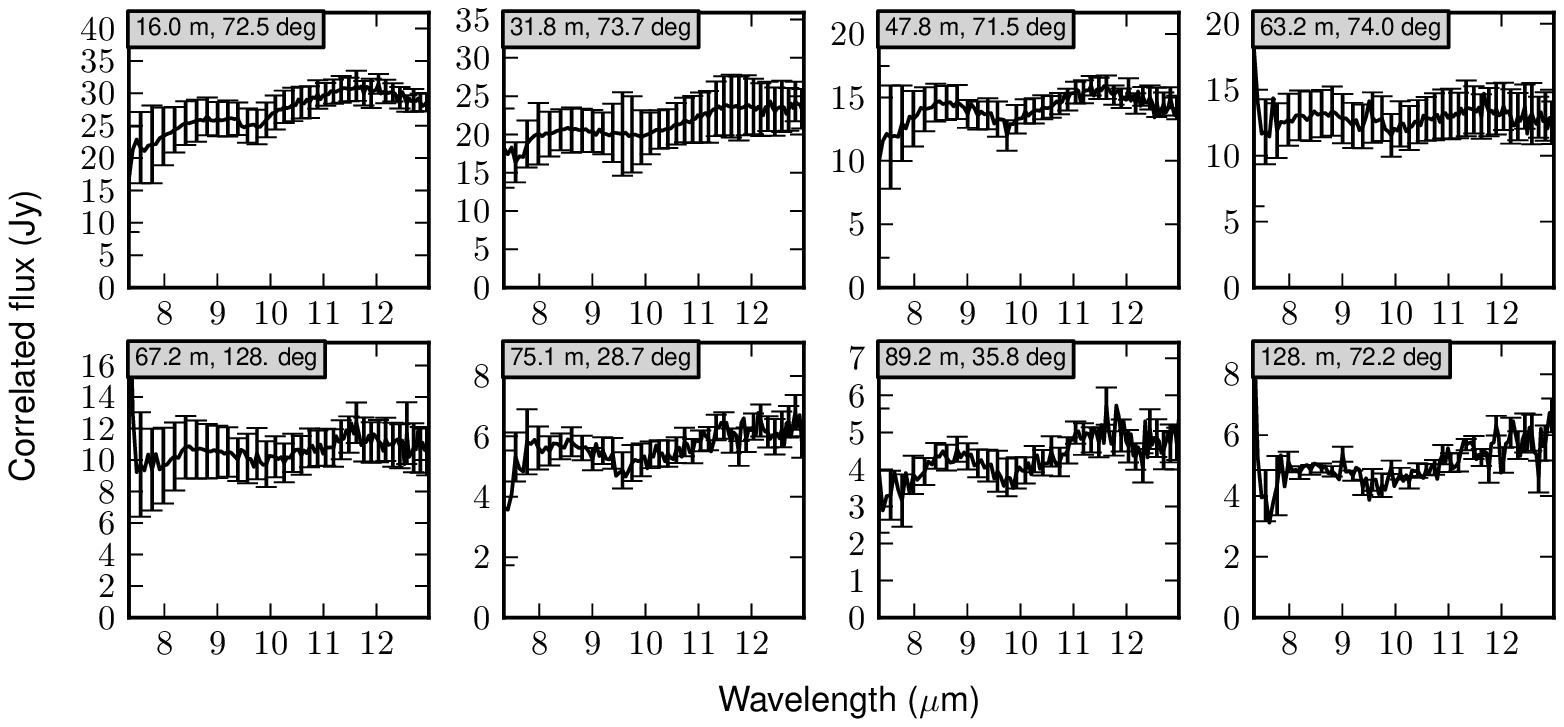}
    \caption{\cfluxcap{VY~Mon}}
    \label{fig_cflux_vymon}
  \end{center}
\end{figure*}
\end{appendix}

\end{document}